\documentclass[aps,prd,twocolumn,showpacs,nofootinbib,floatfix,superscriptaddress]{revtex4-1}

\newcommand{\Rmnum}[1]{\expandafter\@slowromancap\romannumeral #1@}
\makeatother
\usepackage[colorlinks,linkcolor=blue,anchorcolor=blue,citecolor=blue,urlcolor=blue,breaklinks=true]{hyperref}

\usepackage[font=normal]{subfig}
\usepackage{epsfig}
\usepackage{natbib}
\usepackage{epstopdf}
\usepackage{mathrsfs}
\usepackage{slashed}
\usepackage{amsmath}
\usepackage{verbatim}
\usepackage{graphicx}
\usepackage{amssymb}
\usepackage{psfrag}
\usepackage{array}
\usepackage{subfig}

\usepackage{caption}
\usepackage{soul}

\newcommand{\bef}{\begin{figure}[htb]}
\newcommand{\eef}{\end{figure}}

\def\bea#1\eea{\begin{align}#1\end{align}}

\begin{document}
\title{Parton Energy Loss and the Generalized Jet Transport Coefficient}

\author{Yuan-Yuan Zhang}
\affiliation{Key Laboratory of Quark and Lepton Physics (MOE) and Institute of Particle Physics, Central China Normal University, Wuhan 430079, China}
\affiliation{Nuclear Science Division, Lawrence Berkeley National Laboratory, Berkeley, CA 94720, USA}

\author{Guang-You Qin}
\affiliation{Key Laboratory of Quark and Lepton Physics (MOE) and Institute of Particle Physics, Central China Normal University, Wuhan 430079, China}

\author{Xin-Nian Wang}
\email{xnwang@lbl.gov}
\affiliation{Key Laboratory of Quark and Lepton Physics (MOE) and Institute of Particle Physics, Central China Normal University, Wuhan 430079, China}
\affiliation{Nuclear Science Division, Lawrence Berkeley National Laboratory, Berkeley, CA 94720, USA}

\begin{abstract}
We revisit radiative parton energy loss in deeply inelastic scattering (DIS) off a large nucleus within the perturbative QCD approach. We calculate the gluon radiation spectra induced by double parton scattering in DIS without collinear expansion in the transverse momentum of initial gluons as in the original high-twist approach.  The final radiative gluon spectrum can be expressed in terms of the  convolution of hard partonic parts and unintegrated or transverse momentum dependent (TMD) quark-gluon correlations. The TMD quark-gluon correlation can be factorized approximately as a product of initial quark distribution and TMD gluon distribution which can be used to define the generalized or TMD jet transport coefficient. Under the static scattering center and soft radiative gluon approximation,  we recover the result by Gylassy-Levai-Vitev (GLV) in the first order of the opacity expansion. The difference as a result of the soft radiative gluon approximation is investigated numerically under the static scattering center approximation.
\end{abstract}
\maketitle

\section{Introduction}

In high-energy heavy-ion collisions, an energetic parton will undergo multiple scattering in hot quark gluon plasma (QGP) and loses energy along its path. The parton energy loss will lead to the suppression of final energetic jets \cite{Gyulassy:1990ye} and large transverse momentum hadrons \cite{Wang:1991xy} in heavy-ion collisions as compared to proton-proton collisions. This phenomenon known as jet quenching has been observed in experiments at the Relativistic Heavy-ion Collider (RHIC) \cite{Adcox:2001jp,Adler:2002xw} and the Large Hadron Collider (LHC) \cite{Aad:2010bu,Chatrchyan:2011sx,Aamodt:2010jd} and has been used to extract properties of the QGP that is produced in high-energy heavy-ion collisions \cite{Burke:2013yra}. Similar processes of multiple parton scattering and parton energy loss also occur in deeply inelastic scattering (DIS) off a large nucleus. The phenomenon can also be used to study properties of cold nuclear matter as probed by energetic quarks \cite{Wang:2002ri,Arleo:2003jz}. For recent reviews on jet quenching theory and phenomenology see Refs.~\cite{Gyulassy:2003mc,Kovner:2003zj,Majumder:2010qh,Qin:2015srf}.

Since the first attempt to calculate radiative energy loss for a propagating parton in a dense QCD medium \cite{Gyulassy:1993hr}, several studies based on perturbative QCD (pQCD) have been carried out to calculate radiative parton energy loss induced by multiple scattering. The studies by Baier-Dokshitzer-Mueller-Peigne-Schiff and Zakharov (BDMPS-Z)~\cite{Zakharov:1996fv,Baier:1996kr,Baier:1996sk} consider soft gluon radiation as a result of multiple scatterings while  Gyulassy-Levai-Vitev (GLV) and Wiedemann ~\cite{Gyulassy:1999zd,Gyulassy:2000fs,Wiedemann:2000za} assumed the leading order in the opacity expansion for medium-induced gluon radiation. Both of these studies assume the medium as a series of static scattering centers as in the Gyulassy-Wang (GW) model \cite{Gyulassy:1993hr}. Arnold, Moore and Yaffe (AMY)~\cite{Arnold:2001ba,Arnold:2002ja}  employed the hard thermal loop improved pQCD at finite temperature to calculate the scattering and gluon radiation rate in a weakly coupled thermal QGP medium.  The high-twist (HT) approach~\cite{Guo:2000nz,Wang:2001ifa,Zhang:2003yn,Schafer:2007xh} uses the twist-expansion technique in a collinear factorized formalism in which information of the medium is embedded in the high-twist parton correlation matrix elements. In the latest SCET\textsubscript{G} formalism \cite{Ovanesyan:2011xy,Ovanesyan:2011kn}, the standard soft collinear effective theory (SCET) is supplemented with Glauber modes of gluon exchange for parton interaction between a fast parton and static scattering centers to study multiple parton scattering and medium-induced gluon splitting. The relations between some of the above different studies of parton propagation and energy loss have been discussed in detail in Refs.~\cite{Arnold:2008iy,CaronHuot:2010bp,Mehtar-Tani:2019tvy} and numerically compared in Ref.~\cite{Armesto:2011ht}. 

In most of these approaches to parton propagation and energy loss, there are several common approximations. Under the eikonal approximation, energy of the propagating parton $E$ and radiated gluon's energy $\omega$ are considered larger than the transverse momentum transfer $k_{\perp}$ in the scattering, $E,\omega \gg k_{\perp}$.  The energy of a radiative gluon is often considered larger than its transverse momentum $\omega \gg l_{\perp}$ which is known as the approximation of collinear radiation. The mean free path for the propagating parton is assumed larger than the Debye screening length, $\lambda \gg 1 /\mu_D$, which determines the range of interaction in a thermal medium. In addition, a few other approximations, for example  soft radiated gluon approximation $E \gg \omega$ in BDMPS-Z and GLV studies and the large angle approximation $l_{\perp} \gg k_{\perp}$ in the HT approach, are also made in some of the approaches.  Most of the studies except AMY take into account both vacuum and medium-induced radiations and their interference. In BDMPS-Z, GLV study and the SCET\textsubscript{G} approach, the medium is modeled as a collection of static scattering centers. Interactions between the propagating parton and medium, therefore, do not involve energy and longitudinal momentum transfer. In these approaches, the elastic scattering, the radiative processes and the corresponding energy loss are calculated separately. Attempts have been made to improve these theoretical approaches. For example,  GLV calculation has been extended beyond soft radiation approximation \cite{Blagojevic:2018nve} and with a dynamic medium through the hard thermal loop resummed gluon propagator \cite{Djordjevic:2008iz} and beyond first order in opacity expansion \cite{Sievert:2019cwq}. The HT approach has been extended to include longitudinal momentum diffusion \cite{Majumder:2009ge,Qin:2014mya}. Further improvements such as effects of color (de)coherence, angular order \cite{MehtarTani:2011tz,Armesto:2011ir,Caucal:2018dla}  and overlapping formation time in sequential gluon emissions \cite{Arnold:2015qya} have also been studied. 

In the HT formalism~\cite{Guo:2000nz,Wang:2001ifa,Zhang:2003yn,Schafer:2007xh}, the collinear expansion of the hard partonic part 
in the transverse momentum $k_\perp$  of the initial gluon requires $l_{\perp} \gg  k_{\perp}$. Here $\vec{l}_{\perp}$ denotes the transverse momentum of the radiative gluon while $\vec{k}_{\perp}$ is the transverse momentum of the initial gluon or transverse momentum transfer carried by the gluon exchange in the parton-medium scattering. The scattering and radiation amplitudes are then factorized. The initial transverse momentum $k_\perp$ can be integrated, giving rise to the collinear factorized parton distributions and correlations. In this study, we will consider multiple parton scattering and medium-induced gluon radiation in DIS off a large nucleus without collinear expansion in the transverse momentum of the  initial or exchanged gluons. The gluon radiation spectrum due to multiple parton scattering can be expressed in terms of hard partonic parts and the unintegrated or transverse momentum dependent (TMD) quark-gluon correlation functions. The dynamic picture of the parton-medium interaction emerges explicitly with the energy and longitudinal momentum exchange between the propagating parton and medium. We denote this study as the generalized high-twist (GHT) study in order to relate to the original HT formalism \cite{Guo:2000nz,Wang:2001ifa,Zhang:2003yn,Schafer:2007xh} even though the concept of twist expansion in this TMD approach is no longer valid. Since only double and triple parton scattering amplitudes are considered, this is very similar to the leading order contribution of the opacity expansion in the GLV study. We will study the similarity and difference between GLV result and ours. We will show that under soft gluon radiation and static scattering center approximations, we can recover the GLV results. We also study numerically the effect of the soft gluon radiation and static scattering center approximations. During the study presented in this paper, a similar effort in extending the HT approach to a dynamic medium has been completed in Refs.~\cite{Zhang:2018kkn,Zhang:2018nie}. This study assumes a static Yukawa potential model for an exchange gluon field with longitudinal and transverse momentum transfer. Using the light cone expression for four momentum transfer, they assume the minus component of four momentum transfer is of the same order as transverse component, and much larger than the plus component. In our current study, we follow the high-twist approach and assume the plus component and transverse component are much larger than the minus component. We describe the gluon field from the nucleus in terms of a general TMD quark-gluon correlation function which can be reduced to the same result  ( Eq. (49) in Ref ~\cite{Zhang:2018kkn} ) with the static potential assumption.

The remainder of the paper is organized as follows. In Section~\ref{sec-leadingtwist-notations}, we lay out notations and conventions using the single scattering in DIS as an example.  The calculation of the radiative gluon spectrum induced by multiple parton scattering is described in Section~\ref{medium-induced} with details for one example diagram. The full results of a complete list of diagrams are provided in Appendix~\ref{append-spectrum}. We also show how to calculate the radiative gluon spectrum using helicity amplitude method with the soft gluon approximation in Appendix~\ref{append-helicity}. The relation between the unintegrated gluon distribution function $\phi(x,\vec{k}_{\perp})$ and transverse momentum dependent (TMD) jet transport parameter $\hat{q}(\vec{k}_{\perp})$ is also discussed. In Section \ref{sec-approx}, we discuss the results on radiative gluon spectrum in our study under various approximations and compare to the result from the GLV calculation. A summary and some further remarks are presented in Section \ref{sec-sum}.

\section{Single Scattering}
\label{sec-leadingtwist-notations}
The cross section of unpolarized semi-inclusive DIS (SIDIS) process,  
\begin{equation}
e(l_1) + A(p) \rightarrow e(l_2) + h(l_h) +\mathcal{Z},
\end{equation}
as shown in Fig.~\ref{fig:SIDIS}, can be expressed as
\begin{equation}
d \sigma = \frac{e^4}{2s}\frac{\sum_q e_q^2}{q^4}\int \frac{d^4 l_2}{(2\pi)^4} 2\pi \delta(l_2^2) L_{\mu\nu} W^{\mu\nu},
\end{equation}
where the Mandelstam variable $s =  (l_1 + Ap)^2$ is the total invariant center of mass energy squared for the lepton-nucleus system, $p$ is the four-momentum per nucleon in a large nucleus with atomic number $A$, and $q$ is the four-momentum of the intermediate virtual photon. The leptonic tensor is 
\begin{equation}
L_{\mu\nu} = \frac{1}{2}{\rm Tr}[\gamma\cdot l_1 \gamma_{\mu} \gamma\cdot l_2  \gamma_{\nu}  ],
\end{equation}
where 1/2 is the spin average factor of the initial lepton. The unpolarized semi-inclusive hadronic tensor is,
\begin{equation}
\begin{split}
&E_{l_h} \frac{d W^{\mu\nu}}{d^3 l_h} =    \frac{1}{2}  \sum_Z  \langle A| J^{\mu}(0)|\mathcal{Z}, h\rangle \langle h,  \mathcal{Z} |J^{\nu}(0)|A\rangle\\
&\quad \quad  \times (2\pi)\delta^4(Ap-q-p_\mathcal{Z}-l_h) \\
&=\int d^4y e^{-iq\cdot y}  \sum_\mathcal{Z}  \langle A| J^{\mu}(y)|\mathcal{Z}, h \rangle \langle h, \mathcal{Z}|J^{\nu}(0)|A\rangle .
\end{split}
\end{equation}

\bef
  \includegraphics[scale=0.45]{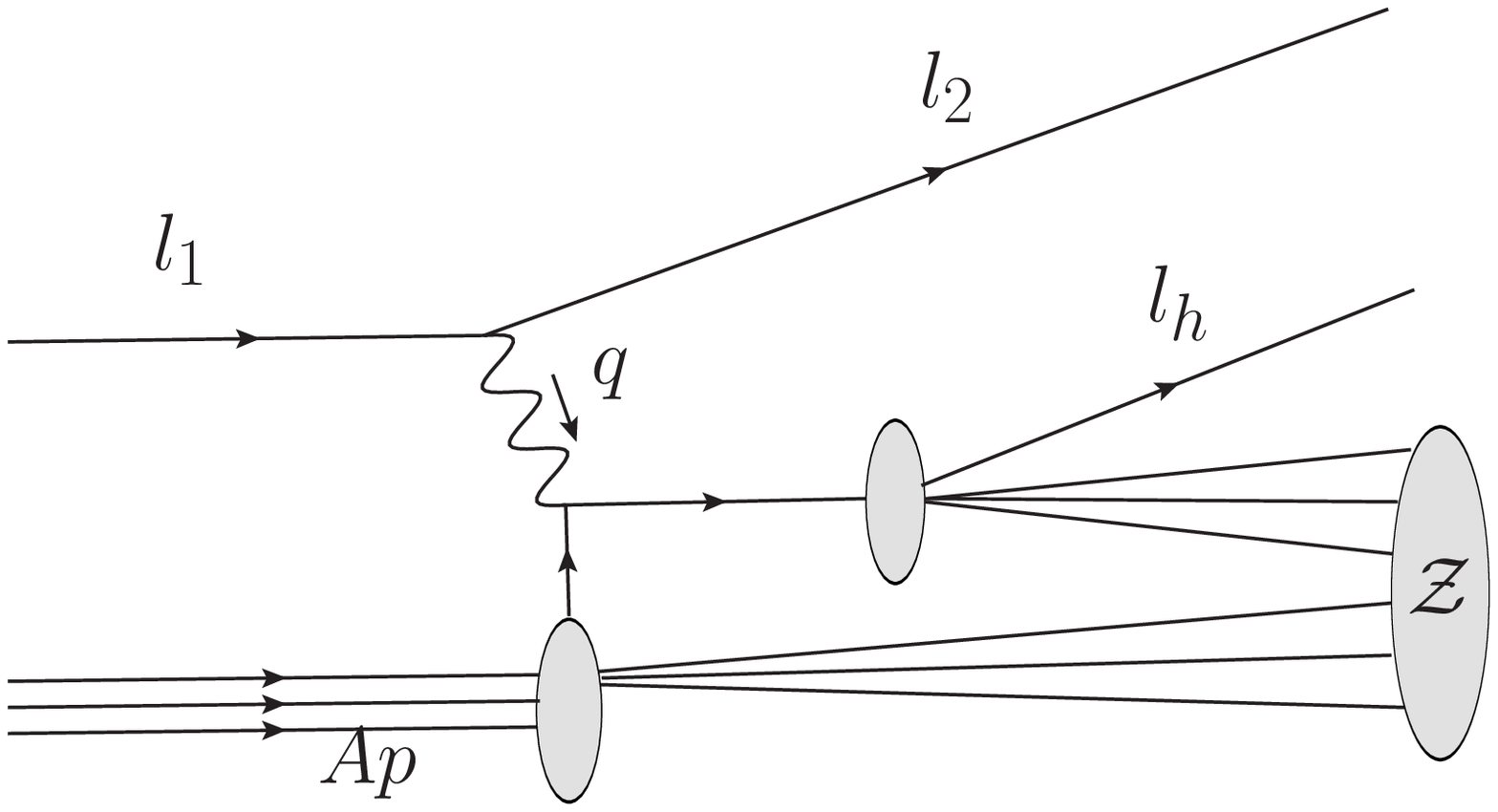}
  \caption{Semi-inclusive DIS process.}
  \label{fig:SIDIS}
\eef

 where the hadronic current is defined as $J^{\mu}(0)=\bar{\psi}_q(0)\gamma^{\mu}\psi_q(0)$. This unpolarized semi-inclusive hadronic tensor, which is also preferred to as the leading-twist hadronic tensor, can be illustrated diagrammatically in Fig.~\ref{fig:LeadingTwist}. 
 
\bef 
  \includegraphics[scale=0.45]{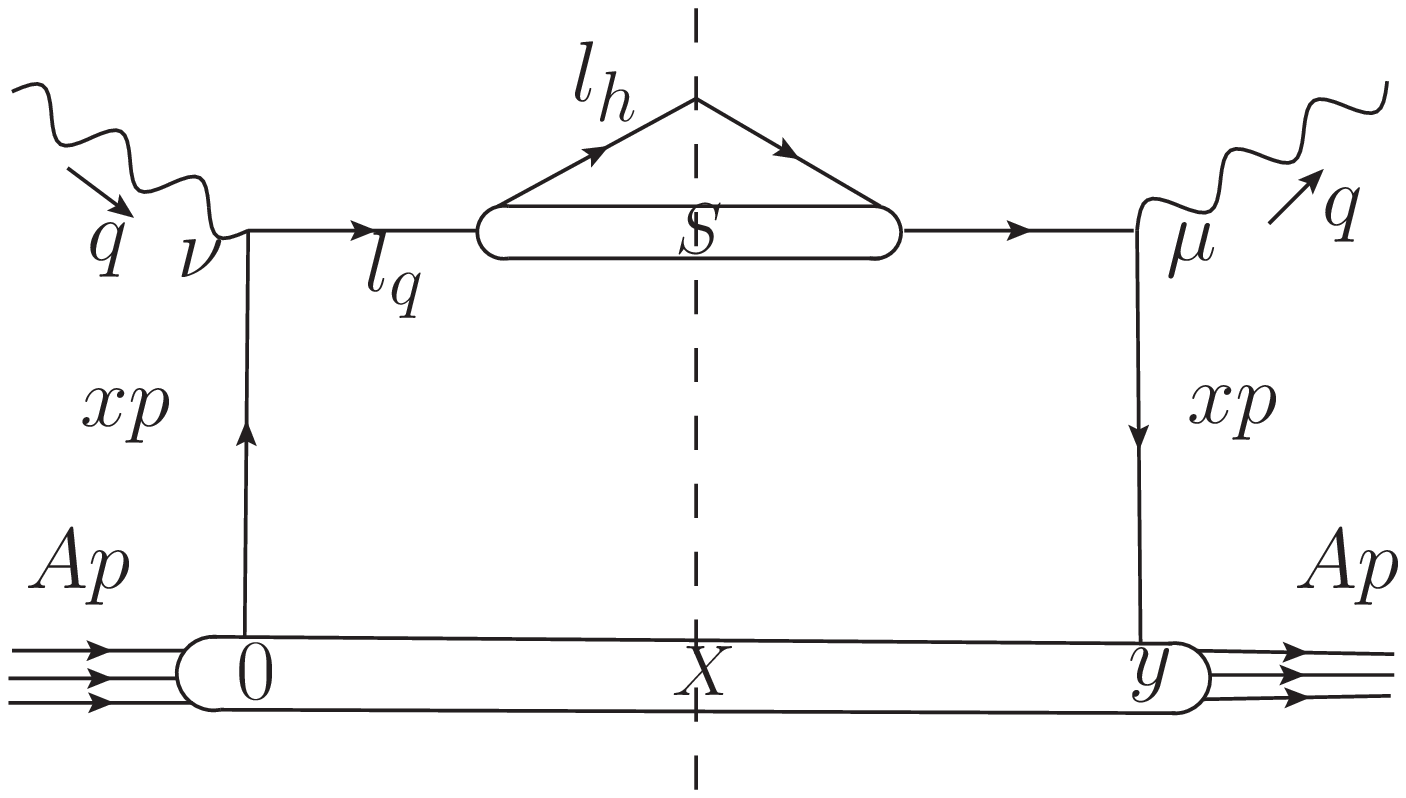}
  \caption{Leading twist hadronic tensor in DIS process.}
  \label{fig:LeadingTwist}
\eef

The intermediate state $\mathcal{Z}$ has two parts, $\mathcal{X}$ and $\mathcal{S}$, where $\mathcal{X}$ represents the spectators in nuclei while $\mathcal{S}$ denotes the remaining hadronic states from the hadronization of the final quark.  The four-momentum of the virtual photon and the initial nucleon are $q=[-Q^2/2q^-, q^-,\vec{0}_{\perp}]$ and $p=[p^+,0,\vec{0}_{\perp}]$, respectively. The Bjorken variable is defined as $x_B = Q^2/2p^+q^-$. The momentum fraction of the struck quark is $x$. The fraction of light-cone momentum carried by the observed hadron with momentum $l_h$ is $z_h = l_h^-/q^-$.  Under the collinear approximation, one can expand the hard partonic part of the $\gamma^*+q$ scattering in the initial transverse momentum of the quark. The leading term of the expansion gives rise to the leading twist  unpolarized semi-inclusive hadronic tensor in a factorized form,
\begin{equation}
\frac{d W^{\mu\nu}_{S(0)}}{d z_h} = \int dx f_q^A(x) H^{\mu\nu}_{(0)}(x) D_{q\rightarrow h}(z_h),
\end{equation}
where the lower index $S(0)$ denotes that the quark originated from the nucleus only undergoes a single scattering with the virtual photon without corrections from the strong interaction. The nuclear quark distribution function is  defined as,
 \begin{equation}
 f_q^A(x) = \int \frac{dy^-}{2\pi} e^{-ixp^+y^-} \frac{1}{2} \langle A|\bar{\psi}_q(y^-)\gamma^+ \psi_q(0)|A \rangle ,
  \end{equation}
  and the definition of the quark fragmentation function is,
 \begin{equation}
 \begin{split}
D_{q \rightarrow h} (z_h) = &\dfrac{z_h^3}{2}  \sum_{S}\int \dfrac{d^4 l_q}{(2\pi)^4} \int d^4 y e^{il_q\cdot y}\\
&\times {\rm Tr}\left[\dfrac{\gamma^{-}}{2l_h^-} \langle 0|\psi(y)|h,\mathcal{S}\rangle \langle \mathcal{S},h|\bar{\psi}(0)|0 \rangle \right]\\
= & \dfrac{z_h}{2}\sum_{S}\int \dfrac{dy^+}{2\pi} e^{il_h^-y^+/z_h}\\ 
&\times {\rm Tr}\left[\dfrac{\gamma^{-}}{2}\langle 0|\psi(y^+)|h,\mathcal{S}\rangle \langle  \mathcal{S},h|\bar{\psi}(0)|0 \rangle\right].
\end{split}
\end{equation}
The hard partonic part is,
\begin{equation}
\begin{split}
H^{\mu\nu}_{(0)} (x)= &\frac{1}{2} {\rm Tr}[\gamma\cdot p \gamma^{\mu} \gamma \cdot (q+xp)\gamma^{\nu}  ] 2\pi \delta[(q+xp)^2] \\
=&  \frac{1}{2} {\rm Tr}[\gamma\cdot p \gamma^{\mu} \gamma \cdot (q+xp)\gamma^{\nu}  ] \frac{2\pi}{2p^+q^-} \delta(x-x_B)
\end{split}
\end{equation}
Soft eikonal gluons attached to the nucleus target and the final state hadrons can be summed as gauge links in the quark distribution function and parton fragmentation function. They are omitted here for brevity of the notation.

The next-to-leading (NLO) order corrections in the strong coupling constant to the fragmentation process in SIDIS are from the final state radiation, as shown in Figs.~\ref{fig:FinalRadiationQ} and  \ref{fig:FinalRadiationG}.

\bef
  \captionsetup{justification=raggedright,singlelinecheck=false}
  \includegraphics[scale=0.45]{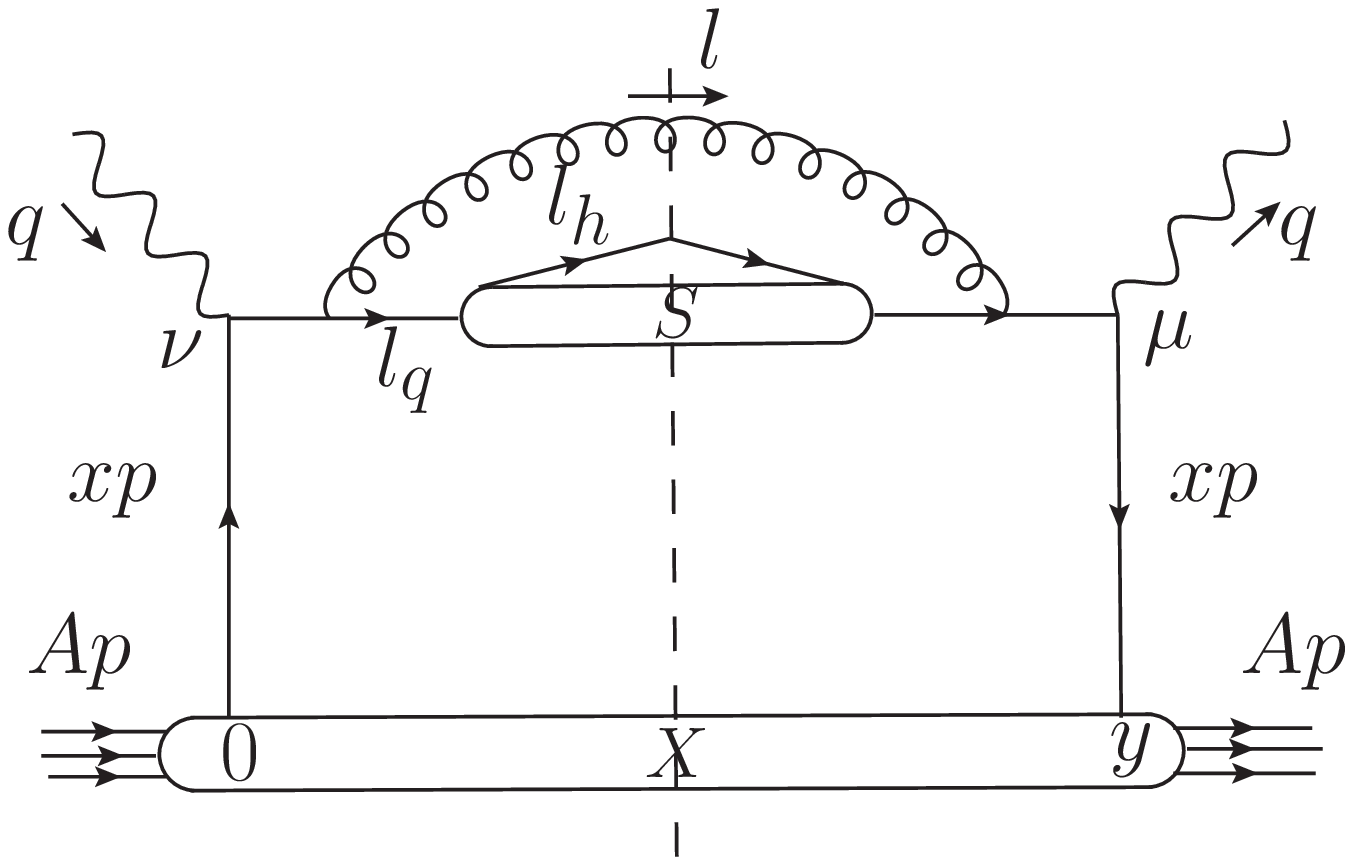}
  \caption{Next-to-leading order contribution to leading twist hadronic tensor with quark fragmentation.}
  \label{fig:FinalRadiationQ}
\eef

If the identified final hadron comes from the quark, as in Fig.~\ref{fig:FinalRadiationQ}, the NLO correction to the hadronic tensor is 
\begin{equation}
\begin{split}
\frac{d W^{\mu\nu}_{S(1)q}}{d z_h} = &\int dx  f_q^A(x)   H^{\mu\nu}_{(0)}(x)  \frac{\alpha_s}{2\pi} C_F    \\
&\times \int_{z_h}^{1}\frac{dz}{z} \int_0^{\mu^2}  \frac{d l_{\perp}^2}{l_{\perp}^2} \frac{1+z^2}{1-z} D_{q\rightarrow h}(z_h/z),
\end{split}
\label{eq:dw1q}
\end{equation}
where the lower index $S(1)q$ denotes the NLO radiative correction to the hadronic tensor from single photon-quark scattering, with the final hadron from the fragmentation of the quark. The fraction of momentum carried by the final state quark $l_q$ is $z = l_q^-/q^-$. The factorization scale is $\mu^2$, which separates the perturbative hard partonic part from the non-perturbative fragmentation processes (fragmentation function).
  
\bef
  \captionsetup{justification=raggedright,singlelinecheck=false}
  \includegraphics[scale=0.45]{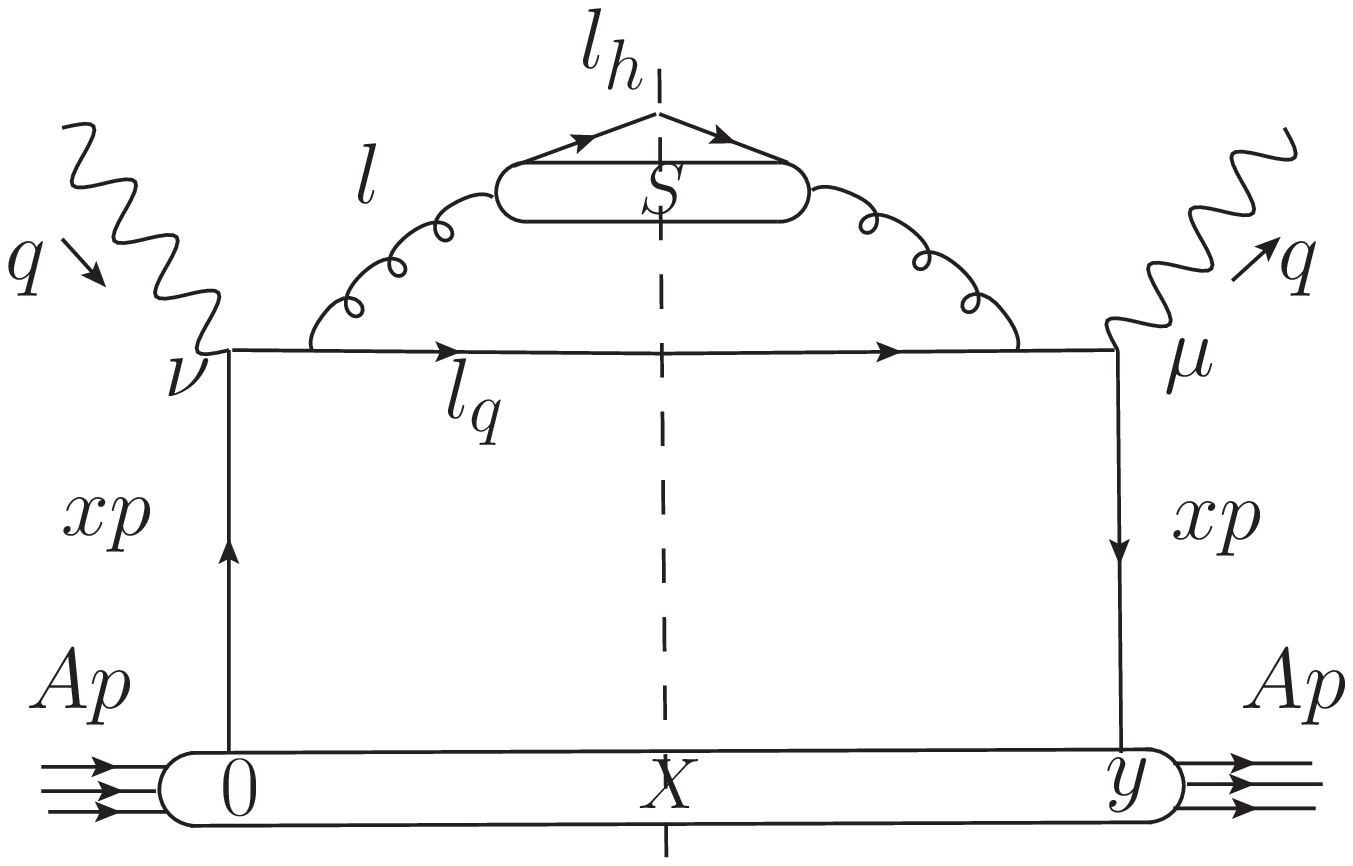}
  \caption{Next-to-leading order contribution to leading twist hadronic tensor with gluon fragmentation.}
   \label{fig:FinalRadiationG}
\eef
 
When the identified final hadron comes from the radiative gluon, as in Fig.~\ref{fig:FinalRadiationG}, the NLO radiative correction to the hadronic tensor is,
\begin{equation}
\begin{split}
\frac{d W^{\mu\nu}_{S(1)g}}{d z_h} = &\int dx  f_q^A(x)   H^{\mu\nu}_{(0)} (x) \frac{\alpha_s}{2\pi}  C_F  \\
& \times  \int_{z_h}^{1} \frac{dz}{z} \int_0^{\mu^2} \frac{d l_{\perp}^2}{l_{\perp}^2} \frac{1+(1-z)^2}{z} D_{g\rightarrow h}(z_h/z).
\end{split}
\end{equation} 
Note that $z =l^-/q^-$ here is the momentum fraction of the radiated gluon and the gluon fragmentation function is
 \begin{equation}
 \begin{split}
 D_{g \rightarrow h} (z_h)  = & \frac{z_h^2}{2} \sum_{S} \int \frac{d^4 l}{(2\pi)^4}   \int d^4 y e^{il\cdot y} \\
 &\times \langle  0|A^{\alpha}(0)|h,\mathcal{S} \rangle  \langle  \mathcal{S},h|A^{\beta}(y)|0\rangle   \epsilon_{\alpha\beta}(l)\\
  = & - \dfrac{z_h^2}{2l_h^-}  \sum_{\mathcal{S}}\int \dfrac{dy^+}{2\pi} e^{il_h^- y^+/z_h}\\
  &\times \langle  0|F^{-\alpha}(y^+)|h,\mathcal{S} \rangle \langle  \mathcal{S},h|F^{-}_{\quad \alpha}(0)|0 \rangle .
 \end{split}
 \end{equation}
 
\bef
\captionsetup{justification=raggedright,singlelinecheck=false}
\includegraphics[width=2.6in]{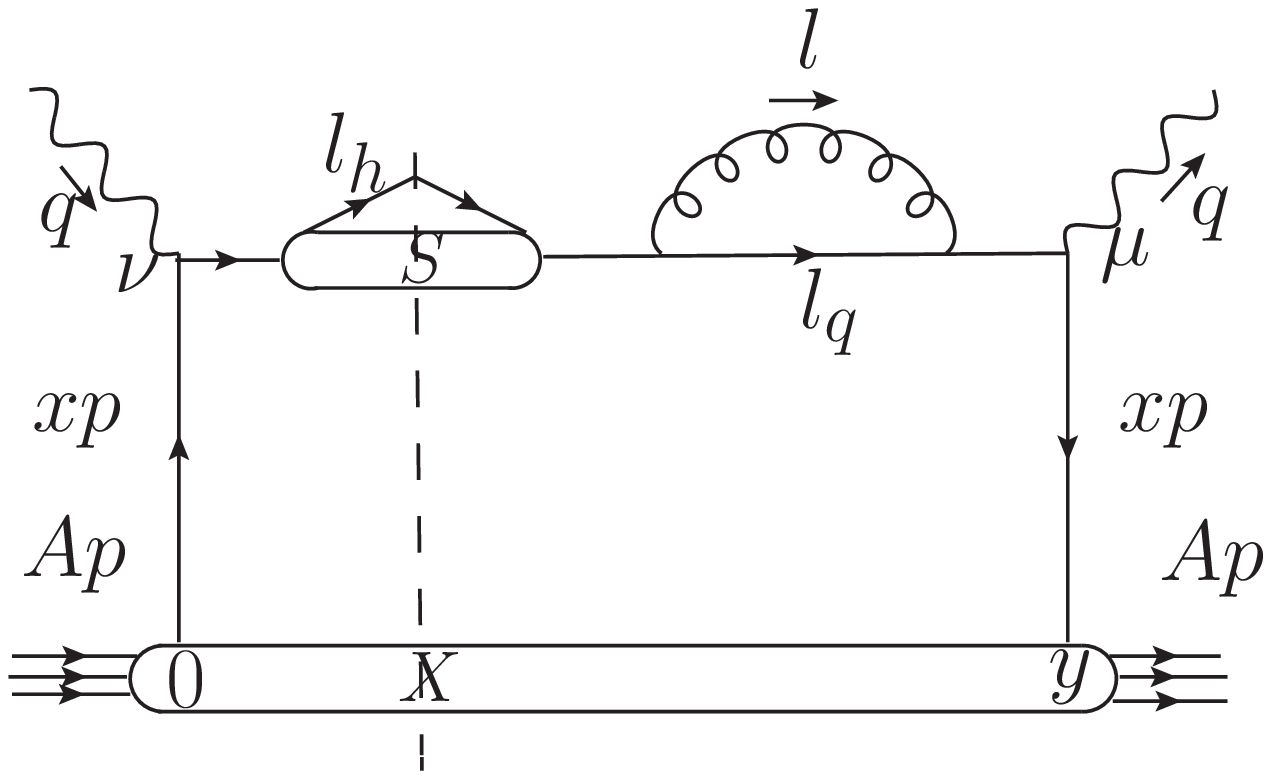} 
\includegraphics[width=2.6in]{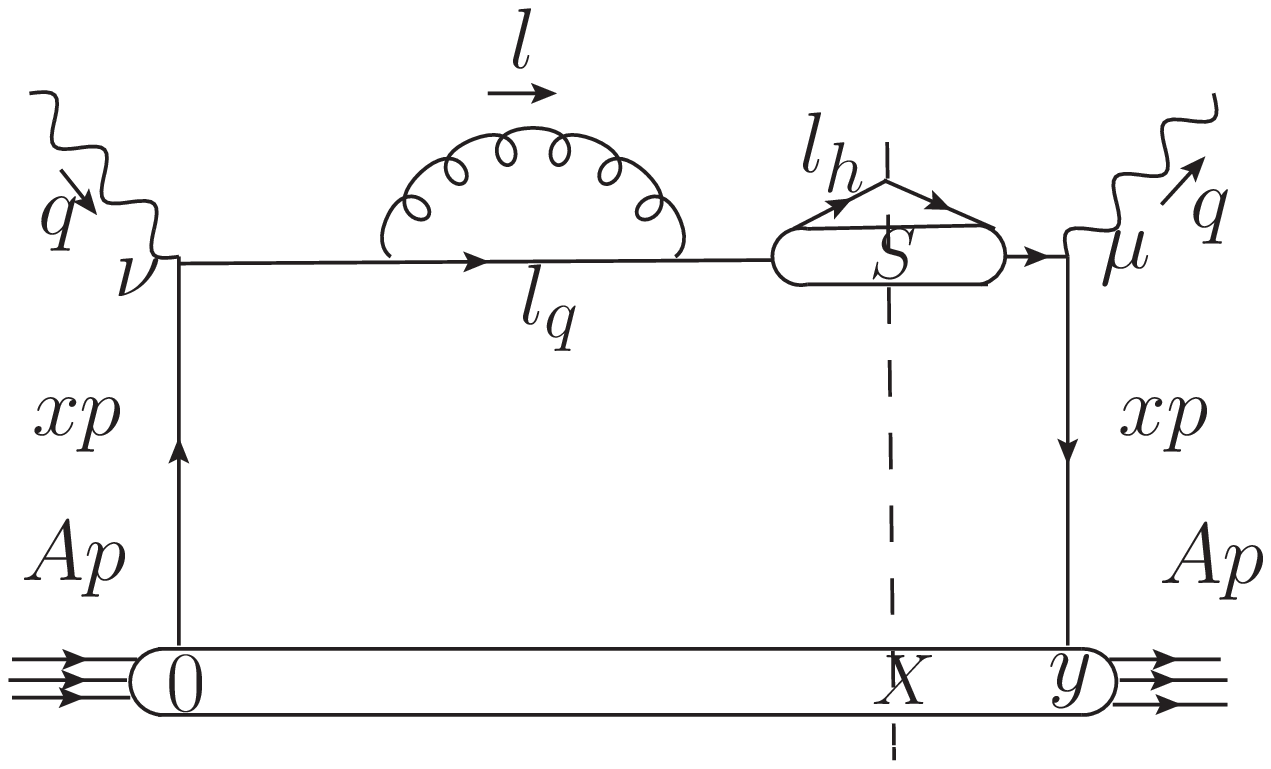} 
\caption{Virtual correction at next-to-leading order.}
\label{fig:Virtual_Single}
\eef

 There are both infrared and collinear divergences in the above radiative corrections to the hadronic tensor of SIDIS. The infrared divergence is in the  splitting function of the hadronic tensor $d W^{\mu\nu}_{S(1)q}/d z_h$ in Eq.~(\ref{eq:dw1q}) for identified hadrons originated from the quark, when the momentum fraction of the final quark $z$ approaches $1$.  To deal with this, one has to include the virtual corrections at NLO to the hadronic tensor as shown in Fig.~\ref{fig:Virtual_Single}.  These virtual corrections from the sum of the two diagrams in Fig.~\ref{fig:Virtual_Single} are
\begin{eqnarray}
\dfrac{dW^{\mu \nu}_{S(v)}}{dz_h}  &=& -\int dx f_q^A(x)   H^{\mu\nu}_{(0)} (x) \dfrac{\alpha_s}{2\pi}  C_F  \nonumber \\
& \times& \int_0^1 dz  \int_0^{\mu^2} \frac{d l_{\perp}^2}{ l_{\perp}^2}\dfrac{1+z^2}{1-z} D_{q \rightarrow h} (z_h)   , 
\end{eqnarray}
where each of the diagram contributes one half. The lower index $S(v)$ denotes the virtual correction at NLO to the hadronic tensor of DIS process which also has both infrared and collinear divergences.

When summed together, the infrared divergences in the radiative and virtual corrections cancel. The remaining collinear divergences can be absorbed into the renormalized fragmentation function $D_{q \rightarrow h} (z_h, \mu^2)$. The leading twist hadronic tensor for SIDIS process, including the final state radiation and virtual correction, can be written as
\begin{equation}
\begin{split}
\dfrac{dW^{\mu \nu}_{S}}{dz_h} =  &\dfrac{dW^{\mu \nu}_{S(0)}}{dz_h}  + \dfrac{dW^{\mu \nu}_{S(1)q}}{dz_h}+\dfrac{dW^{\mu \nu}_{S(1)g}}{dz_h}  +\dfrac{dW^{\mu \nu}_{S(v)}}{dz_h} \\
=& \int dx f_q^A(x)  H^{\mu\nu}_{(0)}  D_{q \rightarrow h} (z_h, \mu^2),
\end{split}
\end{equation}
and the renormalized quark fragmentation function is defined as  
\begin{equation}
\begin{split}
D_{q \rightarrow h} (z_h, \mu^2) &=  D_{q \rightarrow h} (z_h) 
+ \frac{\alpha_s}{2\pi} C_F \int_{z_h}^{1} \frac{dz}{z} \int_0^{\mu^2} \frac{d l_{\perp}^2}{l_{\perp}^2} \\
&\times \left\lbrace  \left[ \frac{1+z^2}{(1-z)_{+}}   + \frac{3}{2}\delta(1-z)     \right] D_{q\rightarrow h}(z_h/z) \right.  \\
&\left. +C_F \frac{1+(1-z)^2}{z}  D_{g\rightarrow h}(z_h/z) \right\rbrace, \\
\end{split}
\end{equation}
which satisfies the DGLAP equation \cite{Gribov:1972ri,Dokshitzer:1977sg,Altarelli:1977zs}.

\section{Medium Induced Gluon Radiation}
\label{medium-induced}

In the process of SIDIS off a nucleus target, the outgoing quark may undergo secondary scatterings with another parton in the nucleus which in turn can induce gluon radiation. Such secondary scatterings are especially important when the initial quark and the second medium parton originate from two different nucleons inside the nucleus. In this case, the corresponding contributions to the hadronic tensor are enhanced by the size of the nucleus $A^{1/3}$. We will only consider contributions with nuclear enhancement and neglect those without, for example, when the initial quark and medium parton are from the same nucleon inside the nucleus. 


\subsection{Double Scattering}
\label{double-scattering}
There are many contributions from gluon radiation induced by double parton scattering to the hadronic tensor of the SIDIS processes. We will focus on the processes that have two gluon exchanges between the propagating quark and the nucleus in the cut diagram. Processes with double quark scattering have been discussed in detail in Ref.~\cite{Schafer:2007xh}. We first illustrate the procedures to calculate the semi-inclusive hadronic tensor from medium induced gluon radiation, using the central cut-diagram in Fig.~\ref{fig:C11w} as an example. Calculations of other cut diagrams are given in Appendix \ref{append-spectrum}. 

\begin{figure}
  \captionsetup{justification=raggedright,singlelinecheck=false}
  \includegraphics[scale=0.4]{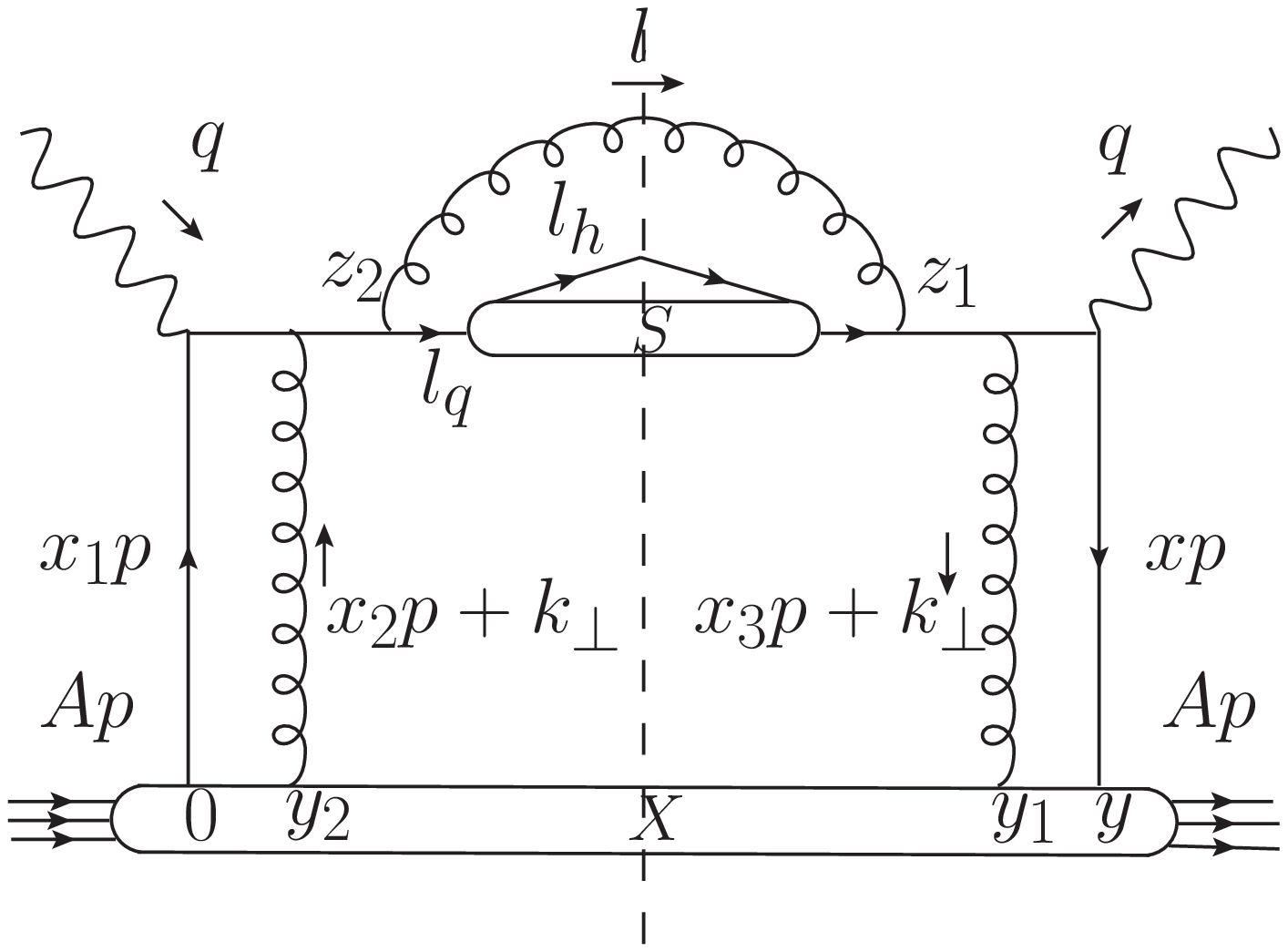}
  \caption{ Central-cut digram of double scattering and induced gluon radiation.}
  \label{fig:C11w}
\end{figure}

We choose the covariant gauge for the gluon field from the beam nucleus while the axial gauge ($A^-=0$) for the final radiated gluon. The semi-inclusive hadronic tensor for the central cut-diagram in Fig.~\ref{fig:C11w} can be written down as, 
\begin{widetext}
\begin{equation}
\begin{split}
W^{\mu \nu {\rm Fig}\ref{fig:C11w}}_{D(1)q}= & \int d^4 y e^{iq\cdot y}  \int d^4 y_1 \int d^4 y_2 \int d^4 z_1 \int d^4 z_2 \int \dfrac{d^4 l}{(2\pi)^4}  2\pi \delta(l^2)  \int \dfrac{d^4 l_h}{(2\pi)^4}  2\pi \delta(l_h^2) \sum_{\mathcal{X},\mathcal{S}} \langle A | \bar{\psi}(y)   \\
&\times \gamma^{\mu} \psi(y)  \bar{\psi}(y_1) (-ig)\gamma^{\sigma} A_{\sigma}(y_1) \psi (y_1) \bar{\psi}(z_1) (-ig)  \gamma^{\alpha}A_{\alpha} \psi(z_1) | l, l_h, \mathcal{S}, p_\mathcal{X}\rangle \langle p_\mathcal{X}, \mathcal{S}, l_h,l | \bar{\psi}(z_2)\\
&\times   (ig) \gamma^{\beta} A_{\beta} \psi(z_2)\bar{\psi}(y_2) (ig)\gamma^{\rho} A_{\rho}(y_2)  \psi (y_2)\bar{\psi}(0) \gamma^{\nu} \psi(0) |A \rangle \frac{{\rm Tr}[t_a t_c t_c t_a]}{N_c (N_c^2-1)},
\end{split}
\end{equation}
where the lower index ``$D(1)q$" denotes radiative corrections to the double scattering process and the identified hadron is from the fragmentation of the final quark. One can carry out the integrations over $z_1$ and $z_2$ which lead to  the energy-momentum conservation at the vertices of gluon radiation. Factoring out the fragmentation function and noticing that the dominant components of the initial gluon field in covariant gauge are $A_{\sigma}(y_1) \approx (p_{\sigma}/p^+) A^+(y_1)$, $A_{\rho}(y_2) \approx (p_{\rho}/p^+) A^+(y_2)$ \cite{Luo:1994np}, the hadronic tensor can be rewritten as,
\begin{equation}
\begin{split}
\frac{d W^{\mu \nu {\rm Fig}\ref{fig:C11w}}_{D(1)q} }{d z_h}= & \int_{z_h}^1 \frac{dz}{z} D_{q\rightarrow h} (z_h/z) \int \frac{dy^-}{2\pi} dy_1^- dy_2^-  \int d^2\vec{y}_{12\perp}\int \frac{d^2 k_{\perp}}{(2\pi)^2} e^{i\vec{k}_{\perp}\cdot \vec{y}_{12\perp}} \int dx \dfrac{dx_1}{2\pi} \dfrac{dx_2}{2\pi}   \\
&\times e^{-ix_1p^+ y^- - ix_2p^+ y_1^- -i(x-x_1-x_2)p^+ y_2^-} \langle A | \bar{\psi}(y^-)  \dfrac{\gamma^{+}}{2} A^{+}(y_1^-,\vec{y}_{1\perp})A^+(y_2^-, \vec{y}_{2\perp}) \psi(0)| A \rangle\\
&\times  \int \dfrac{d^4 l}{(2\pi)^4} \dfrac{1}{2} Tr[p\cdot \gamma \gamma^{\mu}p_{\sigma}p_{\rho} \hat{H}^{\sigma \rho} \gamma^{\nu}] 2\pi \delta(l_q^2) 2\pi \delta(l^2)  \delta (1-z-\dfrac{l^-}{q^-}),
\end{split}
\end{equation}

where $\vec{y}_{12\perp} = \vec{y}_{1\perp} - \vec{y}_{2\perp}$,
\begin{equation}
\begin{split}
 \hat{H}^{\sigma \rho} =& g^4 \frac{C_F}{2N_c}\frac{\gamma\cdot(x_1 p+q)}{(x_1 p+q)^2-i\epsilon} \gamma^{\sigma} \frac{\gamma\cdot(x_1 p+x_2 p+q+k_{\perp})}{(x_1 p+x_2 p+q+k_{\perp})^2-i\epsilon} \gamma^{\alpha} \gamma \cdot l_q \gamma^{\beta}  \\
 &\times \frac{\gamma\cdot(x_1 p+x_2 p+q+k_{\perp})}{(x_1 p+x_2 p+q+k_{\perp})^2+i\epsilon}\gamma^{\rho} \frac{\gamma\cdot(x p+q)}{(x p+q)^2+i\epsilon} \epsilon_{\alpha\beta} (l) ,
\end{split}
\end{equation}
and polarization sum tensor of the final gluon is,
\begin{equation}
\epsilon_{\alpha\beta} (l) = -g_{\alpha \beta} + \dfrac{n_{\alpha}l_{\beta}+n_{\beta}l_{\alpha}}{n\cdot l} + n^2\frac{l_{\alpha}l_{\beta}}{(n\cdot l)^2} ,
\end{equation}
 here one sum over the physical polarizations, where $n = [1,0^-, \vec{0}_{\perp}]$ for axial gauge. Using this polarization sum tensor, the contribution from diagrams of initial gluon radiation is zero.

  Following the notations in Ref.~\cite{Wang:2001ifa} for the high-twist approach to parton energy loss, the hadronic tensor can be expressed as

 \begin{equation}
 \begin{split}
 \frac{d W^{\mu \nu {\rm Fig}\ref{fig:C11w}}_{D(1)q} }{d z_h} =  &\int_{z_h}^1 \frac{dz}{z} D_{q\rightarrow h} (z_h/z) \int \frac{dy^-}{2\pi} dy_1^- dy_2^-  \int d^2 \vec{y}_{12\perp} \int \frac{d^2 k_{\perp}}{(2\pi)^2} e^{i\vec{k}_{\perp}\cdot \vec{y}_{12\perp}} \\
 & \times \dfrac{1}{2} \langle A | \bar{\psi}(y^-)  \gamma^{+}  A^{+}(y_1^-,\vec{y}_{1\perp})A^+(y_2^-, \vec{y}_{2\perp}) \psi(0)| A \rangle  (\overline{H}_{C11}^D)^{\mu \nu} , \\
\end{split}
 \end{equation}
 with partonic hard part as
 \begin{equation}
 \begin{split}
(\overline{H}_{C11}^D)^{\mu \nu} =  & \int dx \dfrac{dx_1}{2\pi} \dfrac{dx_2}{2\pi} e^{-ix_1p^+ y^-}e^{ - ix_2p^+ y_1^- } e^{-i(x-x_1-x_2)p^+ y_2^-}\\
 &\times \int \dfrac{d^4 l}{(2\pi)^4} 2\pi \delta(l^2) 2\pi \delta(l_q^2)  \dfrac{1}{2} Tr[p\cdot \gamma \gamma^{\mu}p_{\sigma}p_{\rho} \hat{H}^{\sigma \rho} \gamma^{\nu}]   \delta (1-z-\dfrac{l^-}{q^-}).
 \end{split}
 \end{equation}

Using the pole structure of the propagators in $\hat{H}^{\sigma \rho}$ 
under the contour integration and $\delta(l_q^2)$ from the on-shell condition of the cut quark line, one can carry out the integrations over $x$, $x_1$ and $x_2$ in $I_C$, 

\begin{equation}
\begin{split}
I_C = & \int dx \dfrac{dx_1}{2\pi} \dfrac{dx_2}{2\pi}  e^{-ix_1p^+ y^- -ix_2p^+ y_1^- -i(x-x_1-x_2)p^+ y_2^-} \frac{1}{(x_1 p+q)^2-i\epsilon}\frac{1}{(x_1 p+x_2 p+q+k_{\perp})^2-i\epsilon}\\
&\times \frac{1}{(x_1 p+x_2 p+q+k_{\perp})^2+i\epsilon} \frac{1}{(x p+q)^2+i\epsilon} \delta(l_q^2)\\
 = &\frac{1}{(2p^+q^-)^5z} \frac{e^{-ix_Bp^+y^-}e^{-i(x_L+x_D)p^+(y_1^- -y_2^-)}}{(x_L+x_D-x_D^0)^2}\theta(y_2^-)\theta(y_1^- -y^-),
\end{split}
\end{equation}

where 
\begin{equation}
x_L = \frac{l_{\perp}^2}{2p^+q^- z(1-z)},   \qquad x_D = \frac{k_{\perp}^2-2\vec{k}_{\perp}\cdot \vec{l}_{\perp}}{2p^+q^-z},  \qquad   x_D^0 = \frac{k_{\perp}^2}{2p^+q^-}.
\end{equation}

Under collinear approximation, one has 
 \begin{equation}
 p_{\sigma} \hat{H}^{\sigma \rho} p_{\rho} \approx \gamma \cdot(xp+q) \dfrac{1}{4q^-} {\rm Tr}[\gamma^- p_{\sigma} \hat{H}^{\sigma \rho} p_{\rho}].
 \end{equation}
 
Taking the trace ${\rm Tr}[\gamma^- p_{\sigma} \hat{H}^{\sigma \rho} p_{\rho}]$ and integrate over $l^+$ and $l^-$, we get the partonic hard part,

\begin{equation}
\begin{split}
 &(\overline{H}_{C11}^D)^{\mu \nu}=  \int dx H_{(0)}^{\mu\nu}(x)   \int d l_{\perp}^2 \frac{\alpha_s}{2\pi} C_F \frac{1+z^2}{1-z} \frac{2\pi \alpha_s}{N_c} \frac{ e^{-ixp^+y^-}e^{-i(x_L+x_D)p^+(y_1^- -y_2^-)} }{[\vec{l}_{\perp} -(1-z)\vec{k}_{\perp}]^2} \theta(y_2^-)\theta(y_1^- -y^-),
\end{split}
\end{equation}
where
\begin{equation}
\begin{split}
H_{(0)}^{\mu\nu}(x)=&\dfrac{1}{2} {\rm Tr}[p\cdot \gamma \gamma^{\mu} \gamma \cdot(x_Bp+q) \gamma^{\nu}] \frac{2\pi}{2p^+q^-} \delta(x-x_B).
\end{split}
\end{equation}
The contribution to the hadronic tensor from this central cut-diagram reads

\begin{equation}
\begin{split}
\frac{d W^{\mu \nu {\rm Fig}\ref{fig:C11w}}_{D(1)q} }{d z_h}=  &\int_{z_h}^1 \frac{dz}{z}  D_{q\rightarrow h} (z_h/z) \int dx H_{(0)}^{\mu\nu}(x)    \int \frac{dy^-}{2\pi} \int dy_1^- \int dy_2^- \int d^2\vec{y}_{12\perp} \int \frac{d^2 k_{\perp}}{(2\pi)^2} \int d l_{\perp}^2  \\
&\times  e^{i\vec{k}_{\perp}\cdot \vec{y}_{12\perp}} e^{-ixp^+y^-}e^{-i(x_L+x_D)p^+(y_1^- -y_2^-)} \langle A | \bar{\psi}(y^-)   \dfrac{\gamma^{+}}{2}A^{+}(y_1^-,\vec{y}_{1\perp})A^+(y_2^-, \vec{y}_{2\perp}) \psi(0)| A \rangle \\
&\times \frac{1}{[\vec{l}_{\perp} -(1-z)\vec{k}_{\perp}]^2} \frac{\alpha_s}{2\pi} C_F \frac{1+z^2}{1-z} \frac{2\pi \alpha_s}{N_c}
\end{split}
\label{eq:central 11}
\end{equation}

Following the same procedures, one can calculate contributions to the hadronic tensor from all the cut diagrams for double parton scattering at $\alpha_s^2$ order whose results are given in Appendix \ref{append-spectrum}. Summing up these contributions, including central, left and right cut diagrams, the hadronic tensor from double parton scattering with hadrons from the fragmentation of the final quark can be expressed as,
 \begin{equation}
 \begin{aligned}
 \frac{d W^{\mu \nu}_{D(1)q} }{d z_h} =  &\int_{z_h}^1 \frac{dz}{z} D_{q\rightarrow h} (z_h/z) \int \frac{dy^-}{2\pi} dy_1^- dy_2^- \int \frac{d^2 k_{\perp}}{(2\pi)^2} e^{i\vec{k}_{\perp}\cdot \vec{y}_{12\perp} }  \langle A | \bar{\psi}(y^-)   \frac{\gamma^{+}}{2}A^{+}(y_1^-,\vec{y}_{1\perp})\\
 &\times A^+(y_2^-, \vec{y}_{2\perp}) \psi(0)| A \rangle \left[ (\overline{H}_{C}^D)^{\mu \nu}+  (\overline{H}_{L}^D)^{\mu \nu} + (\overline{H}_{R}^D)^{\mu \nu} \right], \\
 \label{eq:radiated_gluon}
\end{aligned}
 \end{equation}
with partonic hard parts from central, left and right cut diagrams, 
\begin{equation}
 \begin{split}
 (\overline{H}_{C}^D)^{\mu \nu} =   &\int dx H_{(0)}^{\mu\nu}(x)   \int d l_{\perp}^2 \frac{\alpha_s}{2\pi} \frac{1+z^2}{1-z} \frac{2\pi \alpha_s}{N_c}  H_C^D \theta(y_2^-)\theta(y_1^- -y^-),\\
 ~\\
 (\overline{H}_{L}^D)^{\mu \nu} =   &\int dx H_{(0)}^{\mu\nu}(x)   \int d l_{\perp}^2 \frac{\alpha_s}{2\pi} \frac{1+z^2}{1-z} \frac{2\pi \alpha_s}{N_c}  H_L^D \theta(y_2^- -y_1^-)\theta(y_1^- -y^-),\\
 ~\\
 (\overline{H}_{R}^D)^{\mu \nu} =   &\int dx H_{(0)}^{\mu\nu} (x)  \int d l_{\perp}^2 \frac{\alpha_s}{2\pi} \frac{1+z^2}{1-z} \frac{2\pi \alpha_s}{N_c}  H_R^D \theta(y_1^- -y_2^-) \theta(y_2^-),
  \label{eq:radiated_gluon_CLR}
\end{split}
 \end{equation}

where
\begin{align}
\begin{split}
 H_C^D = &\left\lbrace \left[ \frac{C_A}{(l_{\perp}-k_{\perp})^2} e^{i\frac{z}{1-z}x_Dp^+(y_1^-  -y_2^-)}  e^{-i(x+x_L+\frac{x_D}{1-z})p^+y^- } - \frac{C_A}{l_{\perp}^2} e^{-i\frac{z}{1-z}x_Dp^+(y_1^-  -y_2^-)} e^{-i(x+x_L) p^+y^-} \right]  \right.\\
&+\left[  \left( \frac{C_F}{l_{\perp}^2} +C_A\frac{\vec{k}_{\perp}\cdot \vec{l}_{\perp}}{l_{\perp}^2(\vec{l}_{\perp}-\vec{k}_{\perp})^2 } +   \frac{C_F}{[\vec{l}_{\perp}-(1-z)\vec{k}_{\perp}]^2}+ \frac{1}{N_c}\frac{\vec{l}_{\perp}\cdot [\vec{l}_{\perp} -(1-z)\vec{k}_{\perp}]}{l_{\perp}^2[\vec{l}_{\perp}-(1-z)\vec{k}_{\perp}]^2} \right. \right.  \\
&\left. \left. -C_A  \frac{(\vec{l}_{\perp} - \vec{k}_{\perp})\cdot [\vec{l}_{\perp} -(1-z)\vec{k}_{\perp}]}{(l_{\perp}-k_{\perp})^2[\vec{l}_{\perp}-(1-z)\vec{k}_{\perp}]^2} \right)e^{-i(x_L+x_D)p^+(y_1^-  -y_2^-)}   e^{-ixp^+y^-} \right]\\
& +\left[  \left( - \frac{C_A}{(\vec{l}_{\perp}-\vec{k}_{\perp})^2}
+ \frac{C_A}{2}\frac{\vec{l}_{\perp}\cdot(\vec{l}_{\perp}-\vec{k}_{\perp})}{l_{\perp}^2(\vec{l}_{\perp}-\vec{k}_{\perp})^2}
+  \frac{C_A}{2}\frac{(\vec{l}_{\perp} - \vec{k}_{\perp})\cdot [\vec{l}_{\perp} -(1-z)\vec{k}_{\perp}]}{(l_{\perp}-k_{\perp})^2[\vec{l}_{\perp}-(1-z)\vec{k}_{\perp}]^2} 
 \right)\right.\\
 &\left. \times e^{-i(x_L+x_D,\vec{k}_{\perp}) p^+(y_1^- -y_2^-)} e^{i(x_L+\frac{x_D}{1-z})p^+y_1^-} e^{-i(x+x_L+\frac{x_D}{1-z})p^+y^-} \right] \\
& +\left[  \left( - \frac{C_A}{(\vec{l}_{\perp}-\vec{k}_{\perp})^2}
+ \frac{C_A}{2}\frac{\vec{l}_{\perp}\cdot(\vec{l}_{\perp}-\vec{k}_{\perp})}{l_{\perp}^2(\vec{l}_{\perp}-\vec{k}_{\perp})^2}
+  \frac{C_A}{2}\frac{(\vec{l}_{\perp} - \vec{k}_{\perp})\cdot [\vec{l}_{\perp} -(1-z)\vec{k}_{\perp}]}{(l_{\perp}-k_{\perp})^2[\vec{l}_{\perp}-(1-z)\vec{k}_{\perp}]^2} 
 \right)\right.   \\
& \left. \left. \times e^{i\frac{z}{1-z}x_D p^+(y_1^- -y_2^-)} e^{-i(x_L+\frac{x_D}{1-z})p^+y_1^-}e^{-ixp^+y^-} \right]  \right\rbrace, \\
 \label{eq:radiated_gluon_cfull}
\end{split}
\end{align}

\begin{align}
\begin{split}
H^D_L  = &\left\lbrace \left[ \frac{C_A}{2} \frac{\vec{l}_{\perp}\cdot(\vec{l}_{\perp}-\vec{k}_{\perp})}{l_{\perp}^2(\vec{l}_{\perp}-\vec{k}_{\perp})^2} e^{-i\frac{z}{1-z}x_Dp^+(y_1^- -y_2^-)} e^{i\frac{x_D}{1-z}p^+y_1^-}e^{-i(x+x_L+\frac{x_D}{1-z})p^+y^-}\right.\right.\\
&\left.   -\frac{C_A}{2} \frac{\vec{l}_{\perp}\cdot(\vec{l}_{\perp}-\vec{k}_{\perp})}{l_{\perp}^2(\vec{l}_{\perp}-\vec{k}_{\perp})^2} e^{i\frac{z}{1-z}x_Dp^+(y_1^- -y_2^-)} e^{-i\frac{x_D}{1-z}p^+y_1^-} e^{-i(x+x_L)p^+y^-}   \right] \\
&+\left[   -\left(C_F\frac{1}{l_{\perp}^2}+  \frac{1}{2N_c}\frac{\vec{l}_{\perp}\cdot [\vec{l}_{\perp} -(1-z)\vec{k}_{\perp}]}{l_{\perp}^2[\vec{l}_{\perp}-(1-z)\vec{k}_{\perp}]^2}  \right)e^{-i(x_D^0-x_L)p^+(y_1^- -y_2^-)} e^{-ix_Lp^+y_1^-} e^{-ixp^+y^-}\right.  \\
&\left. +  \frac{C_A}{2} \frac{\vec{l}_{\perp}\cdot(\vec{l}_{\perp}-z\vec{k}_{\perp})}{l_{\perp}^2(\vec{l}_{\perp}-z\vec{k}_{\perp})^2}e^{-i(x_D^0-x_L)p^+(y_1^- -y_2^-)} e^{-ix_Lp^+y_1^-}e^{-ixp^+y^-}  \right]\\
& +\left[  -\left(   \frac{1}{2N_c}\frac{\vec{l}_{\perp}\cdot [\vec{l}_{\perp} -(1-z)\vec{k}_{\perp}]}{l_{\perp}^2[\vec{l}_{\perp}-(1-z)\vec{k}_{\perp}]^2} + \frac{C_F}{l_{\perp}^2} - \frac{C_A}{2} \frac{\vec{l}_{\perp}\cdot(\vec{l}_{\perp}-\vec{k}_{\perp})}{l_{\perp}^2(\vec{l}_{\perp}-\vec{k}_{\perp})^2} \right)e^{-i(x_L+x_D)p^+(y_1^-  -y_2^-)}e^{ix_Lp^+y_1^-} e^{-i(x+x_L)p^+y^-} \right. \\
&\left.\left. + \left(- \frac{C_A}{2}\frac{\vec{l}_{\perp}\cdot(\vec{l}_{\perp}-z\vec{k}_{\perp})}{l_{\perp}^2(\vec{l}_{\perp}-z\vec{k}_{\perp})^2} - \frac{C_A}{2}\frac{\vec{l}_{\perp}\cdot(\vec{l}_{\perp}-\vec{k}_{\perp})}{l_{\perp}^2(\vec{l}_{\perp}-\vec{k}_{\perp})^2} + \frac{C_A}{l_{\perp}^2} \right) e^{-i\frac{z}{1-z}x_D p^+(y_1^- -y_2^-) }e^{-ix_Lp^+y_1^-}e^{-ixp^+y^-}\right]  \right\rbrace ,\\
\end{split}
 \label{eq:radiated_gluon_lfull}
\end{align}  

\begin{align}
\begin{split}
H^D_R  = &\left\lbrace \left[  \frac{C_A}{2} \frac{\vec{l}_{\perp}\cdot(\vec{l}_{\perp}-\vec{k}_{\perp})}{l_{\perp}^2(\vec{l}_{\perp}-\vec{k}_{\perp})^2} e^{ix_Dp^+(y_1^- -y_2^-)}e^{-i\frac{x_D}{1-z}p^+y_1^-} e^{-i(x+x_L)p^+y^-}  \right. \right. \\
&\left.- \frac{C_A}{2} \frac{\vec{l}_{\perp}\cdot(\vec{l}_{\perp}-\vec{k}_{\perp})}{l_{\perp}^2(\vec{l}_{\perp}-\vec{k}_{\perp})^2} e^{-ix_Dp^+(y_1^- -y_2^-)}e^{i\frac{x_D}{1-z}p^+y_1^-}e^{-i(x+x_L+\frac{x_D}{1-z})p^+y^-}\right] \\
&+\left[  -\left(C_F\frac{1}{l_{\perp}^2}+  \frac{1}{2N_c}\frac{\vec{l}_{\perp}\cdot [\vec{l}_{\perp} -(1-z)\vec{k}_{\perp}]}{l_{\perp}^2[\vec{l}_{\perp}-(1-z)\vec{k}_{\perp}]^2}  \right)e^{-ix_D^0p^+(y_1^- -y_2^-)} e^{ix_Lp^+y_1^-}e^{-i(x+x_L)p^+y^-} \right.\\
& \left.+  \frac{C_A}{2} \frac{\vec{l}_{\perp}\cdot(\vec{l}_{\perp}-z\vec{k}_{\perp})}{l_{\perp}^2(\vec{l}_{\perp}-z\vec{k}_{\perp})^2}e^{-ix_D^0p^+(y_1^- -y_2^-)}e^{ix_Lp^+y_1^-}e^{-i(x+x_L)p^+y^-}  \right]\\
& +\left[ -\left(   \frac{1}{2N_c}\frac{\vec{l}_{\perp}\cdot [\vec{l}_{\perp} -(1-z)\vec{k}_{\perp}]}{l_{\perp}^2[\vec{l}_{\perp}-(1-z)\vec{k}_{\perp}]^2} + \frac{C_F}{l_{\perp}^2} - \frac{C_A}{2} \frac{\vec{l}_{\perp}\cdot(\vec{l}_{\perp}-\vec{k}_{\perp})}{l_{\perp}^2(\vec{l}_{\perp}-\vec{k}_{\perp})^2} \right)  e^{-ix_Dp^+(y_1^- -y_2^-)}  e^{-ix_Lp^+y_1^-} e^{-ixp^+y^-} \right. \\
&\left.\left. + \left(- \frac{C_A}{2}\frac{\vec{l}_{\perp}\cdot(\vec{l}_{\perp}-z\vec{k}_{\perp})}{l_{\perp}^2(\vec{l}_{\perp}-z\vec{k}_{\perp})^2} - \frac{C_A}{2}\frac{\vec{l}_{\perp}\cdot(\vec{l}_{\perp}-\vec{k}_{\perp})}{l_{\perp}^2(\vec{l}_{\perp}-\vec{k}_{\perp})^2} + \frac{C_A}{l_{\perp}^2} \right)e^{-i(x_L+\frac{z}{1-z} x_D)p^+(y_1^- -y_2^-)}e^{ix_Lp^+y_1^-}e^{-i(x+x_L)p^+y^-}\right]  \right\rbrace . \\ 
\end{split}
 \label{eq:radiated_gluon_rfull}
\end{align}

In order to organize the above contributions, we have reversed the sign of the initial transverse momentum $\vec{k}_{\perp}$ in some diagrams. See Appendix \ref{append-spectrum} for details. One can also get the hadronic tensor using helicity amplitude approximation, which is the same as the full result from the cut diagrams in the soft gluon approximation $z \rightarrow 1$, as was also studied in the high-twist approach in Ref.~\cite{Wang:2001ifa}.  Details of the helicity amplitude calculations are given in Appendix \ref{append-helicity}. One can also obtain virtual corrections from the unitarity requirement which will cancel the infrared divergence in the radiative corrections listed above.

In addition to the above listed contributions, there are also contact contributions that are not enhanced by the nuclear size due to path ordered integration. They are negligible as compared to contributions listed above. There are two sources of contact contributions. One type of contact contributions come from the combination of central, left and right cut diagrams with a common hard partonic part,

 \begin{equation}
 \begin{split}
 \frac{d W^{\mu \nu}_\textsl{\rm contact1} }{d z_h} = & \int_{z_h}^1 \frac{dz}{z} D_{q\rightarrow h} (z_h/z) \int \frac{dy^-}{2\pi} dy_1^- dy_2^- \int d^2\vec{y}_{12\perp}  \int \frac{d^2 k_{\perp}}{(2\pi)^2} e^{i\vec{k}_{\perp}\cdot \vec{y}_{12\perp}} \langle A | \bar{\psi}(y^-)  \\
 &\times \frac{ \gamma^{+}}{2}A^{+}(y_1^-,\vec{y}_{1\perp})A^+(y_2^-, \vec{y}_{2\perp}) \psi(0)| A \rangle \int dx H_{(0)}^{\mu\nu}(x)   \int d l_{\perp}^2 \frac{\alpha_s}{2\pi} \frac{1+z^2}{1-z} \frac{2\pi \alpha_s}{N_c}\\
 & \times H_\textsl{\rm contact1}^D \left[  \theta(y_2^-)\theta(y_1^- -y^-) -  \theta(y_1^- -y_2^-) \theta(y_2^-)  - \theta(y_2^- -y_1^-)\theta(y_1^- -y^-) \right], 
\end{split}
 \end{equation}
with 
\begin{equation}
\begin{split}
H_\textsl{\rm contact1}^D =&\frac{C_F}{l_{\perp}^2} e^{-ix_Dp^+(y_1^- -y_2^-)}e^{-i(x+x_L)p^+y^-}  +  \frac{C_A}{l_{\perp}^2} e^{-i\frac{z}{1-z}x_Dp^+(y_1^- -y_2^-)}  e^{-i(x+x_L)p^+y^-} \\
&   -\frac{C_A}{2} \frac{\vec{l}_{\perp}\cdot(\vec{l}_{\perp}-\vec{k}_{\perp})}{l_{\perp}^2(\vec{l}_{\perp}-\vec{k}_{\perp})^2} e^{i\frac{z}{1-z}x_Dp^+(y_1^- -y_2^-)}   e^{-i\frac{x_D}{1-z}p^+y_1^-} e^{-i(x+x_L) p^+y^-}\\
& - \frac{C_A}{2} \frac{\vec{l}_{\perp}\cdot(\vec{l}_{\perp}-\vec{k}_{\perp})}{l_{\perp}^2(\vec{l}_{\perp}-\vec{k}_{\perp})^2} e^{-ix_Dp^+(y_1^- -y_2^-)} e^{i\frac{x_D}{1-z}p^+y_1^-} e^{-i(x+x_L+\frac{x_D}{1-z})p^+y^- }\\
&+\left( -\frac{C_F}{{l}_{\perp}^2} -\frac{1}{2N_c}\frac{\vec{l}_{\perp}\cdot [\vec{l}_{\perp} -(1-z)\vec{k}_{\perp}]}{l_{\perp}^2[\vec{l}_{\perp}-(1-z)\vec{k}_{\perp}]^2}   
+ \frac{C_A}{2}\frac{\vec{l}_{\perp}\cdot(\vec{l}_{\perp}-\vec{k}_{\perp})}{l_{\perp}^2(\vec{l}_{\perp}-\vec{k}_{\perp})^2}
\right)e^{-i(x_L+x_D)p^+(y_1^- -y_2^-) }e^{ix_Lp^+y_1^-}e^{-i(x+x_L)p^+y^-}\\
&+  \left( -\frac{C_F}{{l}_{\perp}^2} -\frac{1}{2N_c}\frac{\vec{l}_{\perp}\cdot [\vec{l}_{\perp} -(1-z)\vec{k}_{\perp}]}{l_{\perp}^2[\vec{l}_{\perp}-(1-z)\vec{k}_{\perp}]^2}   
+ \frac{C_A}{2}\frac{\vec{l}_{\perp}\cdot(\vec{l}_{\perp}-\vec{k}_{\perp})}{l_{\perp}^2(\vec{l}_{\perp}-\vec{k}_{\perp})^2}
\right) e^{-ix_Dp^+(y_1^- -y_2^-)}e^{-ix_Lp^+y_1^-} e^{-ixp^+y^-}.
\end{split}
\end{equation}

The combination of $\theta$ functions in these contact terms leads to path-ordered integration,

\begin{equation}
\int dy_1^- dy_2^- \Big[\theta(y_1^- -y^-) \theta(y_2^-)- \theta( y_2^- -y_1^-) \theta( y_1^- -y^-)- \theta(y_1^- -y_2^-) \theta(y_2^-) \Big]=- \int_0^{y^-} dy_1^-  \int_0^{y_1^-} dy_2^-,
\end{equation}

that limits the range of both coordinates in the integration within one single nucleon, $0<y_1^-<y_2^-<y^-$. These contributions are not enhanced by the nuclear size and therefore negligible comparing to other terms that are enhanced by the nuclear size. In the Glauber limit $k_\perp\rightarrow 0$ the above contact term becomes a part of the gauge link for the NLO correction to the single scattering. Other terms in the collinear expansion of these contact contributions lead to higher-twist terms that are not enhanced by the nuclear size.

The second type of contact contributions come from the integration region of right cut diagrams. The integration regions of $y^-, y_1^-, y_2^-$ for central, left and right cut diagrams are,

\begin{equation}
\begin{split}
\theta_C&\equiv \int d y^- \int dy_1^-  \int dy_2^- \theta(y_1^- -y^-) \theta(y_2^-) =  \int  d y^- \int_{y^-}^{\infty} dy_1^-  \int_0^{\infty} dy_2^-   \\
&= \int d y^- \int_{y^-}^{\infty} dy_1^- \int_{-\infty}^{y_1} d y_{12}^- ,\\
\theta_L &\equiv \int d y^- \int dy_1^-  \int dy_2^-  \theta( y_2^- -y_1^-) \theta( y_1^- -y^-)=  \int d y^-   \int_{y^-}^{\infty} dy_1^-   \int_{y_1^-}^{\infty} dy_2^- \\
& =   \int d y^- \int_{y^-}^{\infty} dy_1^- \int_{-\infty}^{0} d y_{12}^-  ,\\
\theta_R&\equiv \int d y^-  \int dy_1^-  \int dy_2^- \theta(y_1^- -y_2^-) \theta(y_2^-)= \int d y^-  \int_{y_2^-}^{\infty} dy_1^-  \int_0^{\infty} dy_2^-  \\
& =   \int d y^- \int_{0}^{\infty} dy_1^- \int_{0}^{y_1} d y_{12}^- ,
\end{split}
\end{equation}
respectively, where $y_{12}^- = y_1^- -y_2^-$. If the integration region for $|y_{12}^-|<r_N$ is limited by the size of the nucleon $r_N$, the three integration regions become
\begin{equation}
\begin{split}
 \theta_C=&   \int d y^- \int_{y^-}^{\infty} dy_1^- \int_{-r_N}^{r_N} d y_{12}^-  ,\\
\theta_L=&    \int d y^- \int_{y^-}^{\infty} dy_1^- \int_{-r_N}^{0} d y_{12}^-  ,\\
\theta_R  =&   \int d y^- \int_{0}^{\infty} dy_1^- \int_{0}^{r_N} d y_{12}^- \\
=  & \int d y^- \int_{y^-}^{\infty} dy_1^- \int_{0}^{r_N} d y_{12}^-  +  \left[ \int d y^- \int_0^{y^-} dy_1^-  \int_0^{y_1^-} dy_2^-\right].
\end{split}
\label{eq:integration_region}
\end{equation}
respectively. The integration in the square  brackets for the contributions from the right-cut diagrams is path-ordered $0<y_1^-<y_2^-<y^-$ for the second type of contact contribution,

 \begin{equation}
 \begin{split}
 \frac{d W^{\mu \nu}_\textsl{\rm contact2} }{d z_h} = & \int_{z_h}^1 \frac{dz}{z} D_{q\rightarrow h} (z_h/z) \int \frac{dy^-}{2\pi}  \int_0^{y^-} dy_1^- \int_0^{y_1^-} dy_2^- \int d^2\vec{y}_{12\perp} \int \frac{d^2 k_{\perp}}{(2\pi)^2} e^{i\vec{k}_{\perp}\cdot \vec{y}_{12\perp} }  \dfrac{1}{2} \langle A | \bar{\psi}(y^-) \\
 &\times  \gamma^{+}A^{+}(y_1^-,\vec{y}_{1\perp})A^+(y_2^-, \vec{y}_{2\perp}) \psi(0)| A \rangle \int dx H_{(0)}^{\mu\nu} (x)  \int d l_{\perp}^2 \frac{\alpha_s}{2\pi} \frac{1+z^2}{1-z} \frac{2\pi \alpha_s}{N_c} H_\textsl{R}^D  \\
\end{split}
 \end{equation}
The summation of these two types of contact contributions reads,
 \begin{equation}
 \begin{split}
 \frac{d W^{\mu \nu}_\textsl{contact} }{d z_h} = & \int_{z_h}^1 \frac{dz}{z} D_{q\rightarrow h} (z_h/z) \int \frac{dy^-}{2\pi}  \int_0^{y^-} dy_1^- \int_0^{y_1^-} dy_2^- \int d^2\vec{y}_{12\perp} \int \frac{d^2 k_{\perp}}{(2\pi)^2} e^{i\vec{k}_{\perp}\cdot \vec{y}_{12\perp}}\dfrac{1}{2} \langle A | \bar{\psi}(y^-)  \\
 &\times   \gamma^{+}A^{+}(y_1^-,\vec{y}_{1\perp})A^+(y_2^-, \vec{y}_{2\perp}) \psi(0)| A \rangle \int dx H_{(0)}^{\mu\nu}(x)   \int d l_{\perp}^2 \frac{\alpha_s}{2\pi} \frac{1+z^2}{1-z} \frac{2\pi \alpha_s}{N_c}  H_\textsl{\rm contact}^D,
\end{split}
 \end{equation}
where
\begin{equation}
\begin{split}
H_\textsl{contact}^D =   & H_\textsl{R}^D - H_\textsl{contact1}^D \\
=&\frac{C_A}{2} \frac{\vec{l}_{\perp}\cdot(\vec{l}_{\perp}-\vec{k}_{\perp})}{l_{\perp}^2(\vec{l}_{\perp}-\vec{k}_{\perp})^2} e^{ix_Dp^+(y_1^- -y_2^-)}e^{-i\frac{x_D}{1-z}p^+y_1^-} e^{-i(x+x_L)p^+y^-}\\
& + \frac{C_A}{2} \frac{\vec{l}_{\perp}\cdot(\vec{l}_{\perp}-\vec{k}_{\perp})}{l_{\perp}^2(\vec{l}_{\perp}-\vec{k}_{\perp})^2} e^{i\frac{z}{1-z}x_Dp^+(y_1^- -y_2^-)}   e^{-i\frac{x_D}{1-z}p^+y_1^-} e^{-i(x+x_L) p^+y^-}  \\
& - \frac{C_F}{l_{\perp}^2} e^{-ix_Dp^+(y_1^- -y_2^-)}e^{-i(x+x_L)p^+y^-} -  \frac{C_A}{l_{\perp}^2} e^{-i\frac{z}{1-z}x_Dp^+(y_1^- -y_2^-)}  e^{-i(x+x_L)p^+y^-}\\
& -\left(C_F\frac{1}{l_{\perp}^2}+  \frac{1}{2N_c}\frac{\vec{l}_{\perp}\cdot [\vec{l}_{\perp} -(1-z)\vec{k}_{\perp}]}{l_{\perp}^2[\vec{l}_{\perp}-(1-z)\vec{k}_{\perp}]^2}  \right)e^{-ix_D^0p^+(y_1^- -y_2^-)} e^{ix_Lp^+y_1^-}e^{-i(x+x_L)p^+y^-} \\
& +  \frac{C_A}{2} \frac{\vec{l}_{\perp}\cdot(\vec{l}_{\perp}-z\vec{k}_{\perp})}{l_{\perp}^2(\vec{l}_{\perp}-z\vec{k}_{\perp})^2}e^{-ix_D^0p^+(y_1^- -y_2^-)}e^{ix_Lp^+y_1^-}e^{-i(x+x_L)p^+y^-}\\
&+\left(- \frac{C_A}{2}\frac{\vec{l}_{\perp}\cdot(\vec{l}_{\perp}-z\vec{k}_{\perp})}{l_{\perp}^2(\vec{l}_{\perp}-z\vec{k}_{\perp})^2} - \frac{C_A}{2}\frac{\vec{l}_{\perp}\cdot(\vec{l}_{\perp}-\vec{k}_{\perp})}{l_{\perp}^2(\vec{l}_{\perp}-\vec{k}_{\perp})^2} + \frac{C_A}{l_{\perp}^2} \right)e^{-i(x_L+\frac{z}{1-z} x_D)p^+(y_1^- -y_2^-)}e^{ix_Lp^+y_1^-}e^{-i(x+x_L)p^+y^-}\\
&+\left( \frac{C_F}{{l}_{\perp}^2}+ \frac{1}{2N_c}\frac{\vec{l}_{\perp}\cdot [\vec{l}_{\perp} -(1-z)\vec{k}_{\perp}]}{l_{\perp}^2[\vec{l}_{\perp}-(1-z)\vec{k}_{\perp}]^2}   
- \frac{C_A}{2}\frac{\vec{l}_{\perp}\cdot(\vec{l}_{\perp}-\vec{k}_{\perp})}{l_{\perp}^2(\vec{l}_{\perp}-\vec{k}_{\perp})^2}
\right)e^{-i(x_L+x_D)p^+(y_1^- -y_2^-) }e^{ix_Lp^+y_1^-}e^{-i(x+x_L)p^+y^-}.
\end{split}
\end{equation}

\subsection{Quark-gluon correlation function and TMD jet transport parameter}
\label{qg-correlation-TMD-qhat}


Before we continue to calculate radiative gluon spectra and parton energy loss, we pause to discuss the quark-gluon correlation function in the contributions to the hadronic tensor from double scattering. 
The generic quark-gluon correlation function in every term in Eq.~(\ref{eq:radiated_gluon}) has the form,
\begin{equation}
\begin{split}
 T_{qg}^A(x,x_1,x_2) =&\int \frac{dy^-}{2\pi} dy_1^- dy_2^- \int d^2\vec{y}_{12\perp}  e^{-ixp^+y^- } e^{-ix_2p^+(y_1^- -y_2^-)} e^{i(x-x_1)p^+y_1^-} e^{i\vec{k}_{\perp}\cdot\vec{y}_{12\perp}}   \\
 &\times \langle A |\bar{\psi}(y^{-}) \frac{\gamma^{+}}{2} A^{+}(y_{1}^{-}, \vec{y}_{1\perp}) A^{+}(y_{2}^{-}, \vec{y}_{2\perp}) \psi(0)| A \rangle \theta(f_1) \theta(f_2),
 \end{split}
\end{equation}
where the $\theta$-functions are different for contributions from central, right and left cut diagrams,

\begin{equation}
\theta(f_1) \theta(f_2)=\left\{\begin{array}{lll}{\theta(y_2^-)\theta(y_1^- -y^-):} & {\text {central,}} \\ {\theta(y_2^- -y_1^-)\theta(y_1^- -y^-):} & {\text {left,}}\\{\theta(y_1^- -y_2^-)\theta(y_2^-):}& {\text {right.}}
\end{array}\right.
\end{equation} 
If we neglect the correlation between parent nucleons of initial quark and medium gluon, and define the effective impact-parameter-dependent nuclear quark distribution function $f_q^A(x,\vec{y_\perp})$ as,
\begin{equation}
\begin{split}
f_q^A(x)=& \int \frac{dy^-}{2\pi} e^{-ixp^+y^-} \langle A |\bar{\psi}(y^-)  \dfrac{\gamma^{+}}{2} \psi(0)| A  \rangle  \\
&\equiv 
 \frac{1}{A} \int dy^- d^2\vec{y}_\perp \rho_A(y^-,\vec{y_\perp}) f_q^A(x,\vec{y}_{\perp}),
 \end{split}
\end{equation}
the correlation function can be factorized as

\begin{equation}
\begin{split}
 T_{qg}^A(x,x_1,x_2)= & \frac{C}{A} \int dy^- d^2\vec{y}_{\perp} \rho_A(y^-,\vec{y}_{\perp}) f_q^A(x,\vec{y_\perp}) \int_{y^-}^{\infty} dy_1^-  \int \frac{dp'^+d^2p'_{\perp}}{(2\pi)^32p'^+} f_A(p'^+, \vec{p'}_{\perp}, y_1^-,\vec{y}_{\perp})\int_{-r_N}^{r_N} dy_{12}^-\\
&\times \int  d^2\vec{y}_{12\perp}  e^{-ix_2p^+y_{12}^- +i\vec{k}_{\perp}\cdot\vec{y}_{12\perp}}  \langle p' | A^{+}(y_1^-,\vec{y}_{12\perp})A^+(y_2^-, \vec{0}_{\perp})| p' \rangle  e^{i(x-x_1)p^+y_1^-},
\end{split}
\end{equation}
\end{widetext}
where the overall factor $C$ depends on the integration region of $y_1^-$ and $y_2^-$ [see Eq.~(\ref{eq:integration_region})] with $C=1$ for central cut diagram, and $C=1/2$ for left and right cut diagrams, $f_A(p'^+, \vec{p'}_{\perp}, y_1^-,\vec{y}_{1\perp})$ is the single nucleon phase space density distribution  \cite{Osborne:2002st,CasalderreySolana:2007sw} and the nucleon density is given by,
\begin{equation}
\begin{split}
\rho_A(y^-,\vec{y}_{\perp}) = \int \frac{dp'^+d^2p'_{\perp}}{(2\pi)^3} f_A(p'^+, \vec{p'}_{\perp}, y^-,\vec{y}_{\perp}),
\end{split}
\end{equation}
which is normalized as,
$$\int dy^-d^2\vec{y}_{\perp} \rho_A(y^-,\vec{y}_{\perp})=A.$$
By converting $k_{\perp}^2A^{+}A^+$ into gluon field strength $F_{\alpha}^{\;+} F^{+\alpha}$ through integration by part and defining the unintegrated gluon distribution function $\phi(x,\vec{k}_{\perp}) $ as

\begin{equation}
\begin{split}
\phi(x,\vec{k}_{\perp}) =& \int \frac{d y_{12}^-}{2\pi p^+} 
\int d^2  \vec{y}_{12\perp} e^{-ixp^+ y_{12}^- + i\vec{k}_{\perp}\cdot \vec{y}_{12\perp}}\\
& \times \langle p|  F_{\alpha}\ ^{+}( y_{12}^-, \vec{y}_{12\perp}) F^{+\alpha}(0, \vec{0}_{\perp}) |p\rangle ,
\label{eq:tmdgluon}
\end{split}
\end{equation}
one can simplify the quark-gluon correlation function as
\begin{equation}
\begin{split}
&T_{qg}^A(x,  x_1,x_2)=  \frac{C}{A} \pi \int dy^- d\vec{y}_{\perp} \rho_A(y^-,\vec{y_\perp}) f_q^A(x,\vec{y}_{\perp}) \\
&\times  \int  dy_1^-   \rho(y_1^-,\vec{y}_{\perp}) e^{i(x-x_1)p^+y_1^-} \frac{\phi(x_2,\vec{k}_{\perp})}{k_{\perp}^2},
\end{split}
\end{equation}

where we assume the nucleon size $|y_2^- -y_1^-|$ is much smaller than the nucleus size, $\rho(y_2)\approx \rho(y_1^-)$ and the averaged momentum of the single nucleon is $p'\approx p$.

The unintegrated or TMD gluon distribution function $\phi(x,\vec{k}_{\perp})$ in the factorized quark-gluon correlation function can also be related to TMD jet transport parameter. The jet transport parameter is defined as the averaged transverse momentum broadening squared per unit length,
\begin{equation}
\hat{q}_R = \langle \rho \int dk_{\perp}^2 \frac{d^2\sigma_R}{d k_{\perp}^2} k_{\perp}^2\rangle,
\label{eq:def-qhat}
\end{equation}
where $\rho$ is the density of the color source, while $\langle d\sigma_R \rangle$ is the differential cross section for scattering between a jet parton in color representation $R$ and medium partons from the color source averaged over the color source momentum. For DIS process, the medium color source is the nucleon inside a nucleus.
\begin{figure}
 \captionsetup{justification=raggedright,singlelinecheck=false}
  \includegraphics[scale=0.4]{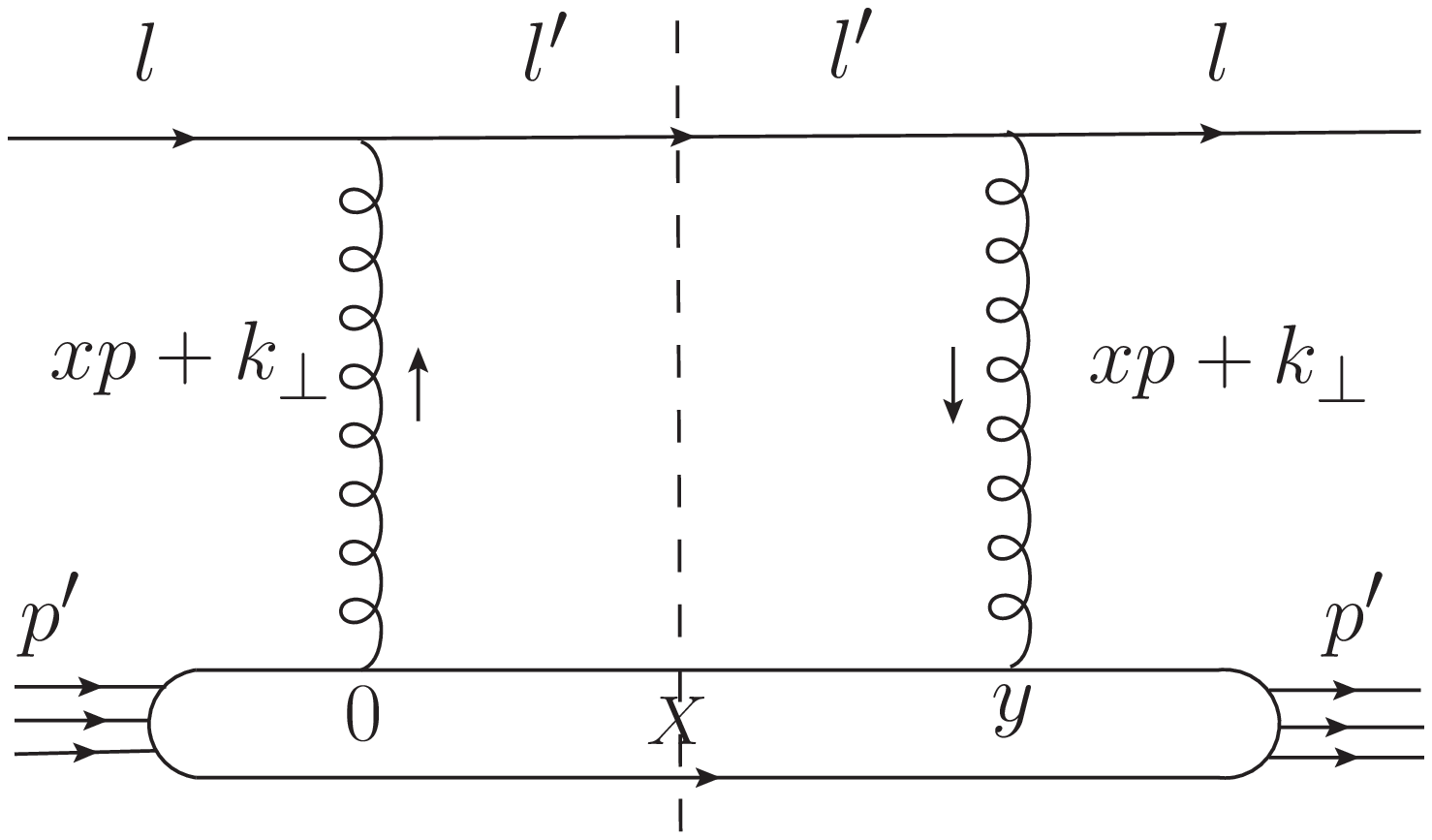}
  \caption{Scattering between jet parton and medium color source}
  \label{fig:qhat}
\end{figure}

In order to relate the medium TMD gluon distribution function to the jet transport parameter $\hat{q}$, we consider jet parton scattering off the medium color source as illustrated in Fig.~\ref{fig:qhat}. For unpolarized jet parton and medium color source, the initial spin is averaged and the final spin is summed. For color fields from the medium color source, the dominant component is $A^+$ and therefore, $A^{\mu} \approx (p^{\mu}/p^+) A^+$.
The momentum of medium color constituent is $p'=[p'^+,0,\vec{0}_{\perp}]$, the jet parton momentum is $l=[0,l^-,0]$, and the momentum transfer of the scattering is $xp'+k_{\perp}$. Note that we consider energy and longitudinal momentum transfer between the jet parton and the medium. The medium color source is therefore dynamic rather than static as in the GW static color-screened Yukawa potential model. Given the above kinematics and assumptions about the medium, the differential cross section for scattering  between the jet parton and medium is
\begin{widetext}
\begin{equation}
\begin{split}
&d \sigma_R = \int dx \delta(x-\frac{k_{\perp}^2}{2p'^+l^-}) \frac{C_2(R)}{N_c^2 -1}\frac{g^2}{2p'^+} \int \frac{d^2 k_{\perp}}{(2\pi)^2}   \int d y_1^-  e^{-ixp'^+y_1^-+i\vec{k}_{\perp}\cdot \vec{y}_{1\perp}}  \langle p'|A^+(y_1^-,\vec{y}_{1\perp})A^+(0)|p' \rangle.
\end{split}
\end{equation}
\end{widetext}
where the quadratic Casimir of the jet parton is denoted as $C_2(R)$ ($C_A=N_c$ for a gluon and $C_F=(N_c^2-1)/2N_c$ for quark), the color of the jet parton and the medium gluon are both averaged. Averaging over the momentum of the color source, we assume the momentum of the color source can be approximated by its average value $p'\approx \langle p' \rangle\equiv p$. 
According to the definition of jet transport parameter in Eq.~(\ref{eq:def-qhat}), one obtains from the above cross section,
\begin{equation}
\begin{split}
\hat{q}_R(y) = & \int \frac{d^2 k_{\perp}}{(2\pi)^2} \hat{q}_R(\vec{k}_{\perp},y), \\
\hat{q}_R(\vec{k}_{\perp},y)=& \int dx \delta(x-\frac{k_{\perp}^2}{2p^+l^-})  \frac{4\pi^2 \alpha_s C_2(R)}{N_c^2-1} \rho(y)  \phi(x, \vec{k}_{\perp}),
\end{split}
\end{equation}
where the unintegrated or TMD gluon distribution $\phi(x,\vec{k}_{\perp})$ \cite{CasalderreySolana:2007sw} is defined in Eq.~(\ref{eq:tmdgluon}). The TMD jet transport parameter $\hat{q}(\vec{k}_{\perp})$ should depend on the jet parton energy $l^-$ and the average momentum of the color source $p^+$ through $x$ in $\phi(x,\vec{k}_{\perp})$. It is also proportional to the local density of color source $\rho(y)$. In the limiting case when the energy transfer is small, i.e. $x\approx 0$, the jet transport parameter becomes,  
\begin{equation}
\begin{split}
\hat{q}_R(y) \approx \frac{4\pi^2 \alpha_s C_2(R)}{N_c^2-1} \rho(y) \int \frac{d^2 k_{\perp}}{(2\pi)^2}  \phi(0, \vec{k}_{\perp}).
\end{split}
\label{eq:qhat_nucleon}
\end{equation}

\bef
\captionsetup{justification=raggedright,singlelinecheck=false}
\includegraphics[scale=0.5]{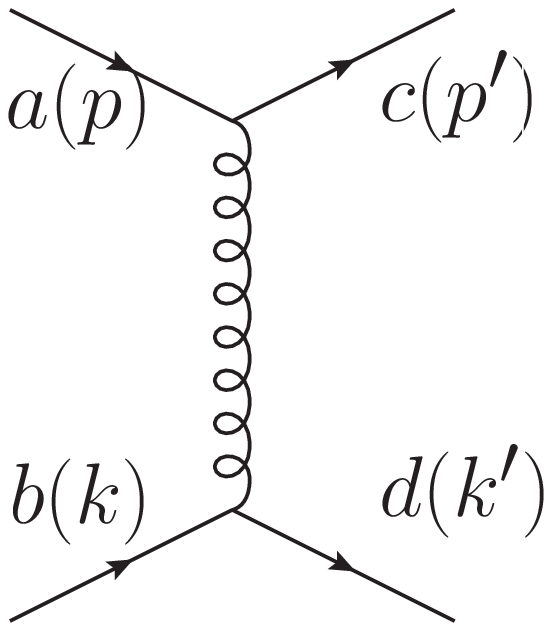} 
\includegraphics[scale=0.5]{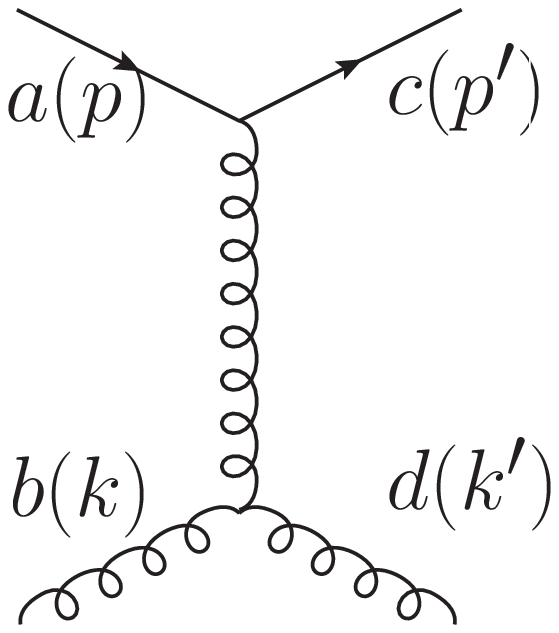} 
\caption{Elastic quark-quark and  quark-gluon scattering.}
\label{fig:Elastic}
\eef

Under small angle scattering approximation, the elastic cross section for jet and medium parton scattering, shown in Fig.~\ref{fig:Elastic} for quark-quark and quark-gluon scattering as an example, can be written as,
\begin{equation}
\begin{split}
d\sigma = & \frac{C_2(R)C_2(T)}{N_c^2-1}\frac{4\pi\alpha_s^2}{t^2}dt,
\end{split}
\end{equation}
where $C_2(R)$ and $C_2(T)$ are the quadratic Casimirs for jet and medium parton, respectively, and  $t=(p-p')^2=(k-k')^2$ is one of the Mandelstam variables. Including the Debye screening mass $\mu_D^2$ in the exchange gluons, it can be related to the transverse momentum $t\approx-k_\perp^2-\mu_D^2$ if one neglects the energy and longitudinal momentum transfer. According to Eq.~(\ref{eq:def-qhat}),  the jet transport parameter $\hat{q}$ for the jet parton scattering with a medium of partons with density $\rho$ in color representation $T$ is,
\begin{equation}
\hat{q}_R =  \rho \int dk_{\perp}^2\frac{C_2(R)C_2(T)}{N_c^2-1}\frac{4\pi\alpha_s^2}{(k_{\perp}^2+\mu_D^2)^2} k_{\perp}^2.
\label{eq:qhat_elastic}
\end{equation}
The corresponding unintegrated gluon distribution function with zero longitudinal momentum and energy transfer is,
\begin{equation}
\begin{split}
\frac{\phi(0, \vec{k}_{\perp})}{k_{\perp}^2} =  C_2(T)  \frac{4 \alpha_s}{(k_{\perp}^2+\mu_D^2)^2}.
\label{eq:phi(0,kperp)}
\end{split}
\end{equation}

\subsection{Radiative gluon spectrum}
\label{sec: gluon-spectrum}

With the factorized quark-gluon correlation function, one can express the differential hadronic tensor from double parton scattering in the SIDIS processes in terms of the TMD medium gluon distribution function or jet transport parameter as, 
\begin{widetext}
\begin{equation}
\begin{split}
\dfrac{dW^{\mu \nu}_{D(1)}}{dz_h}&= \dfrac{dW^{\mu \nu}_{D(1)q}}{dz_h}+\dfrac{dW^{\mu \nu}_{D(1)g}}{dz_h},\\
 \dfrac{dW^{\mu \nu}_{D(1)q}}{dz_h} &=  \frac{1}{A}  \int dx \int dy^- d^2\vec{y}_{\perp} \rho_A(y^-,\vec{y_\perp}) f_q^A(x,\vec{y}_{\perp})   H^{\mu\nu}_{(0)}(x)  \int \frac{dz}{z} D_{g \rightarrow h} (z_h/z) \int d l_{\perp}^2 \dfrac{d N_q}{d l_{\perp}^2 dz},\\
 \dfrac{dW^{\mu \nu}_{D(1)g}}{dz_h} &= \frac{1}{A}  \int dx \int dy^- d^2\vec{y}_{\perp} \rho_A(y^-,\vec{y}_{\perp}) f_q^A(x,\vec{y_\perp})  H^{\mu\nu}_{(0)}(x)  \int \frac{dz}{z} D_{q \rightarrow h} (z_h/z) \int d l_{\perp}^2 \dfrac{d N_g}{d l_{\perp}^2 dz}.
 \end{split}
\end{equation}
\end{widetext}
One can interpret the above as the hadronic tensor for hadron production from quark or gluon fragmentation in which the quark was first knocked out from the nucleus at position $(y^-,\vec{y_\perp})$ and scatters again with another gluon from the nucleus at $(y_1^-,\vec{y_\perp})$ with induced gluon radiations. The radiative gluon spectrum $\dfrac{d N_q}{d l_{\perp}^2 dz}$ and $\dfrac{d N_g}{d l_{\perp}^2 dz}$ has a relation
\begin{equation}
\dfrac{d N_g}{d l_{\perp}^2 dz} = \dfrac{d N_q}{d l_{\perp}^2 dz} (z \to 1-z)
\end{equation}
 The radiative gluon spectrum $d N_g/d l_{\perp}^2 dz$, which depends on the initial production position of the quark $(y^-, \vec{y_\perp})$,  is
\begin{widetext}
\begin{equation}
\begin{split}
\dfrac{d N_g}{d l_{\perp}^2 dz} = &\frac{\pi}{f_q^A(x)}\dfrac{\alpha_s}{2\pi}\dfrac{1+(1-z)^2}{z}\dfrac{2\pi\alpha_s}{N_c}    
 \int  \dfrac{d ^2k_{\perp}}{(2\pi)^2}  \int_{y^-}^{\infty} d y_1^- \rho_A(y_1^-,\vec{y}_{\perp})  \left[ \tilde{H}^D_C + \frac{1}{2}\tilde{H}^D_L + \frac{1}{2} \tilde{H}^D_R  \right],
\end{split}
\label{eq:gluon_spectrum}
\end{equation}
with 
\begin{equation}
\begin{split}
\tilde{H}^D_C =&\left\lbrace \left[ \left( \frac{C_A}{(\vec{l}_{\perp}-\vec{k}_{\perp})^2}f_q^A(x+x_L+\frac{x_D}{1-z}) - \frac{C_A}{l_{\perp}^2} f_q^A(x+x_L) \right)  \frac{ \phi(x_D,\vec{k}_{\perp})}{k_{\perp}^2} \right]  \right.\\
&+\left[  \left( \frac{C_F}{l_{\perp}^2} +C_A\frac{\vec{k}_{\perp}\cdot \vec{l}_{\perp}}{l_{\perp}^2(\vec{l}_{\perp}-\vec{k}_{\perp})^2 } +   \frac{C_F}{(\vec{l}_{\perp}-z\vec{k}_{\perp})^2}+ \frac{1}{N_c}\frac{\vec{l}_{\perp}\cdot (\vec{l}_{\perp} -z\vec{k}_{\perp})}{l_{\perp}^2(\vec{l}_{\perp}-z\vec{k}_{\perp})^2} \right. \right.  \\
&\left. \left. -C_A  \frac{(\vec{l}_{\perp} - \vec{k}_{\perp})\cdot (\vec{l}_{\perp} -z\vec{k}_{\perp})}{(l_{\perp}-k_{\perp})^2(\vec{l}_{\perp}-z\vec{k}_{\perp})^2} \right)\frac{  \phi(x_L+\frac{z}{1-z} x_D,\vec{k}_{\perp})}{k_{\perp}^2}f_q^A(x) \right]\\
& +\left[  \left( - \frac{C_A}{(\vec{l}_{\perp}-\vec{k}_{\perp})^2}
+ \frac{C_A}{2}\frac{\vec{l}_{\perp}\cdot(\vec{l}_{\perp}-\vec{k}_{\perp})}{l_{\perp}^2(\vec{l}_{\perp}-\vec{k}_{\perp})^2}
+  \frac{C_A}{2}\frac{(\vec{l}_{\perp} - \vec{k}_{\perp})\cdot (\vec{l}_{\perp} -z\vec{k}_{\perp})}{(l_{\perp}-k_{\perp})^2(\vec{l}_{\perp}-z\vec{k}_{\perp})^2} 
 \right)\right.\\
 &\left. \times \frac{ \phi(x_L+\frac{z}{1-z} x_D,\vec{k}_{\perp})}{k_{\perp}^2} e^{i(x_L+\frac{x_D}{1-z})p^+y_1^-} f_q^A(x+x_L+\frac{x_D}{1-z}) \right] \\
& +\left[  \left( - \frac{C_A}{(\vec{l}_{\perp}-\vec{k}_{\perp})^2}
+ \frac{C_A}{2}\frac{\vec{l}_{\perp}\cdot(\vec{l}_{\perp}-\vec{k}_{\perp})}{l_{\perp}^2(\vec{l}_{\perp}-\vec{k}_{\perp})^2}
+  \frac{C_A}{2}\frac{(\vec{l}_{\perp} - \vec{k}_{\perp})\cdot (\vec{l}_{\perp} -z\vec{k}_{\perp})}{(l_{\perp}-k_{\perp})^2(\vec{l}_{\perp}-z\vec{k}_{\perp})^2} 
 \right)\right.   \\
& \left. \left. \times\frac{ \phi(x_D,\vec{k}_{\perp})}{k_{\perp}^2} e^{-i(x_L+\frac{x_D}{1-z})p^+y_1^-} f_q^A(x) \right]  \right\rbrace,
\end{split}
\end{equation}
\begin{equation}
\begin{split}
\tilde{H}_L^D =&\left\lbrace \left[ \frac{C_A}{2} \frac{\vec{l}_{\perp}\cdot(\vec{l}_{\perp}-\vec{k}_{\perp})}{l_{\perp}^2(\vec{l}_{\perp}-\vec{k}_{\perp})^2} \left(   e^{i\frac{x_D}{1-z}p^+y_1^-} f_q^A(x+x_L+\frac{x_D}{1-z})-   e^{-i\frac{x_D}{1-z}p^+y_1^-} f_q^A(x+x_L)  \right)\frac{\phi(x_D,\vec{k}_{\perp})}{k_{\perp}^2}   \right] \right.\\
&+\left[   -\left(C_F\frac{1}{l_{\perp}^2}+  \frac{1}{2N_c}\frac{\vec{l}_{\perp}\cdot (\vec{l}_{\perp} -z\vec{k}_{\perp})}{l_{\perp}^2(\vec{l}_{\perp}-z\vec{k}_{\perp})^2}  \right)  \frac{\phi(x_D^0-x_L,\vec{k}_{\perp})}{k_{\perp}^2} e^{-ix_Lp^+y_1^-}f_q^A(x)\right.  \\
&\left. +  \frac{C_A}{2} \frac{\vec{l}_{\perp}\cdot[\vec{l}_{\perp}-(1-z)\vec{k}_{\perp}]}{l_{\perp}^2[\vec{l}_{\perp}-(1-z)\vec{k}_{\perp}]^2}\frac{\phi(x_D^0-x_L,\vec{k}_{\perp})}{k_{\perp}^2} e^{-ix_Lp^+y_1^-}f_q^A(x)  \right]\\
& +\left[  -\left(   \frac{1}{2N_c}\frac{\vec{l}_{\perp}\cdot [\vec{l}_{\perp} -z\vec{k}_{\perp}]}{l_{\perp}^2[\vec{l}_{\perp}-z\vec{k}_{\perp}]^2} + \frac{C_F}{l_{\perp}^2} - \frac{C_A}{2} \frac{\vec{l}_{\perp}\cdot(\vec{l}_{\perp}-\vec{k}_{\perp})}{l_{\perp}^2(\vec{l}_{\perp}-\vec{k}_{\perp})^2} \right)  \frac{\phi(x_L+\frac{z}{1-z}x_D,\vec{k}_{\perp})}{k_{\perp}^2}e^{ix_Lp^+y_1^-} f_q^A(x+x_L) \right. \\
&\left.\left. + \left(- \frac{C_A}{2}\frac{\vec{l}_{\perp}\cdot[\vec{l}_{\perp}-(1-z)\vec{k}_{\perp}]}{l_{\perp}^2[\vec{l}_{\perp}-(1-z)\vec{k}_{\perp}]^2} - \frac{C_A}{2}\frac{\vec{l}_{\perp}\cdot(\vec{l}_{\perp}-\vec{k}_{\perp})}{l_{\perp}^2(\vec{l}_{\perp}-\vec{k}_{\perp})^2} + \frac{C_A}{l_{\perp}^2} \right) \frac{ \phi(x_D,\vec{k}_{\perp})}{k_{\perp}^2}e^{-ix_Lp^+y_1^-}f_q^A(x)\right]  \right\rbrace,
\end{split}
\end{equation}
\begin{equation}
\begin{split}
\tilde{H}_R^D  = &\left\lbrace \left[  \frac{C_A}{2} \frac{\vec{l}_{\perp}\cdot(\vec{l}_{\perp}-\vec{k}_{\perp})}{l_{\perp}^2(\vec{l}_{\perp}-\vec{k}_{\perp})^2} \left( e^{-i\frac{x_D}{1-z}p^+y_1^-} f_q^A(x+x_L) - e^{i\frac{x_D}{1-z}p^+y_1^-} f_q^A(x+x_L+\frac{x_D}{1-z}) \right) \frac{  \phi(\frac{z}{1-z}x_D,\vec{k}_{\perp})}{k_{\perp}^2} \right] \right. \\
&+\left[  -\left(C_F\frac{1}{l_{\perp}^2}+  \frac{1}{2N_c}\frac{\vec{l}_{\perp}\cdot (\vec{l}_{\perp} -z\vec{k}_{\perp})}{l_{\perp}^2(\vec{l}_{\perp}-z\vec{k}_{\perp})^2}  \right) \frac{\phi(x_D^0,\vec{k}_{\perp})}{k_{\perp}^2} e^{ix_Lp^+y_1^-}f_q^A(x+x_L)   \right.\\
& \left.+  \frac{C_A}{2} \frac{\vec{l}_{\perp}\cdot[\vec{l}_{\perp}-(1-z)\vec{k}_{\perp}]}{l_{\perp}^2[\vec{l}_{\perp}-(1-z)\vec{k}_{\perp}]^2}\frac{  \phi(x_D^0,\vec{k}_{\perp})}{k_{\perp}^2} e^{ix_Lp^+y_1^-}f_q^A(x+x_L)  \right]\\
& +\left[ -\left(   \frac{1}{2N_c}\frac{\vec{l}_{\perp}\cdot (\vec{l}_{\perp} -z\vec{k}_{\perp})}{l_{\perp}^2(\vec{l}_{\perp}-z\vec{k}_{\perp})^2} + \frac{C_F}{l_{\perp}^2} - \frac{C_A}{2} \frac{\vec{l}_{\perp}\cdot(\vec{l}_{\perp}-\vec{k}_{\perp})}{l_{\perp}^2(\vec{l}_{\perp}-\vec{k}_{\perp})^2} \right)  \frac{  \phi(\frac{z}{1-z}x_D,\vec{k}_{\perp})}{k_{\perp}^2}e^{-ix_Lp^+y_1^-} f_q^A(x) \right. \\
&\left.\left. + \left(- \frac{C_A}{2}\frac{\vec{l}_{\perp}\cdot[\vec{l}_{\perp}-(1-z)\vec{k}_{\perp}]}{l_{\perp}^2[\vec{l}_{\perp}-(1-z)\vec{k}_{\perp}]^2} - \frac{C_A}{2}\frac{\vec{l}_{\perp}\cdot(\vec{l}_{\perp}-\vec{k}_{\perp})}{l_{\perp}^2(\vec{l}_{\perp}-\vec{k}_{\perp})^2} + \frac{C_A}{l_{\perp}^2} \right)\frac{  \phi(x_L+x_D ,\vec{k}_{\perp})}{k_{\perp}^2}e^{ix_Lp^+y_1^-}f_q^A(x+x_L)\right]  \right\rbrace,
\end{split}
\end{equation}
\end{widetext}
Note that the $1/2$ factor before $\tilde{H}_R^D$ and $\tilde{H}_L^D$ is from the constant $C$ in quark-gluon correlation function. One can find the corresponding expression of $\tilde{H}$ for each cut diagram in Appendix \ref{append-spectrum}.

The radiative parton energy loss can be expressed in terms of gluon radiation spectrum as
\begin{equation}
\begin{split}
\Delta E = E \int_0^{\mu^2} d l_{\perp}^2 \int_0^1 dz  \frac{d N_g}{dl_{\perp}^2 dz}z.
\end{split}
\end{equation}

\section{Soft and Static Approximations}
\label{sec-approx}

In order to simplify the final results for the radiative gluon spectrum in this study and compare to past results, we will consider three approximations: static scattering center approximation, soft radiative gluon approximation and the combination of these two. At the end of this section, we will also numerically compare results under these approximations.

\subsection{Static scattering center approximation}
For static scattering center approximation, we consider the energy transfer in the scattering between jet and medium parton  negligible as compared to the hard scattering energy scale, $x_B \gg x_L, {x_D}/{1-z}$ or $Q^2 \gg  {l_{\perp}^2}/{[z(1-z)]} , {k_{\perp}^2}/{[z(1-z)]}$  and $y_1^- - y_2^- \approx y^-$. Under these approximations,
\begin{equation}
\begin{split}
&\phi(x_L+x_D,\vec{k}_{\perp}) \approx  \phi(\frac{z}{1-z}x_D,\vec{k}_{\perp}) \approx  \phi(x_D^0,\vec{k}_{\perp}) \approx  \phi(0,\vec{k}_{\perp}),\\
&f_q^A(x_B+x_L+\frac{x_D}{1-z}) \approx f(x_B+x_L) \approx f_q^A(x_B).
\end{split}
\end{equation}
The radiative gluon spectrum is
\begin{widetext}
\begin{equation}
\begin{split}
\dfrac{d N_g^{\textsl{static}} }{d l_{\perp}^2 dz} = &\pi\dfrac{\alpha_s}{2\pi}\dfrac{1+(1-z)^2}{z}\dfrac{2\pi\alpha_s}{N_c}    
 \int  \dfrac{d ^2k_{\perp}}{(2\pi)^2}  \int d y_1^-  \rho_A(y_1^-,\vec{y}_{\perp})  \left[ (\tilde{H}^D_C)_{\textsl{static}} + \frac{1}{2}(\tilde{H}^D_L)_{\textsl{static}} + \frac{1}{2}(\tilde{H}^D_R)_{\textsl{static}} \right]\frac{\phi(0,\vec{k}_{\perp})}{{k}_{\perp}^2},
\label{eq:gluon_spectrum_static}
\end{split}
\end{equation}

Summing up contributions from central, right and left cut diagrams, one finds that many terms cancel and gets the final gluon spectrum as, 
\begin{equation}
\begin{split}
\dfrac{d N_g^{\textsl{static}} }{d l_{\perp}^2 dz}= &\pi\dfrac{\alpha_s}{2\pi}\dfrac{1+(1-z)^2}{z}\dfrac{2\pi\alpha_s}{N_c}    
 \int  \dfrac{d ^2k_{\perp}}{(2\pi)^2}  \int d y_1^- 
\rho(y_1^-,\vec{y}_{1\perp})  \left\lbrace  C_F \left[ \frac{1}{(\vec{l}_{\perp}-z\vec{k}_{\perp})^2} -  \frac{1}{{l}_{\perp}^2} \right]  \right.\\
&+C_A \left[  \frac{2}{(\vec{l}_{\perp}-\vec{k}_{\perp})^2} - \frac{\vec{l}_{\perp}\cdot(\vec{l}_{\perp}-\vec{k}_{\perp})}{l_{\perp}^2(\vec{l}_{\perp}-\vec{k}_{\perp})^2}
-  \frac{(\vec{l}_{\perp} - \vec{k}_{\perp})\cdot (\vec{l}_{\perp} -z\vec{k}_{\perp})}{(l_{\perp}-k_{\perp})^2(\vec{l}_{\perp}-z\vec{k}_{\perp})^2}\right]   \times (1-\cos[(x_L+\frac{x_D}{1-z})p^+y_1^-])  \\
 &\left. + \frac{1}{N_c}\left[   \frac{\vec{l}_{\perp}\cdot (\vec{l}_{\perp} -z\vec{k}_{\perp})}{l_{\perp}^2(\vec{l}_{\perp}-z\vec{k}_{\perp})^2}  - \frac{1}{l_{\perp}^2}  \right]\times \left(1- \cos[x_Lp^+y_1^-]\right)     \right\rbrace  \frac{\phi(0,\vec{k}_{\perp})}{{k}_{\perp}^2}. \\
\end{split}
\end{equation}

\subsection{Soft radiative gluon approximation}
Under the soft gluon approximation $z\ll 1$, one only keeps the leading terms when $ z \rightarrow 0 $. The splitting function becomes $P_{qg}(z) = [1+(1-z)^2]/z\approx 2/z$. The radiative gluon spectrum is,
\begin{equation}
\begin{split}
\dfrac{d N_g^{\textsl{soft}}}{d l_{\perp}^2 dz} = &\frac{\pi}{f_q^A(x)}\dfrac{\alpha_s}{2\pi}P_{qg}(z)\dfrac{2\pi\alpha_s}{N_c}    
 \int  \dfrac{d ^2k_{\perp}}{(2\pi)^2}  \int d y_1^- 
\rho_A(y_1^-,\vec{y}_{\perp}) \\
& \times\left[ (\tilde{H}^D_C)_{\textsl{soft}}+ \frac{1}{2} (\tilde{H}^D_L)_{\textsl{soft}} +  \frac{1}{2}  (\tilde{H}^D_R)_{\textsl{soft}} \right] ,\\
\label{eq:gluon_spectrum_soft}
\end{split}
\end{equation}
where
\begin{equation}
\begin{split}
(\tilde{H}^D_C)_{\textsl{soft}} =&\left\lbrace \left[ \left(  \frac{C_A}{(\vec{l}_{\perp}-k_{\perp})^2}  f_q^A(x+x_L+\frac{x_D}{1-z}) - \frac{C_A}{l_{\perp}^2}f_q^A(x+x_L) \right) \frac{ \phi(\frac{z}{1-z}x_D,\vec{k}_{\perp})}{k_{\perp}^2}\right]  \right.\\
&+\left[   C_A\frac{k_{\perp}^2}{l_{\perp}^2(\vec{l}_{\perp}-\vec{k}_{\perp})^2 } \frac{  \phi(x_L+\frac{z}{1-z}x_D,\vec{k}_{\perp})}{k_{\perp}^2}f_q^A(x) \right]\\
& +\left[  -C_A \frac{\vec{k}_{\perp}\cdot \vec{l}_{\perp} }{l_{\perp}^2(\vec{l}_{\perp}-\vec{k}_{\perp})^2}
 \frac{ \phi(x_L+\frac{z}{1-z}x_D,\vec{k}_{\perp})}{k_{\perp}^2} e^{i(x_L+\frac{x_D}{1-z})p^+y_1^-} f_q^A(x+x_L+\frac{x_D}{1-z}) \right] \\
&  \left. +\left[ -C_A \frac{\vec{k}_{\perp}\cdot \vec{l}_{\perp} }{l_{\perp}^2(\vec{l}_{\perp}-\vec{k}_{\perp})^2} \frac{ \phi(x_D,\vec{k}_{\perp})}{k_{\perp}^2} e^{-i(x_L+\frac{x_D}{1-z})p^+y_1^-} f_q^A(x) \right]  \right\rbrace ,\\ 
\end{split}
\end{equation}

\begin{equation}
\begin{split}
(\tilde{H}^D_L)_{\textsl{soft}}  = &\left\lbrace \left[ \frac{C_A}{2} \frac{\vec{l}_{\perp}\cdot(\vec{l}_{\perp}-\vec{k}_{\perp})}{l_{\perp}^2(\vec{l}_{\perp}-\vec{k}_{\perp})^2}  \left( e^{i\frac{x_D}{1-z}p^+y_1^-} f_q^A(x+x_L+\frac{x_D}{1-z}) -   e^{-i\frac{x_D}{1-z}p^+y_1^-} f_q^A(x+x_L)   \right)\frac{\phi(x_D,\vec{k}_{\perp})}{k_{\perp}^2}  \right] \right. \\
&+\left[ -  \frac{C_A}{2}\frac{1}{l_{\perp}^2}\frac{\phi(x_D^0-x_L,\vec{k}_{\perp})}{k_{\perp}^2} e^{-ix_Lp^+y_1^-}f_q^A(x)+  \frac{C_A}{2} \frac{\vec{l}_{\perp}\cdot(\vec{l}_{\perp}-\vec{k}_{\perp})}{l_{\perp}^2(\vec{l}_{\perp}-\vec{k}_{\perp})^2}\frac{\phi(x_D^0-x_L,\vec{k}_{\perp})}{k_{\perp}^2} e^{-ix_Lp^+y_1^-}f_q^A(x)  \right]\\
&\left. +\left[  - \frac{C_A}{2}\frac{{k}_{\perp}^2 -\vec{l}_{\perp}\cdot \vec{k}_{\perp}}{l_{\perp}^2(\vec{l}_{\perp}-\vec{k}_{\perp})^2} \frac{\phi(x_L+\frac{z}{1-z} x_D,\vec{k}_{\perp})}{k_{\perp}^2}e^{ix_Lp^+y_1^-} f_q^A(x+x_L)  +C_A  \frac{{k}_{\perp}^2 -\vec{l}_{\perp}\cdot \vec{k}_{\perp}}{l_{\perp}^2(\vec{l}_{\perp}-\vec{k}_{\perp})^2} \frac{ \phi(x_D,\vec{k}_{\perp})}{k_{\perp}^2}e^{-ix_Lp^+y_1^-}f_q^A(x)\right]  \right\rbrace, \\
\end{split}
\end{equation}

\begin{equation}
\begin{split}
(\tilde{H}^D_R)_{\textsl{soft}}  =&\left\lbrace \left[  \frac{C_A}{2} \frac{\vec{l}_{\perp}\cdot(\vec{l}_{\perp}-\vec{k}_{\perp})}{l_{\perp}^2(\vec{l}_{\perp}-\vec{k}_{\perp})^2} \left(  e^{-i\frac{x_D}{1-z}p^+y_1^-} f_q^A(x+x_L)-  e^{i\frac{x_D}{1-z}p^+y_1^-} f_q^A(x+x_L+\frac{x_D}{1-z}) \right) \frac{\phi(\frac{z}{1-z}x_D,\vec{k}_{\perp})}{k_{\perp}^2}  \right]  \right. \\
&+\left[  -\frac{C_A}{2}\frac{1}{l_{\perp}^2} \frac{\phi(x_D^0,\vec{k}_{\perp})}{k_{\perp}^2} e^{ix_Lp^+y_1^-}f_q^A(x+x_L) +  \frac{C_A}{2} \frac{\vec{l}_{\perp}\cdot(\vec{l}_{\perp}-\vec{k}_{\perp})}{l_{\perp}^2(\vec{l}_{\perp}-\vec{k}_{\perp})^2}\frac{  \phi(x_D^0,\vec{k}_{\perp})}{k_{\perp}^2} e^{ix_Lp^+y_1^-}f_q^A(x+x_L)  \right]\\
&\left. +\left[ - \frac{C_A}{2}\frac{{k}_{\perp}^2 -\vec{l}_{\perp}\cdot \vec{k}_{\perp}}{l_{\perp}^2(\vec{l}_{\perp}-\vec{k}_{\perp})^2}   \frac{  \phi(\frac{z}{1-z}x_D,\vec{k}_{\perp})}{k_{\perp}^2}e^{-ix_Lp^+y_1^-} f_q^A(x)  + C_A  \frac{{k}_{\perp}^2 -\vec{l}_{\perp}\cdot \vec{k}_{\perp}}{l_{\perp}^2(\vec{l}_{\perp}-\vec{k}_{\perp})^2}\frac{  \phi(x_L+x_D ,\vec{k}_{\perp})}{k_{\perp}^2}e^{ix_Lp^+y_1^-}f_q^A(x+x_L)\right]  \right\rbrace .\\ 
\end{split}
\end{equation}

\subsection{Static scattering center + soft gluon approximation}
 Under the static scattering center + soft gluon approximation, 
%
the radiative gluon spectrum is, 
\begin{equation}
\begin{split}
\dfrac{d N_g^{\textsl{static+soft}}}{d l_{\perp}^2 dz} =& \pi\dfrac{\alpha_s}{2\pi}P_{qg}(z)\dfrac{2\pi\alpha_s}{N_c}    
 \int  \dfrac{d ^2k_{\perp}}{(2\pi)^2}  \int d y_1^- \rho_A(y_1^-,\vec{y}_{\perp})\\
 &\left[ (\tilde{H}^D_C)_{\textsl{static+soft}} + \frac{1}{2}(\tilde{H}^D_L)_{\textsl{static+soft}} + \frac{1}{2}(\tilde{H}^D_R)_{\textsl{static+soft}} \right] \frac{\phi(0,\vec{k}_{\perp})}{k_{\perp}^2},
\label{eq:gluon_spectrum_static_soft}
\end{split}
\end{equation}

Similarly as in the static scattering center approximation, many terms in the central, right and left cut diagrams cancel. One can get a simple expression for the radiative gluon spectrum,
\begin{equation}
\begin{split}
\dfrac{d N_g^{\textsl{static+soft}}  }{d l_{\perp}^2 dz} = & \pi\dfrac{\alpha_s}{2\pi}P_{qg}(z)\dfrac{2\pi\alpha_s}{N_c}    
 \int  \dfrac{d ^2k_{\perp}}{(2\pi)^2}  \int d y_1^- \rho_A(y_1^-,\vec{y}_{\perp}) C_A \frac{2\vec{k}_{\perp}\cdot\vec{l}_{\perp}}{l_{\perp}^2(\vec{l}_{\perp}-\vec{k}_{\perp})^2} \left( 1-\cos[(x_L+\frac{x_D}{1-z})p^+y_1^-]\right)  \frac{\phi(0,\vec{k}_{\perp})}{k_{\perp}^2}.
\end{split}
\end{equation}
The above result is very similar to the GLV result under the first opacity approximation \cite{Gyulassy:1999zd,Gyulassy:2000fs}. To have an exact comparison, we also consider a static screened potential model for scattering between the jet parton and the medium parton. Under this model, we will substitute the unintegrated gluon distribution with zero longitudinal momentum and energy transfer $\phi(0,\vec{k}_{\perp})$ from Eq.~(\ref{eq:phi(0,kperp)}) in the above expression and obtain,
%
\begin{equation}
\begin{split}
\dfrac{d N_g^{\textsl{static+soft}} }{d l_{\perp}^2 dz} = &8\pi\alpha_s^3 \frac{C_2(T)C_A}{N_c}P_{qg}(z) 
 \int  \dfrac{d ^2k_{\perp}}{(2\pi)^2}  \int d y_1^-  \rho_A(y_1^-,\vec{y}_{\perp}) \\ 
&\times \frac{\vec{k}_{\perp}\cdot\vec{l}_{\perp}}{l_{\perp}^2(\vec{l}_{\perp}-\vec{k}_{\perp})^2} \left( 1-\cos[(x_L+\frac{x_D}{1-z})p^+y_1^-]\right)  \frac{1}{(k_{\perp}^2+\mu_D^2)^2}.
\end{split}
\label{eq:High Twist approximation to GLV}
\end{equation}

The radiative gluon number distribution from the GLV result in the first order  opacity approximation can be cast in a similar expression \cite{Wang:2001cs},
\begin{equation}
\begin{split}
\frac{d N_g^{\rm GLV}}{d z dl_{\perp}^2} = &8\pi \alpha_s^3\frac{C_2(T)  C_A}{N_c}P_{qg}(z) \int \frac{d^2 k_{\perp}}{(2\pi)^2}  \frac{N}{A_{\perp}}   \int dy_{10} \bar{\rho}(y_{10})\frac{\vec{k}_{\perp}\cdot\vec{l}_{\perp}}{{l}_{\perp}^2(\vec{l}_{\perp} - \vec{k}_{\perp})^2} \left(1-\cos[\omega_1 y_{10}]\right) \frac{1}{(k_{\perp}^2+\mu_D^2)^2},
\end{split}
\label{eq:GLV first order radiation}
\end{equation}
\end{widetext}
where the scattering kernel in static color-screened Yukawa potential is

\begin{equation}
\begin{split}
v(\vec{k}_{\perp}) =\frac{4\pi\alpha_s}{{k}_{\perp}^2+\mu_D^2}.
\end{split}
\end{equation}

The arguments in the cosine function are
\begin{equation}
\begin{split}
\omega_1 =& \frac{El_{\perp}^2}{2w(E-w)} - \frac{l_{\perp}^2}{2w}+\frac{(\vec{l}_{\perp}-\vec{k}_{\perp})^2}{2w} \\
=& \frac{l_{\perp}^2}{2\frac{q^-}{\sqrt{2}}z(1-z)} + \frac{k_{\perp}^2-2\vec{l}_{\perp}\cdot\vec{k}_{\perp}}{2\frac{q^-}{\sqrt{2}}z} \\
=& \sqrt{2}(x_L+x_D)p^+ \approx \sqrt{2}(x_L+\frac{x_D}{1-z})p^+ ,
\end{split}
\end{equation}
\begin{equation}
\begin{split}
y_{10} =& y_1 - y_0 \approx \frac{y_1^-}{\sqrt{2}},\\
\cos[\omega_1 y_{10}] \approx & \cos[(x_L+\frac{x_D}{1-z})p^+y_1^-].
\end{split}
\end{equation}
The density $\bar{\rho}(y_{10})$ in the GLV result is the normalized distribution of $N$ number of scattering centers over the transverse area $A_\perp$. It can be related to the color source density in our calculation as,
\begin{equation}
\frac{N}{A_{\perp}}   \int dy_{10} \bar{\rho}(y_{10}) 
=  \int dy_{1}^-\rho_A(y_{1}^-,\vec{y}_{\perp}).
\end{equation}
Under these approximations, our result in Eq.~(\ref{eq:High Twist approximation to GLV}) recovers that of GLV in Eq.~(\ref{eq:GLV first order radiation}) in the first order opacity approximation.

\subsection{Numerical comparisons of soft and static approximation}
\label{sec-num}

To investigate the effect of soft gluon and static scattering center approximations numerically, we define a dimensionless scaled spectrum $\mathcal{N}_g$  for the induced gluon radiation per mean-free-path,
\begin{equation}
\begin{split}
\dfrac{d N_g }{d l_{\perp}^2 dz}= &
  \int_{y^-}^\infty d y_1^- \left[ \rho_A(y_1^-,\vec{y}_{\perp}) \dfrac{2\pi\alpha_s}{N_c}  \pi \int  \dfrac{d k^2_{\perp}}{(2\pi)^2}  \frac{\phi(0,\vec{k}_{\perp})}{{k}_{\perp}^2} \right]\\
\times&\pi\dfrac{\alpha_s}{2\pi}P_{qg}(z)\frac{C_A}{l_{\perp}^2}\mathcal{N}_g,
\end{split}
\end{equation}
where the azimuthal angle $\varphi$ between the transverse momentum $\vec{k}_\perp$ of the initial medium gluon and $\vec{l}_\perp$ of the radiated gluon is averaged in $\mathcal{N}_g$. According to Eq.~(\ref{eq:qhat_nucleon}), the integrant inside the square brackets in the first line in the above equation  is the inverse of mean-free-path of quark-medium interaction or the scattering rate.

Under the static scattering center approximation, the scaled spectrum is
\begin{widetext}
\begin{equation}
\begin{split}
\mathcal{N}_g^{\textsl{static}}
=&  \int \frac{d\varphi}{2\pi} \frac{l_{\perp}^2}{C_A} \left\lbrace C_F\left[\frac{1}{(\vec{l}_{\perp}-z\vec{k}_{\perp})^2} -  \frac{1}{{l}_{\perp}^2}\right] +C_A \left[  \frac{2}{(\vec{l}_{\perp}-\vec{k}_{\perp})^2} - \frac{\vec{l}_{\perp}\cdot(\vec{l}_{\perp}-\vec{k}_{\perp})}{l_{\perp}^2(\vec{l}_{\perp}-\vec{k}_{\perp})^2}
-  \frac{(\vec{l}_{\perp} - \vec{k}_{\perp})\cdot (\vec{l}_{\perp} -z\vec{k}_{\perp})}{(l_{\perp}-k_{\perp})^2(\vec{l}_{\perp}-z\vec{k}_{\perp})^2}\right]    \right. \\
&\left. \times (1-\cos[(x_L+\frac{x_D}{1-z})p^+y_1^-])+  \frac{1}{N_c} \left[ \frac{\vec{l}_{\perp}\cdot (\vec{l}_{\perp} -z\vec{k}_{\perp})}{l_{\perp}^2(\vec{l}_{\perp}-z\vec{k}_{\perp})^2}  - \frac{1}{l_{\perp}^2}  \right]\times \left(1- \cos[x_Lp^+y_1^-]\right)\right\rbrace
\end{split}
\label{eq:ghtscc}
\end{equation}
 
With both static scattering center and soft gluon approximations (GLV result in the first order opacity expansion), the scaled gluon spectrum is
\begin{equation} 
\mathcal{N}_g^{\textsl{static+soft}}
 =\int \frac{d\varphi}{2\pi} \frac{2\vec{k}_{\perp}\cdot\vec{l}_{\perp}}{(\vec{l}_{\perp}-\vec{k}_{\perp})^2} \left( 1-\cos[(x_L+\frac{x_D}{1-z})p^+y_1^-]\right) .
 \label{eq:glvscc}
\end{equation}
\end{widetext}

The first term in Eq. (\ref{eq:ghtscc}) comes from the induced gluon radiation off the initial gluon (with three-gluon vertex). The spectrum inside the square brackets has a
collinear divergence at $\vec{l}_{\perp} = \vec{k}_{\perp}$ when the intermediate gluon (gluon propagator) is collinear to the initial quark. Note that
\begin{equation}
 x_L+\frac{x_D}{1-z}=\frac{(\vec{l}_{\perp} -\vec{k}_{\perp})^2}{2p^+q^-z(1-z)}.
 \end{equation}
The cosine function from the Landau-Pomeranchuk-Migdal (LPM) interference \cite{Landau:1953gr,Migdal:1956tc} regularizes the divergence at $\vec{l}_\perp =\vec{k}_\perp$ when the formation time of the intermediate gluon $\tau_{f} = 1/[x_L+x_D/(1-z)]p^+ = 2q^-z(1-z)/(\vec{l}_{\perp}-\vec{k}_{\perp})$
 becomes infinite. 

Similarly, the second term in Eq. (\ref{eq:ghtscc}) comes from gluon radiation off the quark lines (both initial and final) during the quark-gluon interaction. The divergence of the spectrum inside the square brackets at $\vec{l}_{\perp} = 0$ is also regularized by the cosine function from LPM interference when the formation time of the final gluon $\tau_f =1/x_Lp^+ = 2q^-z(1-z)/l_{\perp}^2$ is infinite.

In general, terms with $1/l_{\perp}^2$ arise when the final gluon from the initial state radiation during the quark-gluon scattering is emitted from the struck quark after photon-quark scattering. When $\vec{l}_{\perp}=0$, the radiated gluon is collinear to the struck quark. Similarly, terms with $1/(\vec{l}_{\perp} -z\vec{k}_{\perp})^2$ come from the final state gluon radiation come from the final state gluon radiation of quark-gluon scattering. When $\vec{l}_{\perp} =z\vec{k}_{\perp}$, gluon’s momentum $l=[l_{\perp}^{2} /(z q^{-}), z q^{-}, \vec{l}_{\perp}]=[z k_{\perp}^{2} / q^{-}, z q^{-}, z \vec{k}_{\perp}]$ is collinear to the final quark which has a momentum $l_q =  [(\vec{k}_{\perp}-\vec{l}_{\perp})^{2} /\left((1-z) q^{-}\right),(1-z) q^{-}, \vec{k}_{\perp}-\vec{l}_{\perp}]=[(1-z) k_{\perp}^{2} / q^{-},(1-z) q^{-},(1-z) \vec{k}_{\perp}]=(1-z) l / z$. These terms cancel with each other when $z=0$. One is left with gluon spectra in Eq. (\ref{eq:glvscc}) from gluon radiation off the three-gluon vertex and its interference with gluon radiation off quark lines. For finite gluon momentum fraction z, contributions from gluon radiation off the quark lines do not vanish. The collinear divergencies at $\vec{l}_{\perp} = 0$ and $\vec{l}_{\perp} = z\vec{k}_{\perp}$ may be regularized through renormalization of the quark-gluon correlation function and the final quark fragmentation function.

Since the azimuthal angle between $\vec k_\perp$ and $\vec l_\perp$ is averaged over in $\mathcal{N}_g$, the dimensionless scaled spectrum should be a function
of the scaled transverse momentum 
\begin{equation}
\tilde k_{\perp l} \equiv k_\perp/l_\perp,
\end{equation}
scaled propagation length 
\begin{equation}
\tilde y_\tau\equiv \frac{y_1^- l_\perp^2}{2q^-z(1-z)} \equiv \frac{y_1^-}{\tau_f},
\end{equation}
and momentum fraction $z$, where $\tau_f=2q^-z(1-z)/l_\perp^2$ is the gluon formation time.

\begin{figure}
\captionsetup{justification=raggedright,singlelinecheck=false}
\begin{center}
\includegraphics[scale=0.5]{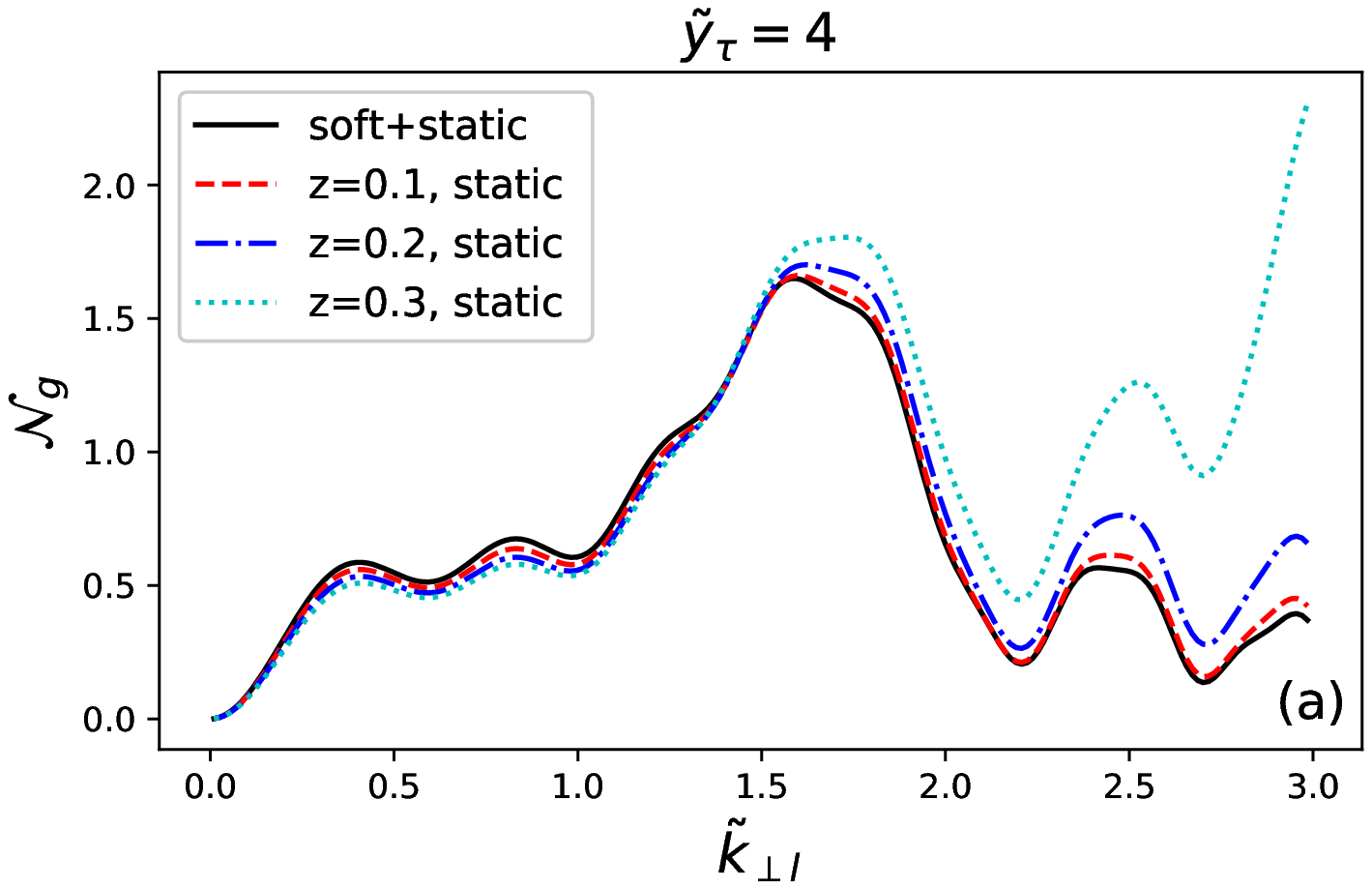} \\
\includegraphics[scale=0.5]{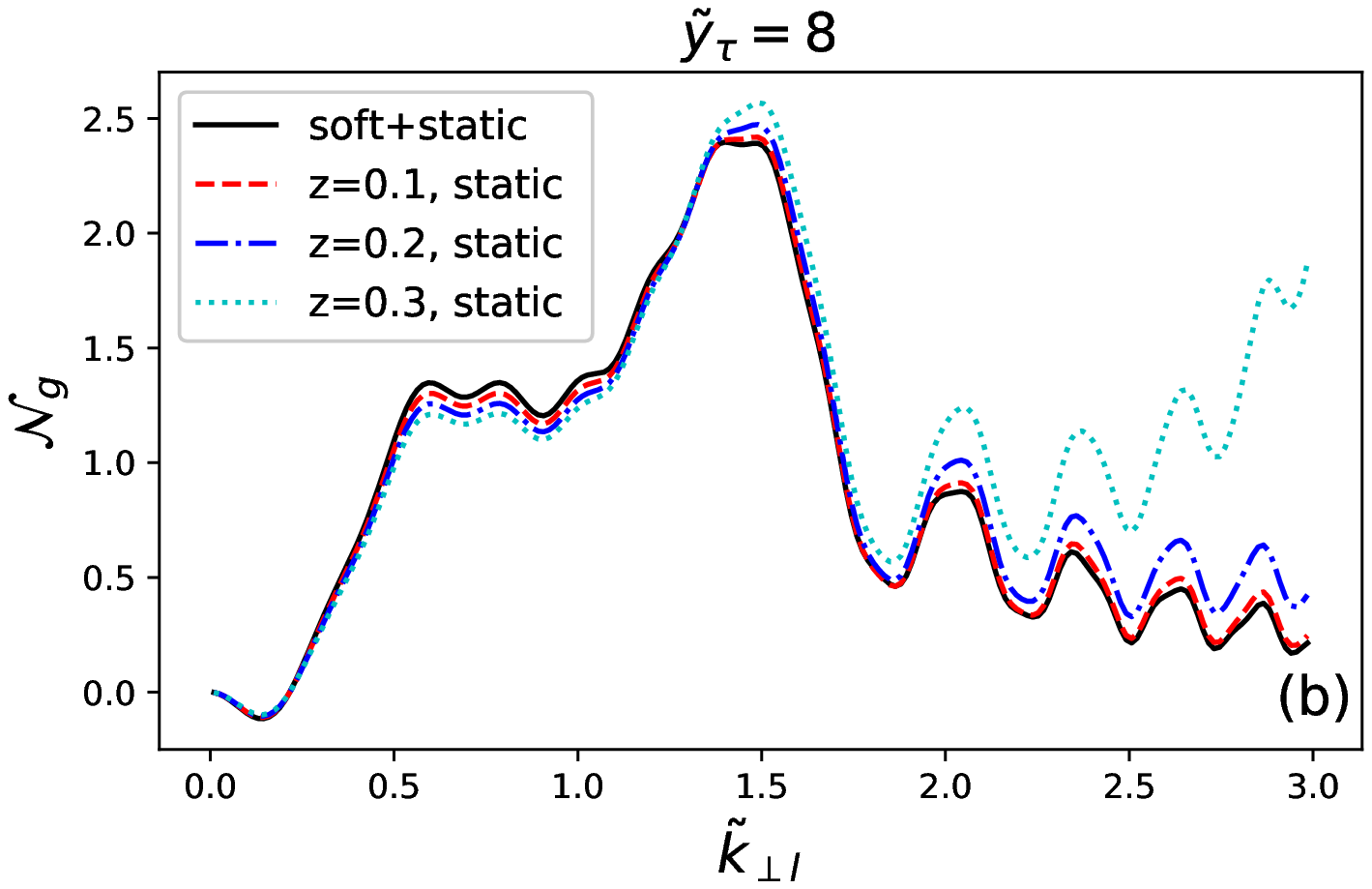} 
\end{center}
 \caption{The scaled gluon spectrum $\mathcal{N}_g$ in GHT calculation with static scattering center approximation (dashed, dot-dashed and dotted lines) and with static+soft gluon approximation (solid lines) with different momentum fraction $z$ as a function of the scaled transverse momentum $\tilde k_{\perp l}$ with fixed  scaled propagation length (a) $\tilde y_\tau=4$ (b) and 8.}
  \label{fig:integration_kernel1}
\end{figure}

\begin{figure}
\captionsetup{justification=raggedright,singlelinecheck=false}
\begin{center}
\includegraphics[scale=0.5]{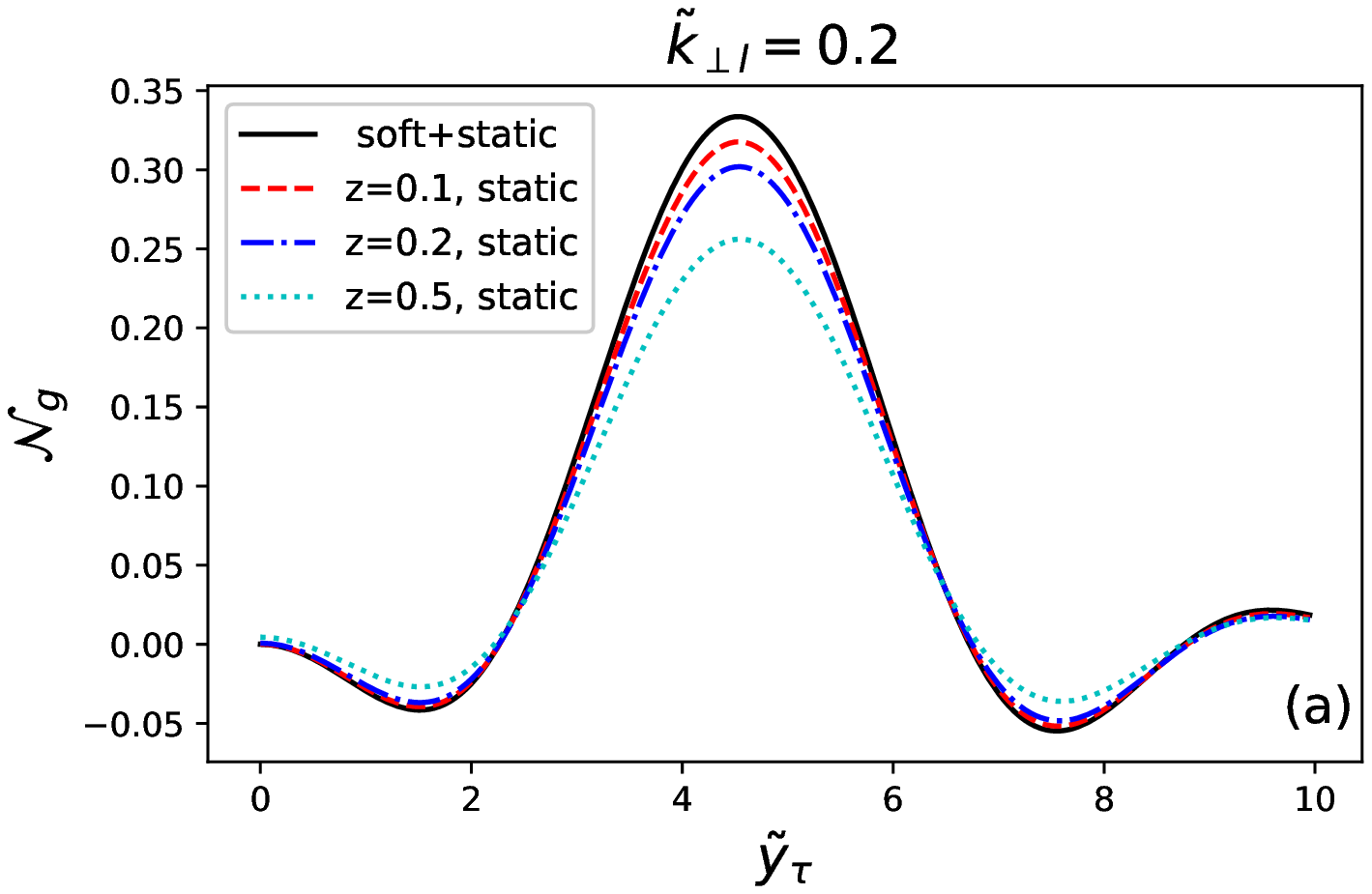} \\
\includegraphics[scale=0.5]{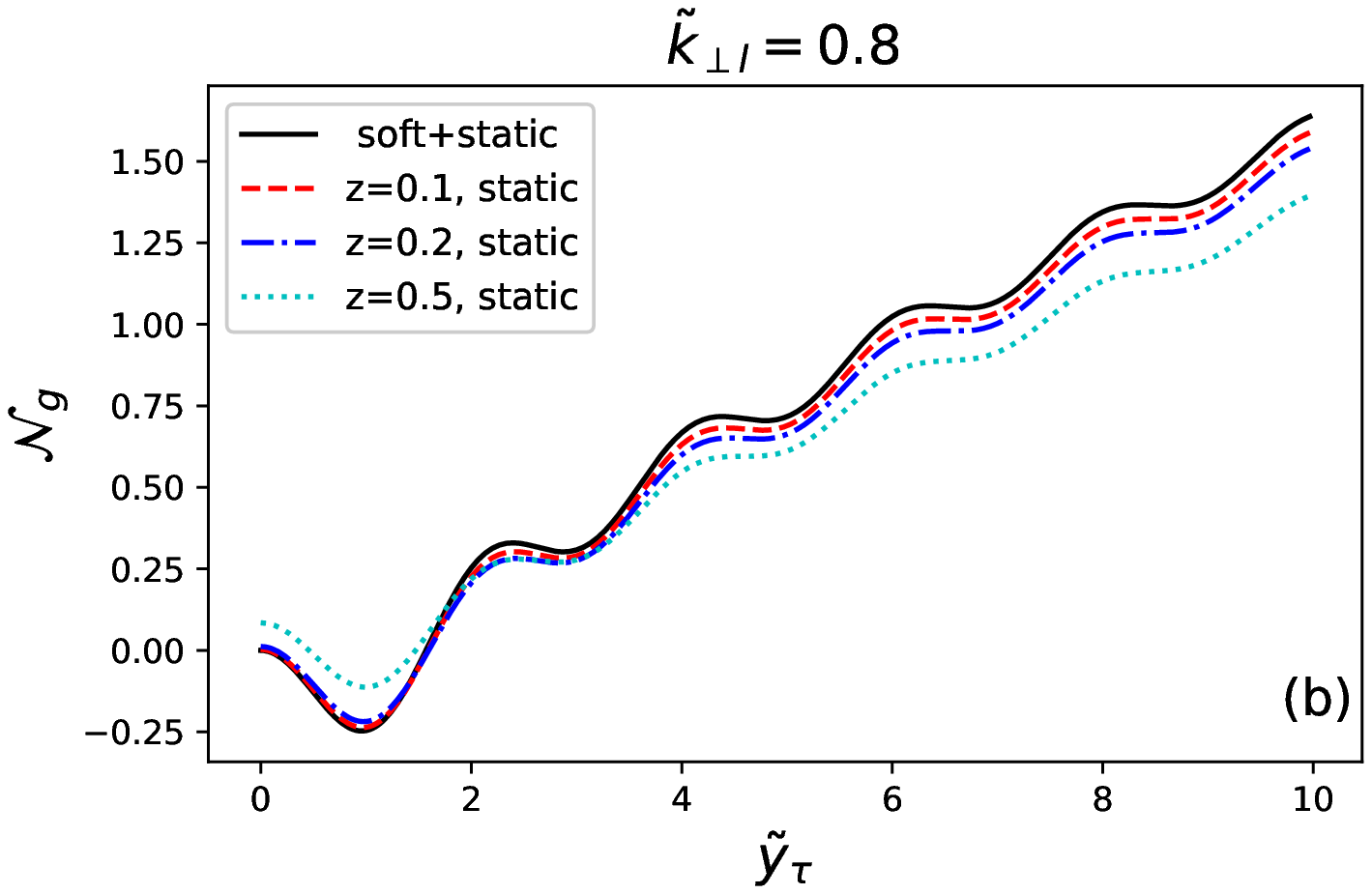} 
\end{center}
 \caption{The GHT scaled gluon spectrum $\mathcal{N}_g$ in this study with static scattering center approximation (dashed, dot-dashed and dotted lines) and with static+soft gluon approximation (solid lines) with different momentum fraction $z$ as a function of the scaled propagation length $\tilde y_\tau$ with fixed scaled transverse momentum $(a) \tilde k_{\perp l}=0.2$  (b) and 0.8. }
  \label{fig:integration_kernel2}
\end{figure}

We plot in Figs.~\ref{fig:integration_kernel1}(a) and \ref{fig:integration_kernel1}(b) the scaled GHT gluon spectrum $\mathcal{N}_g$ with static scattering center approximation (dashed lines) and static scattering center + soft gluon approximation or GLV result (solid lines) as a function of $\tilde k_{\perp l}$ for fixed propagation length $\tilde y_\tau=4,8$ and different momentum fractions $z=0.1,0.2,0.5$. The static+soft approximation or GLV result for fixed scaled propagation length $\tilde y_\tau$ does not depend on momentum fraction $z$ as shown by the solid lines. One can see the difference between GLV and GHT results with static scattering center approximation is very small for small momentum fraction $z\ll 1$. The difference becomes appreciable for large values of $z$ as the scaled transverse momentum $\tilde k_{\perp l}$ approaches $1/z$, the location of the collinear divergence in the final state radiation when the radiated gluon becomes collinear to the final quark. The oscillatory behavior comes from the cosine function in the spectrum due to  the LPM interference.  We also show in Figs.~\ref{fig:integration_kernel2}(a) and \ref{fig:integration_kernel2}(b) the  scaled gluon spectrum $\mathcal{N}_g$ as a function of the scaled propagation length $\tilde y_\tau$ for fixed $\tilde k_{\perp l}=0.2, 0.8$ and different momentum fractions $z=0.1,0.2,0.5$. The difference between GLV and GHT results with static scattering center approximation again becomes appreciable at a large momentum fraction.

\section{Summary}
\label{sec-sum}

We have revisited parton energy loss in the \textit{eA} deeply inelastic scattering process. In our study, one does not carry out collinear expansion of the hard partonic part of the parton-medium scattering and induced gluon radiation as in the collinear factorized approach. The final radiative gluon spectra induced by multiple parton scattering can be expressed in terms of a convolution of the hard partonic part and TMD gluon distribution density inside the nucleus. In general, the final GHT results on radiative gluon spectra can include the effect of a dynamic medium with both energy and transverse momentum transfer between the propagating parton and the medium. We have considered several limits of the final results under the static scattering center approximation, soft gluon approximation and the recombination of the two. Under static scattering center + soft gluon approximation, we recover the GLV result in the first opacity approximation. We have also examined numerically the effect of soft gluon approximation and find the difference between GHT and GLV result with static scattering center approximation appreciable at moderately large momentum fraction and long propagation length.

The TMD gluon distribution density can be related to the TMD jet transport parameter and encodes the properties of the nuclear medium as probed by the propagating parton. This general feature of our study can be used to incorporate different models of the dynamic medium in the calculation of parton energy loss and the jet quenching observables. It can also be incorporated in the Monte Carlo simulation of jet transport and propagation such as the Linear Boltzmann Transport (LBT) model \cite{Li:2010ts,Wang:2013cia,He:2015pra} for both cold and hot QCD medium.

\acknowledgments
We thank Feng Yuan and Bowen Xiao for helpful discussions. This work is supported by the Director, Office of Energy Research, Office of High Energy and Nuclear Physics, Division of Nuclear Physics, of the U.S. Department of Energy (DOE) under grant No. DE-AC02-05CH11231, by the U.S. National Science Foundation under grant Nos. ACI-1550228 within JETSCAPE Collaboration, and by the National Science Foundation of China under grants No. 11861131009, No. 11775095, No. 11890711 and No. 11890714.
 
\appendix

 \section{Hard parts of multiple parton scattering and gluon radiation}
\label{append-spectrum}

 In this Appendix, we list contributions to the SIDIS hadronic tensor from all cut diagrams for gluon radiation induced by multiple scattering. We categorize the diagrams according to the position of the cut-line: central, left and right cut diagrams. The kinematics for the SIDIS process in our convention are
\begin{equation}
\begin{split}
 &p = [p^+, 0, \vec{0}_{\perp}] , \quad q = [-\dfrac{Q^2}{2q^-},q^-,\vec{0}_{\perp}]\\
 &l = [\dfrac{l_{\perp}^2}{2(1-z)q^-}, (1-z)q^-, \vec{l}_{\perp}] \\
 &\epsilon (l) = [\dfrac{\vec{\epsilon}_{\perp}\cdot \vec{l}_{\perp}}{(1-z)q^-}, 0, \vec{\epsilon}_{\perp}]
\end{split}
\end{equation}
where the last line is the  polarization vector for radiative gluon.

With the above kinematics, the final gluon radiation spectrum is given as in Eq.~(\ref{eq:gluon_spectrum}), except that here the momentum fraction are defined as $z = l_q^- / q^-$ carried by the final quark $z=l^-/q^-$ in Eq.~(\ref{eq:gluon_spectrum}) is the momentum fraction carried the gluon]. We list $\tilde{H}^D_C$, $\tilde{H}^D_L$ and $\tilde{H}^D_R$ from each cut-diagram labelled by the type of radiation amplitudes it contains according to the convention given in Appendix \ref{append-helicity}.

 \subsection{Central cut diagram}
 
The $\tilde{H}^D_C$ of central cut diagram 11and 22, see Fig \ref{fig:C11_C22}.
 \begin{equation}
 \begin{split}
 \tilde{H}^D_{C11} =&\frac{C_F}{[\vec{l}_{\perp} -(1-z)\vec{k}_{\perp}]^2}\frac{\phi(x_L+x_D,\vec{k}_{\perp})}{k_{\perp}^2}f_q^A(x) \\
 \end{split}
 \end{equation}
 
 \begin{widetext}

\begin{equation}
\begin{split}
\tilde{H}^D_{C22} &= \frac{C_F}{l_{\perp}^2} \bigg[ \frac{\phi(x_D,\vec{k}_{\perp})}{k_{\perp}^2}  f(x+x_L) - \frac{\phi(x_L+x_D,\vec{k}_{\perp})}{k_{\perp}^2}e^{ix_Lp^+y_1^-} f(x+x_L) -\frac{\phi(x_D,\vec{k}_{\perp})}{k_{\perp}^2}e^{-ix_Lp^+y_1^-}  f(x)\\
&+\frac{\phi(x_L+x_D,\vec{k}_{\perp})}{k_{\perp}^2}f(x)  \bigg]  \\
\end{split}
\end{equation} 
 
 \begin{figure*}[h] 
\centering
{
\includegraphics[scale=0.45]{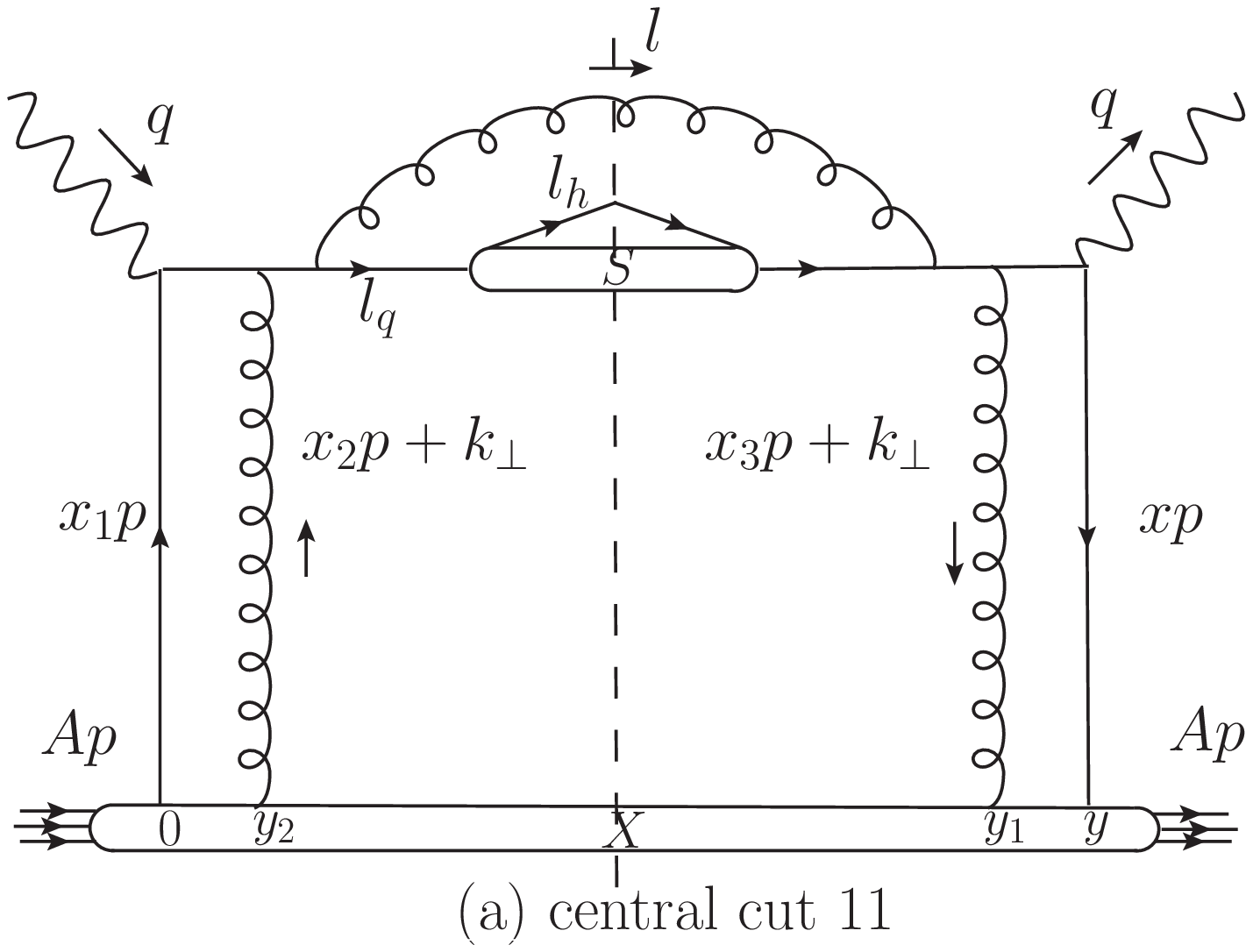} 
  \label{fig:C11}
\includegraphics[scale=0.45]{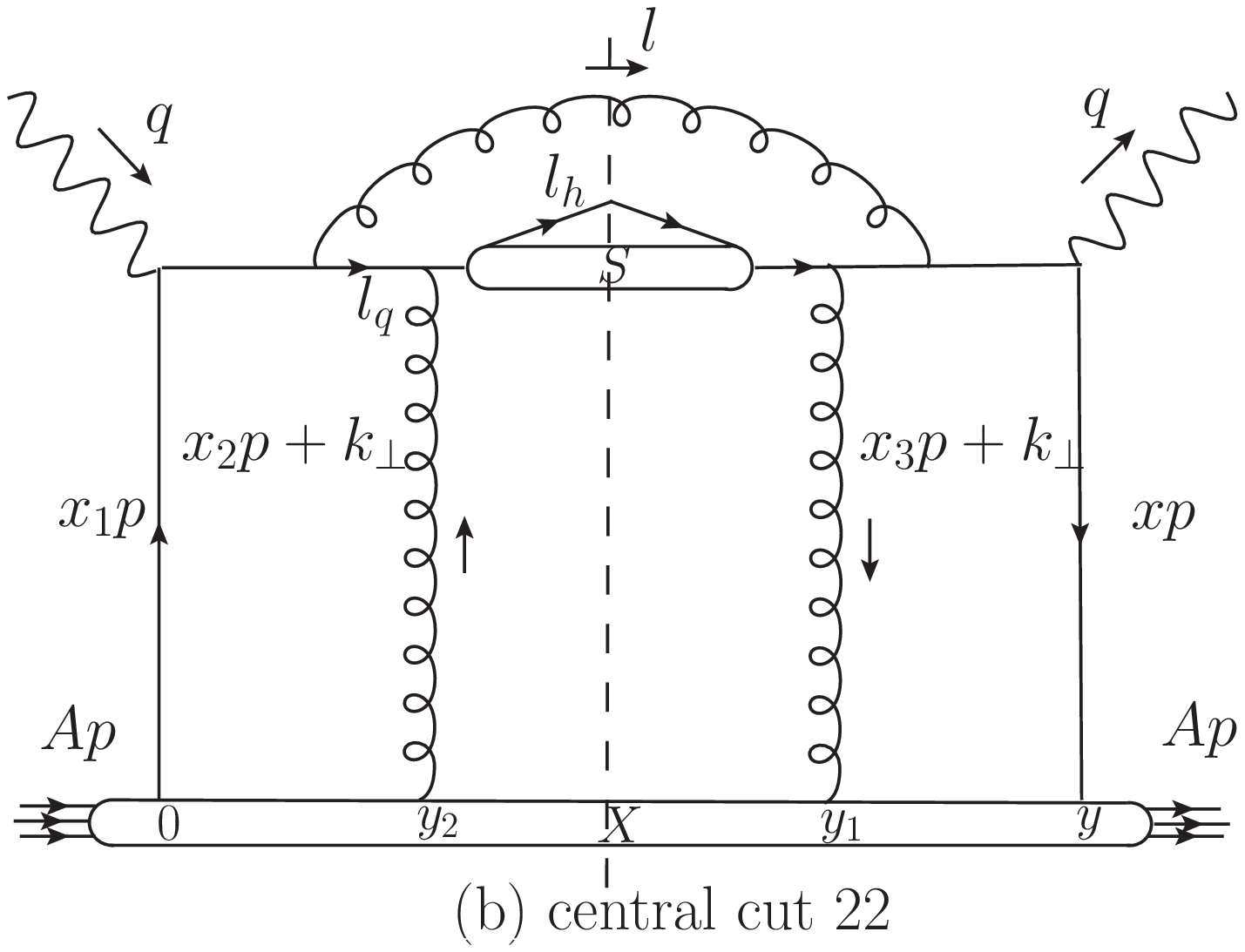} 
  \label{fig:C22}
}
 \caption{(a) Central cut 11 and (b) central cut 22.}
  \label{fig:C11_C22}
\end{figure*}  
 
The $\tilde{H}^D_C$ of central cut diagram 33, see Fig \ref{fig:C33}.
 \begin{equation}
\begin{split}
\tilde{H}^D_{C33} = &\frac{C_A}{(\vec{l}_{\perp}-\vec{k}_{\perp})^2} \left[ \frac{\phi(\frac{z}{1-z}x_D,\vec{k}_{\perp})}{k_{\perp}^2}  f(x+x_L+\frac{x_D}{1-z}) - \frac{\phi(x_L+x_D,\vec{k}_{\perp})}{k_{\perp}^2}e^{i(x_L+\frac{x_D}{1-z})p^+y_1^-}  f(x+x_L+\frac{x_D}{1-z})\right.\\
&\left. -\frac{\phi(\frac{z}{1-z}x_D,\vec{k}_{\perp})}{k_{\perp}^2}e^{-i(x_L+\frac{x_D}{1-z})p^+y_1^-}  f(x)+\frac{\phi(x_L+x_D,\vec{k}_{\perp})}{k_{\perp}^2}f(x)  \right]  \\
\end{split}
\end{equation}
 
 \begin{figure}[h]
  \includegraphics[scale=0.45]{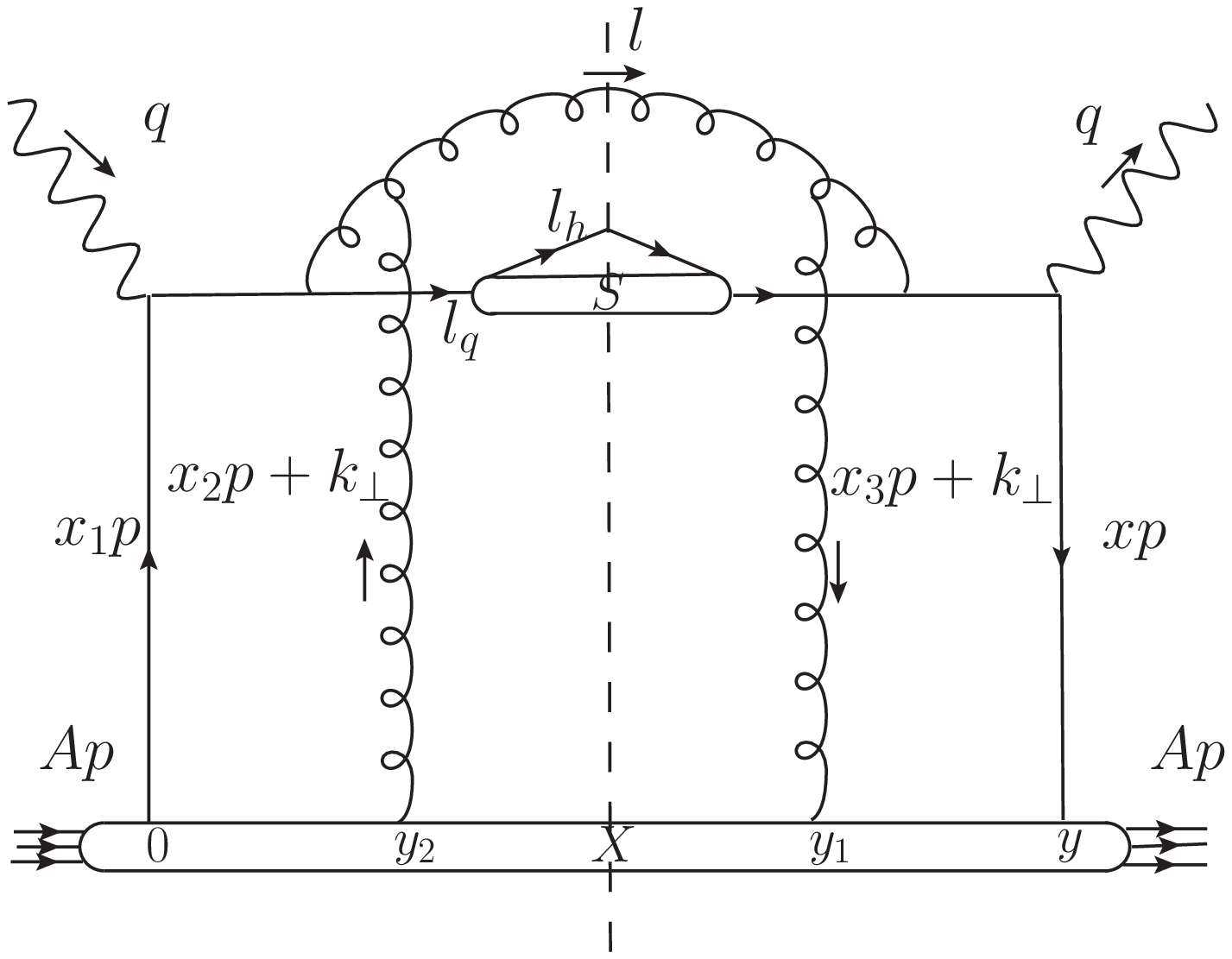}
  \caption{Central cut 33}
  \label{fig:C33}
\end{figure} 

The $\tilde{H}^D_C$ of central cut diagram 12 and 21, see Fig \ref{fig:C12_C21}.
 
 \begin{equation}
\begin{split}
\tilde{H}^D_{C12} = &\frac{1}{2N_c}\frac{\vec{l}_{\perp}\cdot[\vec{l}_{\perp}-(1-z)\vec{k}_{\perp}]}{\vec{l}_{\perp}^2[\vec{l}_{\perp}-(1-z)\vec{k}_{\perp}]^2} \left[ \frac{\phi(x_L+x_D,\vec{k}_{\perp})}{k_{\perp}^2}  f(x) - \frac{\phi(x_L+x_D,\vec{k}_{\perp})}{k_{\perp}^2}e^{ix_Lp^+y_1^-}  f(x+x_L) \right]  \\
\end{split}
\end{equation}

 \begin{equation}
\begin{split}
\tilde{H}^D_{C21} = &\frac{1}{2N_c}\frac{\vec{l}_{\perp}\cdot[\vec{l}_{\perp}-(1-z)\vec{k}_{\perp}]}{\vec{l}_{\perp}^2[\vec{l}_{\perp}-(1-z)\vec{k}_{\perp}]^2} \left[ \frac{\phi(x_L+x_D,\vec{k}_{\perp})}{k_{\perp}^2}  f(x) - \frac{\phi(x_D,\vec{k}_{\perp})}{k_{\perp}^2}e^{-ix_Lp^+y_1^-}  f(x) \right]  \\
\end{split}
\end{equation} 

 \begin{figure*}[h] 
\centering
{
\includegraphics[scale=0.45]{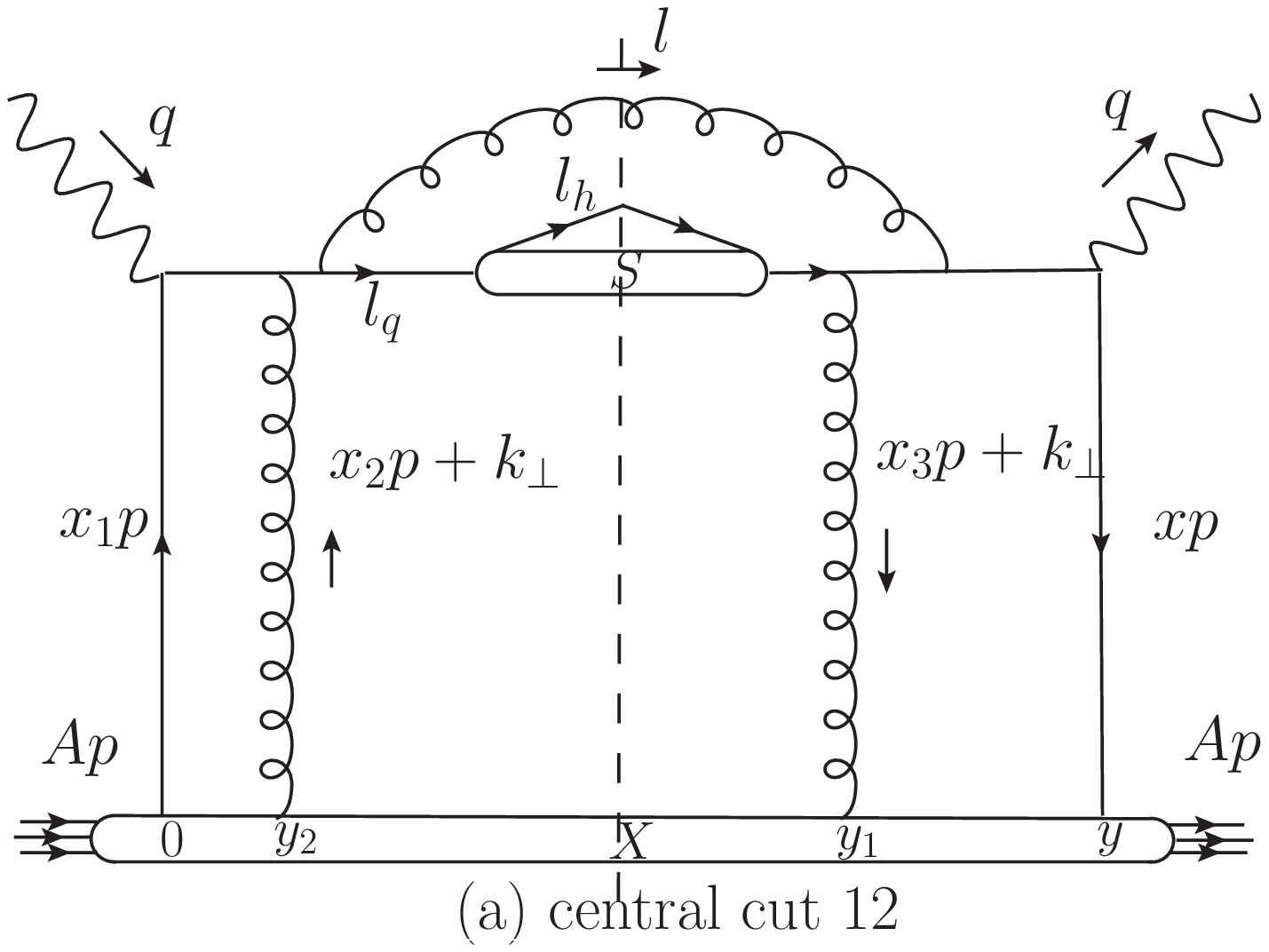} 
\label{fig:C12}
\includegraphics[scale=0.45]{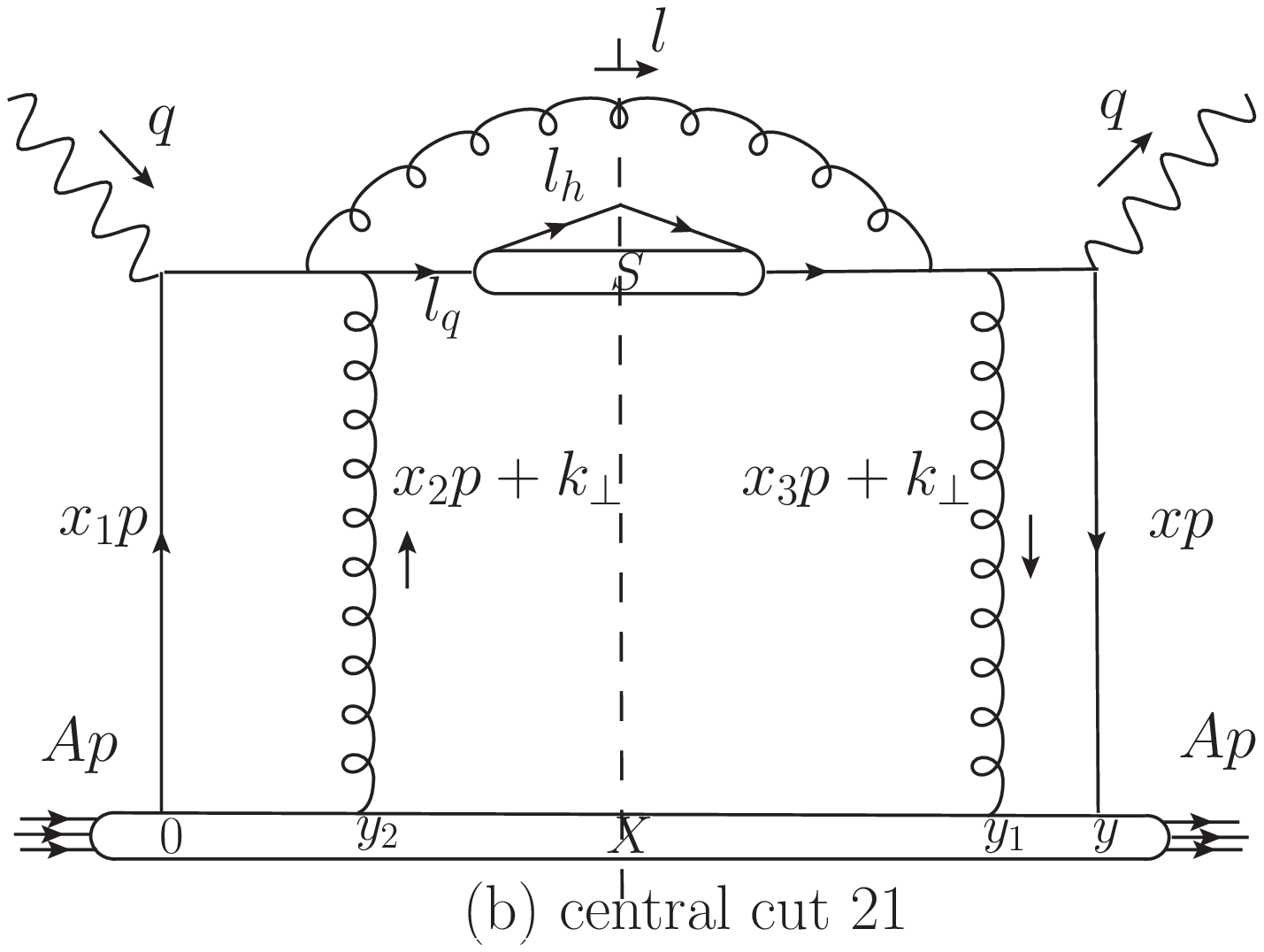} 
\label{fig:C21}
}
 \caption{(a) Central cut 12 and (b) central cut 21.}
  \label{fig:C12_C21}
\end{figure*}  

The $\tilde{H}^D_C$ of central cut diagram 13 and 31, see Fig \ref{fig:C13_C31}.

 \begin{equation}
\begin{split}
\tilde{H}^D_{C13} = &\frac{C_A}{2}\frac{(\vec{l}_{\perp}-\vec{k}_{\perp})\cdot[\vec{l}_{\perp}-(1-z)\vec{k}_{\perp}]}{(\vec{l}_{\perp}-\vec{k}_{\perp})^2[\vec{l}_{\perp}-(1-z)\vec{k}_{\perp}]^2} \left[ \frac{\phi(x_L+x_D,\vec{k}_{\perp})}{k_{\perp}^2}e^{i(x_L+\frac{x_D}{1-z})p^+y_1^-}  f(x+x_L+\frac{x_D}{1-z}) \right. \\
&\left. - \frac{\phi(x_L+x_D,\vec{k}_{\perp})}{k_{\perp}^2}  f(x) \right]  \\
\end{split}
\end{equation} 

 \begin{equation}
\begin{split}
\tilde{H}^D_{C31} = &\frac{C_A}{2}\frac{(\vec{l}_{\perp}-\vec{k}_{\perp})\cdot[\vec{l}_{\perp}-(1-z)\vec{k}_{\perp}]}{(\vec{l}_{\perp}-\vec{k}_{\perp})^2[\vec{l}_{\perp}-(1-z)\vec{k}_{\perp}]^2} \left[ \frac{\phi(\frac{z}{1-z} x_D,\vec{k}_{\perp})}{k_{\perp}^2}e^{-i(x_L+\frac{x_D}{1-z})p^+y_1^-}  f(x) - \frac{\phi(x_L+x_D,\vec{k}_{\perp})}{k_{\perp}^2}  f(x) \right]  \\
\end{split}
\end{equation}

The $\tilde{H}^D_C$ of central cut diagram 23 and 32, see Fig \ref{fig:C23_C32}.
 \begin{equation}
\begin{split}
\tilde{H}^D_{C23} = &-\frac{C_A}{2}\dfrac{\vec{l}_{\perp} \cdot( \vec{l}_{\perp}- \vec{k}_{\perp})}{(\vec{l}_{\perp}-\vec{k}_{\perp})^2{l}_{\perp}^2} \left[ \frac{\phi( x_D,\vec{k}_{\perp})}{k_{\perp}^2}e^{i\frac{x_D}{1-z}p^+y_1^-}  f(x+x_L+\frac{x_D}{1-z})   - \frac{\phi(x_L+x_D,\vec{k}_{\perp})}{k_{\perp}^2} e^{i(x_L+\frac{x_D}{1-z})p^+y_1^-}  f(x+x_L+\frac{x_D}{1-z}) \right.\\
& \left. - \frac{\phi( x_D,\vec{k}_{\perp})}{k_{\perp}^2}e^{-ix_Lp^+y_1^-}  f(x) + \frac{\phi( x_L+x_D,\vec{k}_{\perp})}{k_{\perp}^2} f(x)   \right]  \\
\end{split}
\end{equation}

 \begin{equation}
\begin{split}
\tilde{H}^D_{C32} = &-\frac{C_A}{2}\dfrac{\vec{l}_{\perp} \cdot( \vec{l}_{\perp}- \vec{k}_{\perp})}{(\vec{l}_{\perp}-\vec{k}_{\perp})^2{l}_{\perp}^2} \left[ \frac{\phi( \frac{z}{1-z} x_D,\vec{k}_{\perp})}{k_{\perp}^2}e^{-i\frac{x_D}{1-z}p^+y_1^-}  f(x+x_L)  - \frac{\phi(x_L+x_D,\vec{k}_{\perp})}{k_{\perp}^2} e^{ix_Lp^+y_1^-}  f(x+x_L) \right.\\
& \left. - \frac{\phi(  \frac{z}{1-z} x_D,\vec{k}_{\perp})}{k_{\perp}^2}e^{-i(x_L+\frac{x_D}{1-z})p^+y_1^-}  f(x)  + \frac{\phi( x_L+x_D,\vec{k}_{\perp})}{k_{\perp}^2} f(x)   \right]  \\
\end{split}
\end{equation}

  \begin{figure*}[h] 

\centering
{
\includegraphics[scale=0.45]{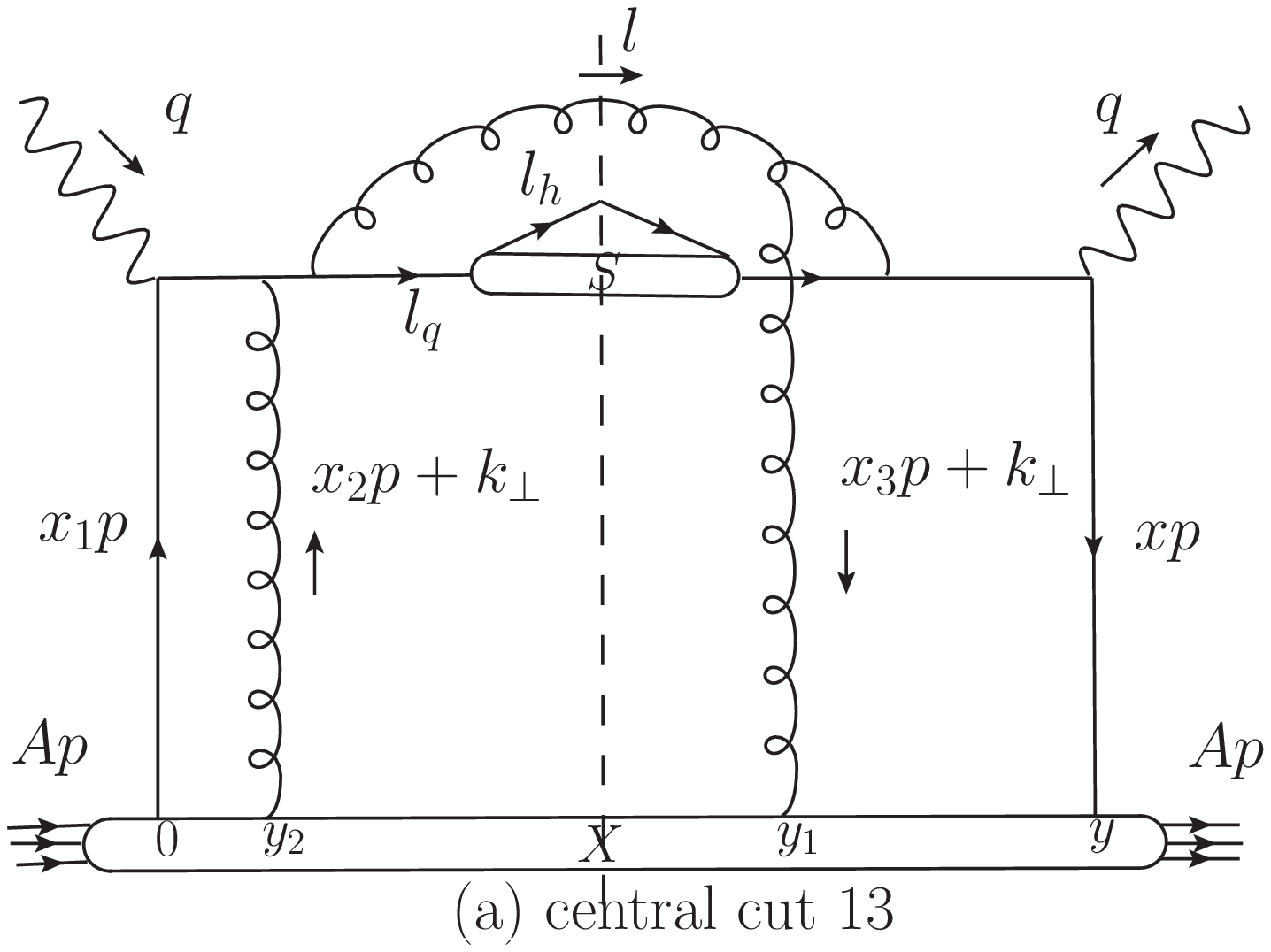} 
  \label{fig:C13}
\includegraphics[scale=0.45]{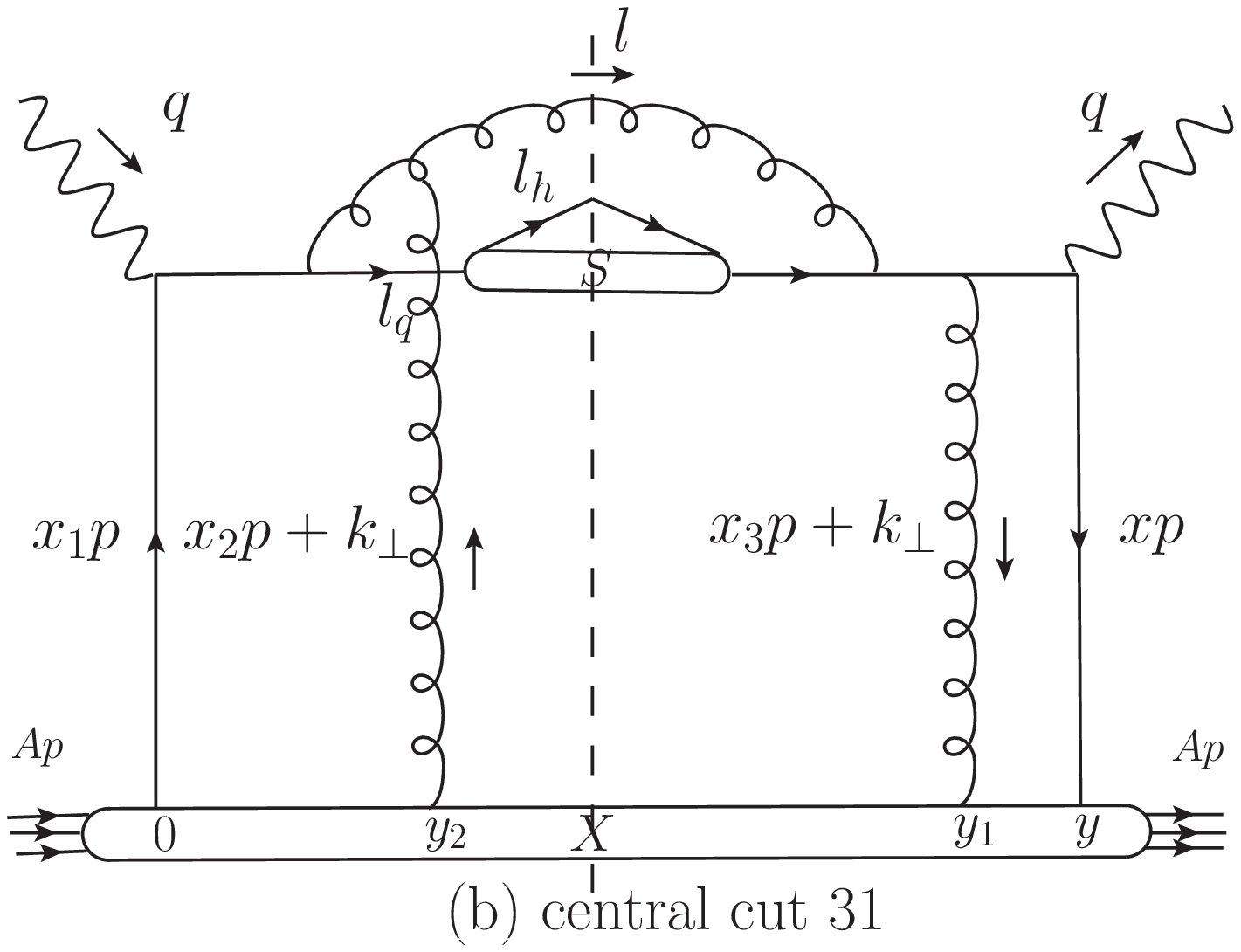} 
  \label{fig:C31}
}
 \caption{(a) Central cut 13 and (b) central cut 31.}
  \label{fig:C13_C31}
\end{figure*}

 \begin{figure*}[h] 

\centering
{
\includegraphics[scale=0.45]{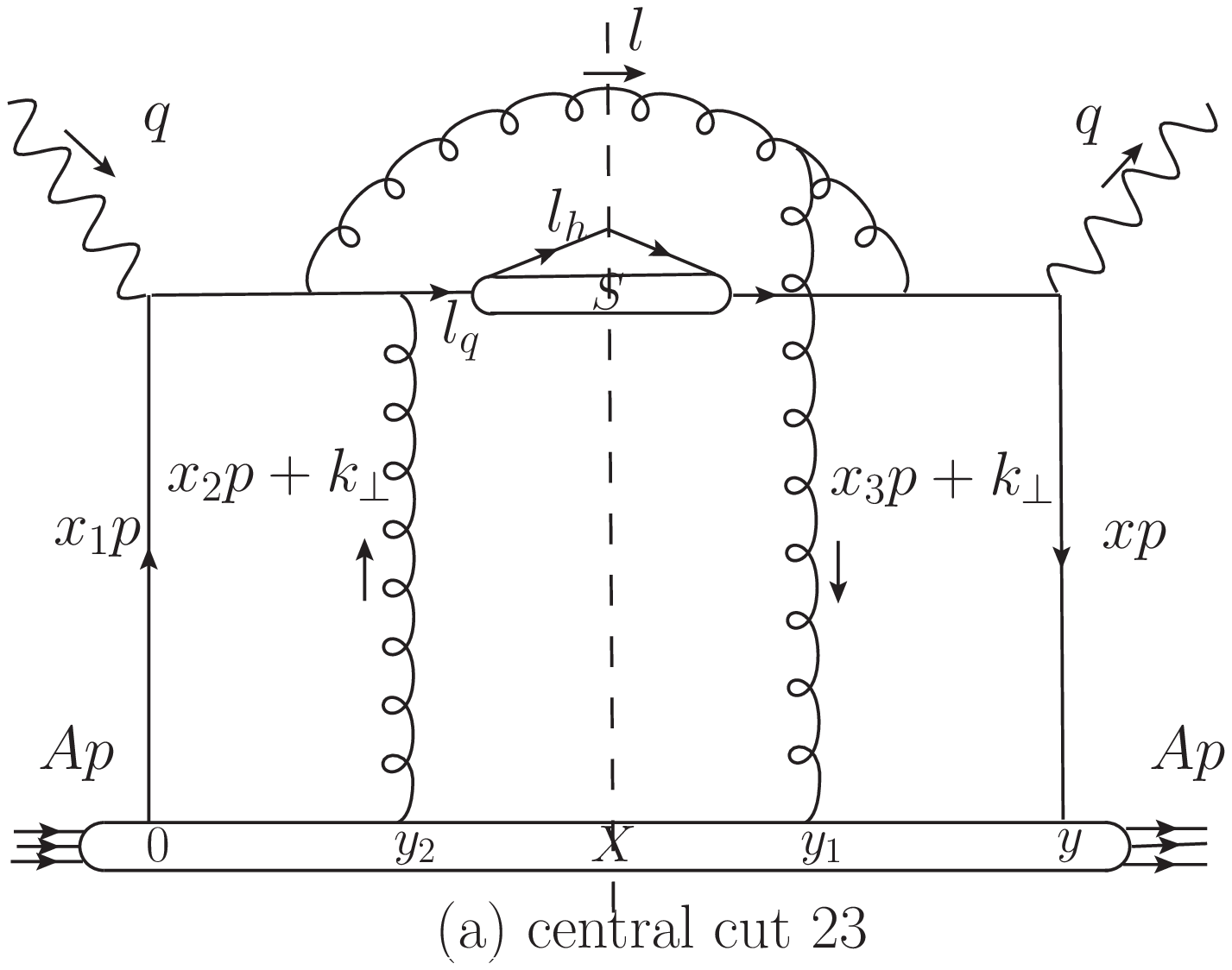} 
  \label{fig:C23}
\includegraphics[scale=0.45]{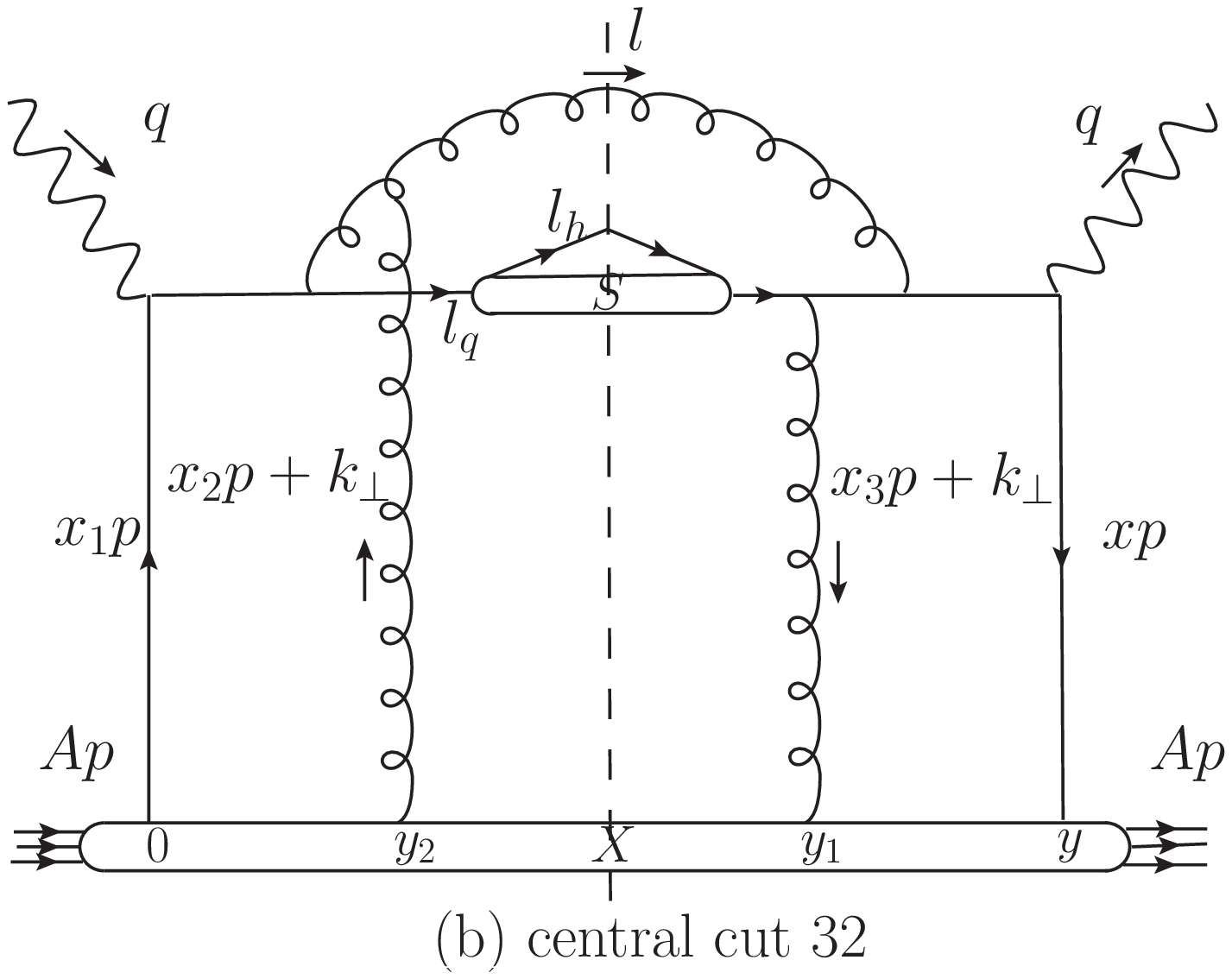} 
  \label{fig:C32}
}
 \caption{(a) Central cut 23 and (b) central cut 32.}
  \label{fig:C23_C32}
\end{figure*}

 \subsection{Right cut diagram}
 The $\tilde{H}^D_{R}$ of right cut diagram 1 and 2, see Fig \ref{fig:R1_R2}.
 \begin{equation}
\begin{split}
\tilde{H}^D_{R1} = &-C_F\frac{1}{{l}_{\perp}^2}\frac{\phi(x_D^0,\vec{k}_{\perp})}{k_{\perp}^2}e^{ix_Lp^+y_1^-}  f(x+x_L)  
\end{split}
\end{equation} 

\begin{equation}
\begin{split}
\tilde{H}^D_{R2} = &\frac{1}{2N_c}\frac{\vec{l}_{\perp}\cdot[\vec{l}_{\perp}-(1-z)\vec{k}_{\perp}]}{l_{\perp}^2[\vec{l}_{\perp}-(1-z)\vec{k}_{\perp}]^2} \left[ \frac{\phi(x_L+x_D,\vec{k}_{\perp})}{k_{\perp}^2}e^{ix_Lp^+y_1^-}  f(x+x_L)  -   \frac{\phi(x_D^0,\vec{k}_{\perp})}{k_{\perp}^2}e^{ix_Lp^+y_1^-}  f(x+x_L)  \right]
\end{split}
\end{equation}

 \begin{figure*}[h] 
\centering
{
\includegraphics[scale=0.55]{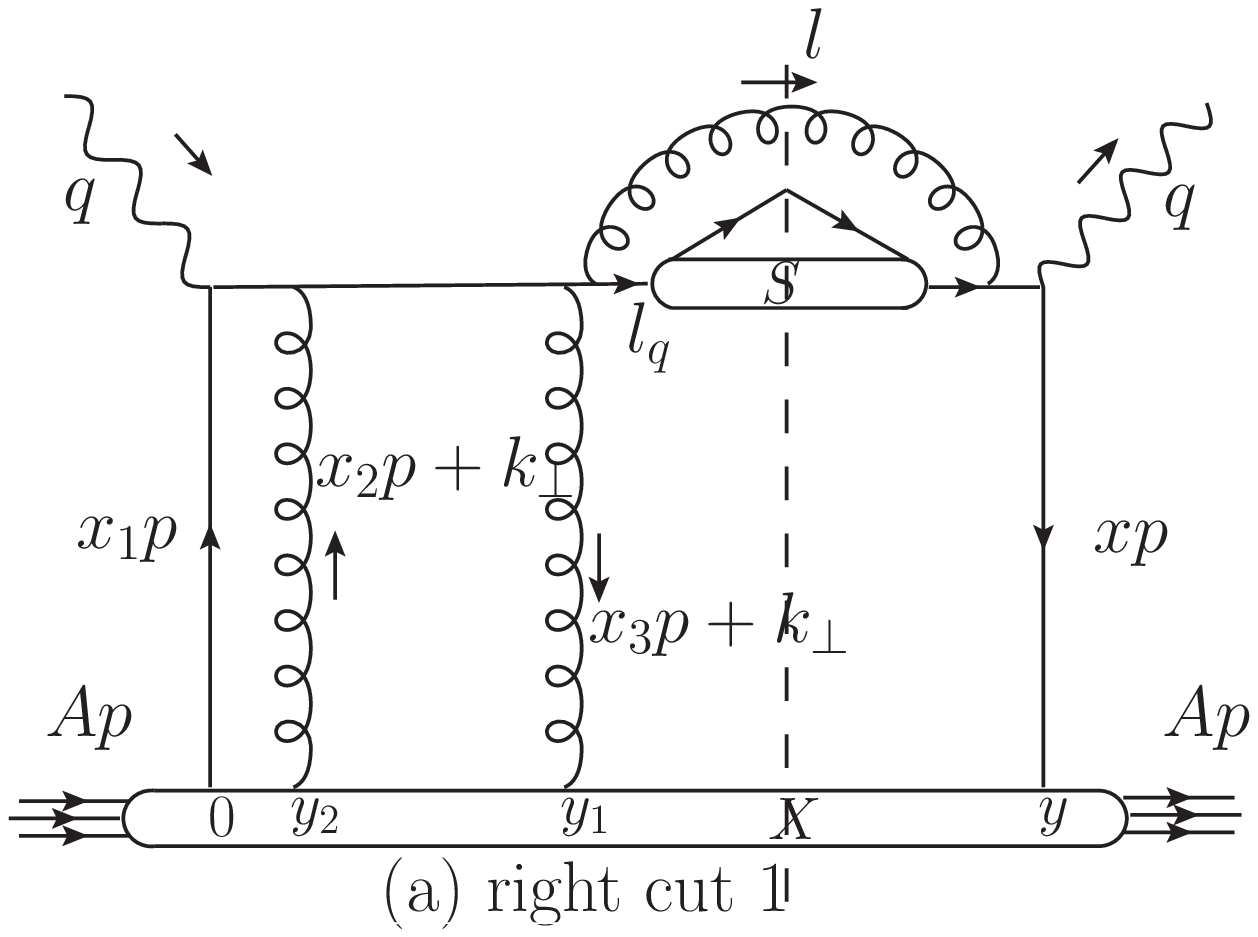} 
  \label{fig:R1}
\includegraphics[scale=0.55]{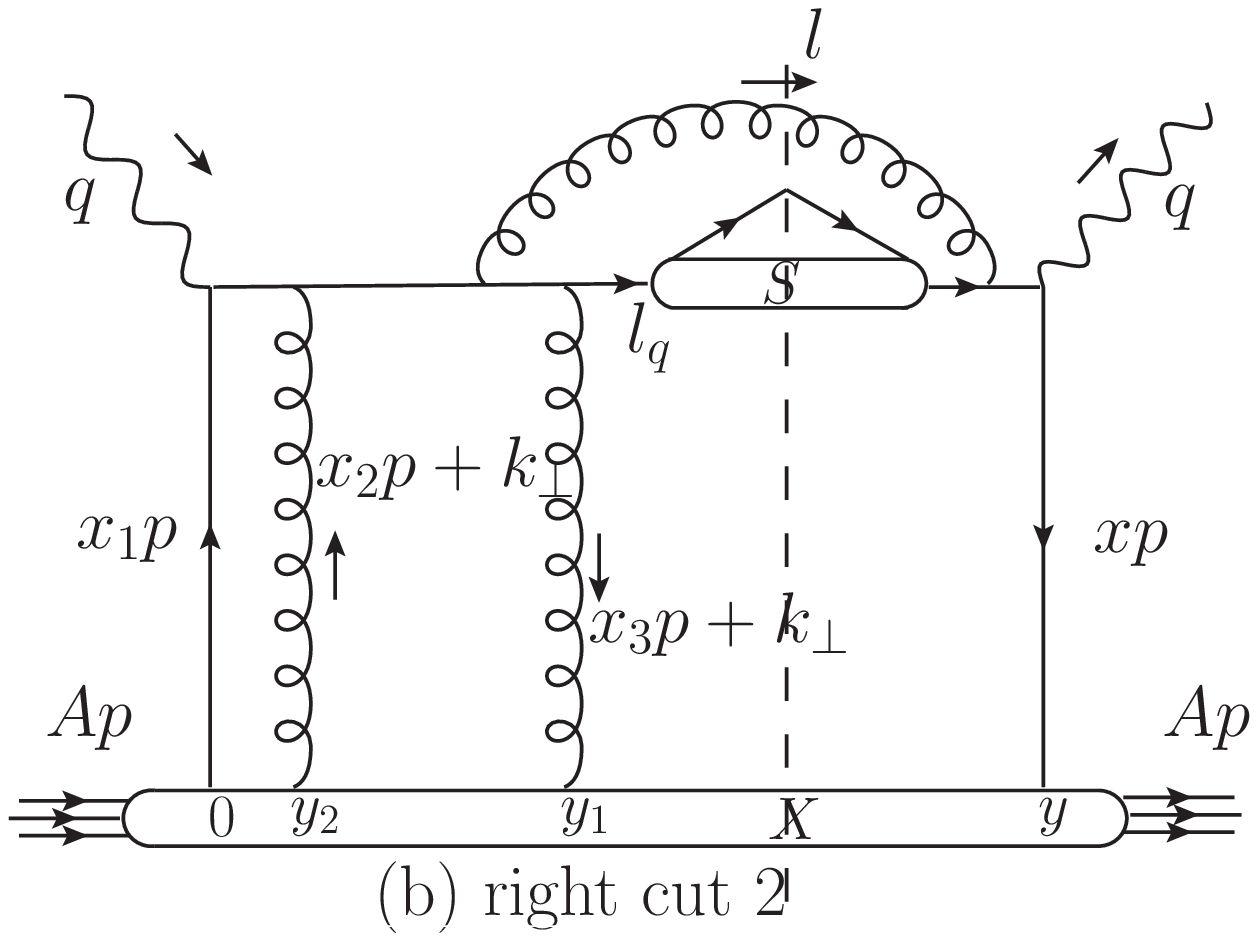} 
  \label{fig:R2}
}
 \caption{(a) Right cut 1 and (b) right cut 2.}
  \label{fig:R1_R2}
\end{figure*}  

The $\tilde{H}^D_{R}$ of right cut diagram 3 and 4, see Fig \ref{fig:R3_R4}.

\begin{equation}
\begin{split}
\tilde{H}^D_{R3} = &C_F\frac{1}{l_{\perp}^2} \left[ \frac{\phi(x_L+x_D,\vec{k}_{\perp})}{k_{\perp}^2}e^{ix_Lp^+y_1^-}  f(x+x_L)  -   \frac{\phi(x_D,\vec{k}_{\perp})}{k_{\perp}^2} f(x+x_L)  \right]
\end{split}
\end{equation} 

\begin{equation}
\begin{split}
\tilde{H}^D_{R4} = &\frac{C_A}{2}\frac{\vec{l}_{\perp}\cdot[\vec{l}_{\perp}-z\vec{k}_{\perp}]}{l_{\perp}^2[\vec{l}_{\perp}-z\vec{k}_{\perp}]^2} \left[ \frac{\phi(x_D^0,\vec{k}_{\perp})}{k_{\perp}^2}e^{ix_Lp^+y_1^-} f(x+x_L)  -   \frac{\phi(x_L+\frac{z}{1-z}x_D,\vec{k}_{\perp})}{k_{\perp}^2}e^{ix_Lp^+y_1^-} f(x+x_L)  \right]
\end{split}
\end{equation} 

 \begin{figure*}[h] 
\centering
{
\includegraphics[scale=0.55]{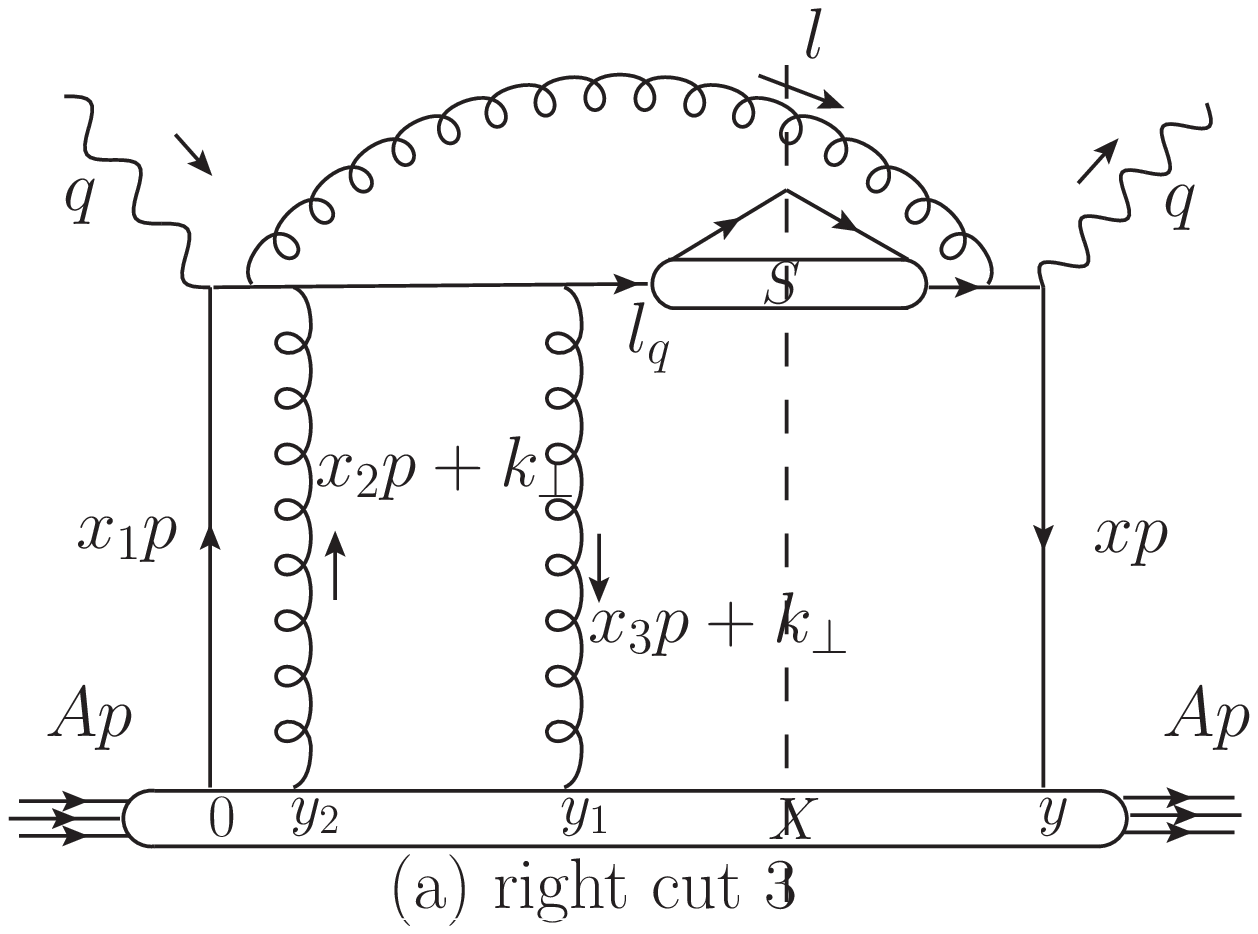} 
  \label{fig:R3}
  \includegraphics[scale=0.55]{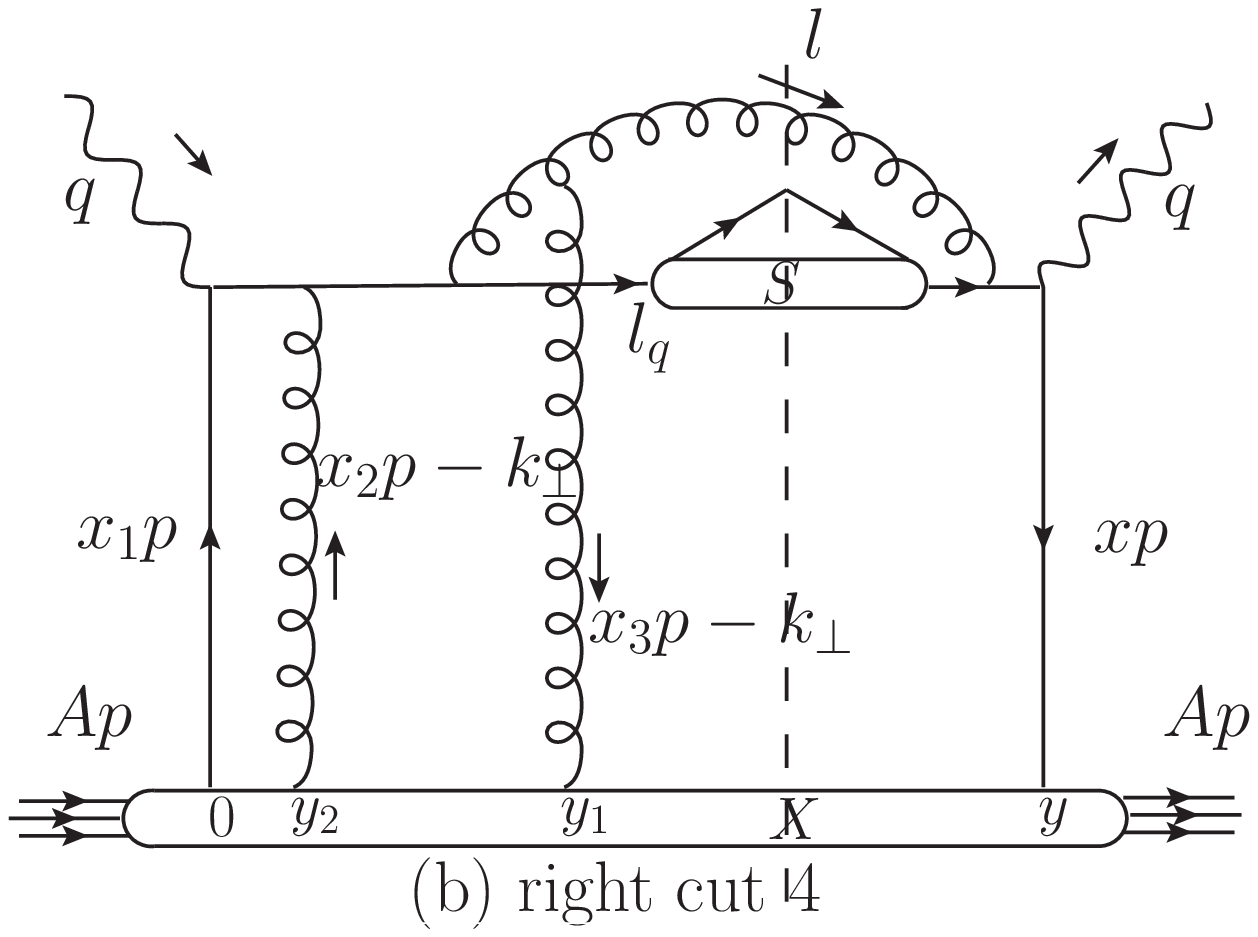} 
  \label{fig:R4}
}
 \caption{(a) Right cut 3 and (b) right cut 4.}
  \label{fig:R3_R4}
\end{figure*}  

The $\tilde{H}^D_{R}$ of right cut diagram 5 and 6, see Fig \ref{fig:R5_R6}.
\begin{equation}
\begin{split}
\tilde{H}^D_{R5} = &\frac{C_A}{2}\frac{\vec{l}_{\perp}\cdot[\vec{l}_{\perp}-\vec{k}_{\perp}]}{l_{\perp}^2[\vec{l}_{\perp}-\vec{k}_{\perp}]^2} \left[ \frac{\phi(\frac{z}{1-z}x_D,\vec{k}_{\perp})}{k_{\perp}^2}e^{-i\frac{x_D}{1-z}p^+y_1^-} f(x+x_L)  -   \frac{\phi(x_L+x_D,\vec{k}_{\perp})}{k_{\perp}^2}e^{ix_Lp^+y_1^-} f(x+x_L)  \right]
\end{split}
\end{equation}

\begin{equation}
\begin{split}
\tilde{H}^D_{R6} = &\frac{C_A}{2}\frac{\vec{l}_{\perp}\cdot[\vec{l}_{\perp}-\vec{k}_{\perp}]}{l_{\perp}^2[\vec{l}_{\perp}-\vec{k}_{\perp}]^2} \left[ \frac{\phi(x_D,\vec{k}_{\perp})}{k_{\perp}^2}e^{-i\frac{x_D}{1-z}p^+y_1^-} f(x+x_L)  -   \frac{\phi(x_L+\frac{z}{1-z}x_D,\vec{k}_{\perp})}{k_{\perp}^2}e^{ix_Lp^+y_1^-} f(x+x_L)  \right]
\end{split}
\end{equation} 

 \begin{figure*}[h] 
\centering
{
\includegraphics[scale=0.55]{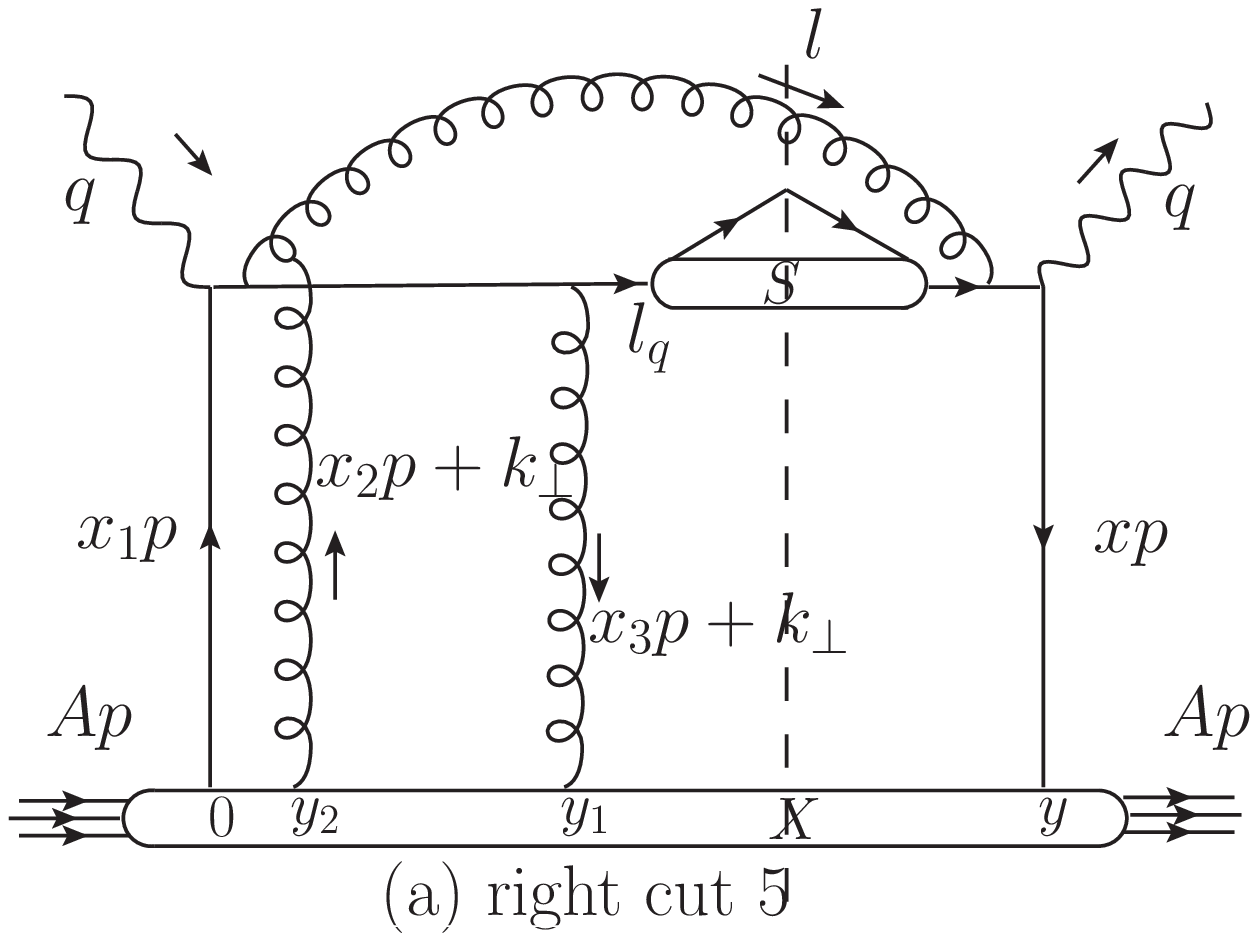} 
  \label{fig:R5}
  \includegraphics[scale=0.55]{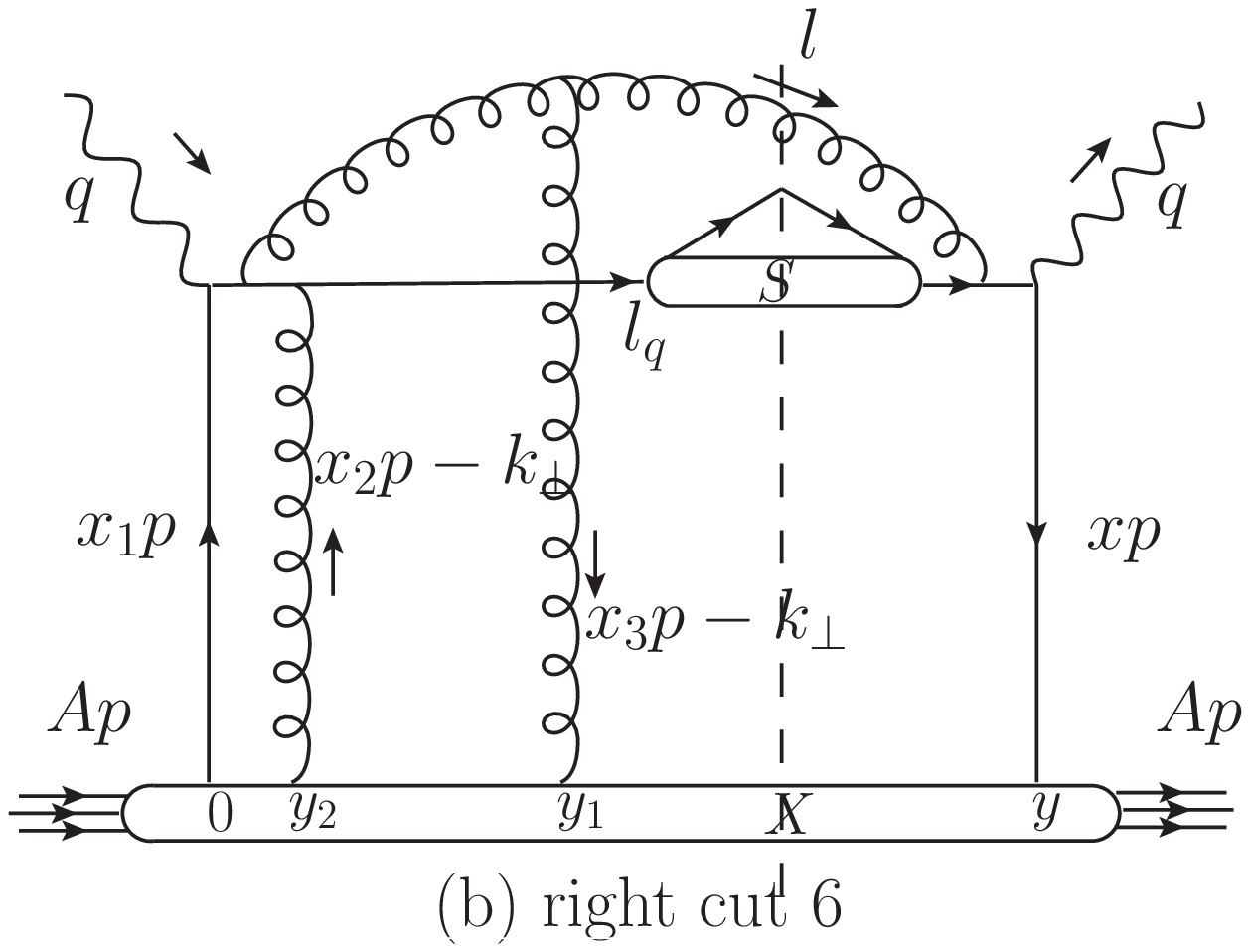} 
  \label{fig:R6}
}
 \caption{ (a) Right cut 5 and (b) right cut 6.}
  \label{fig:R5_R6}
\end{figure*}  

The $\tilde{H}^D_{R}$ of right cut diagram 7, see Fig \ref{fig:R7}.
\begin{equation}
\begin{split}
\tilde{H}^D_{R7} = &C_A\frac{1}{l_{\perp}^2} \left[    \frac{\phi(x_L+\frac{z}{1-z}x_D,\vec{k}_{\perp})}{k_{\perp}^2}e^{ix_Lp^+y_1^-} f(x+x_L)   - \frac{\phi(\frac{z}{1-z}x_D,\vec{k}_{\perp})}{k_{\perp}^2} f(x+x_L)  \right]
\end{split}
\end{equation}  
 
  \begin{figure}[h]
  \centering 
  \includegraphics[scale=0.55]{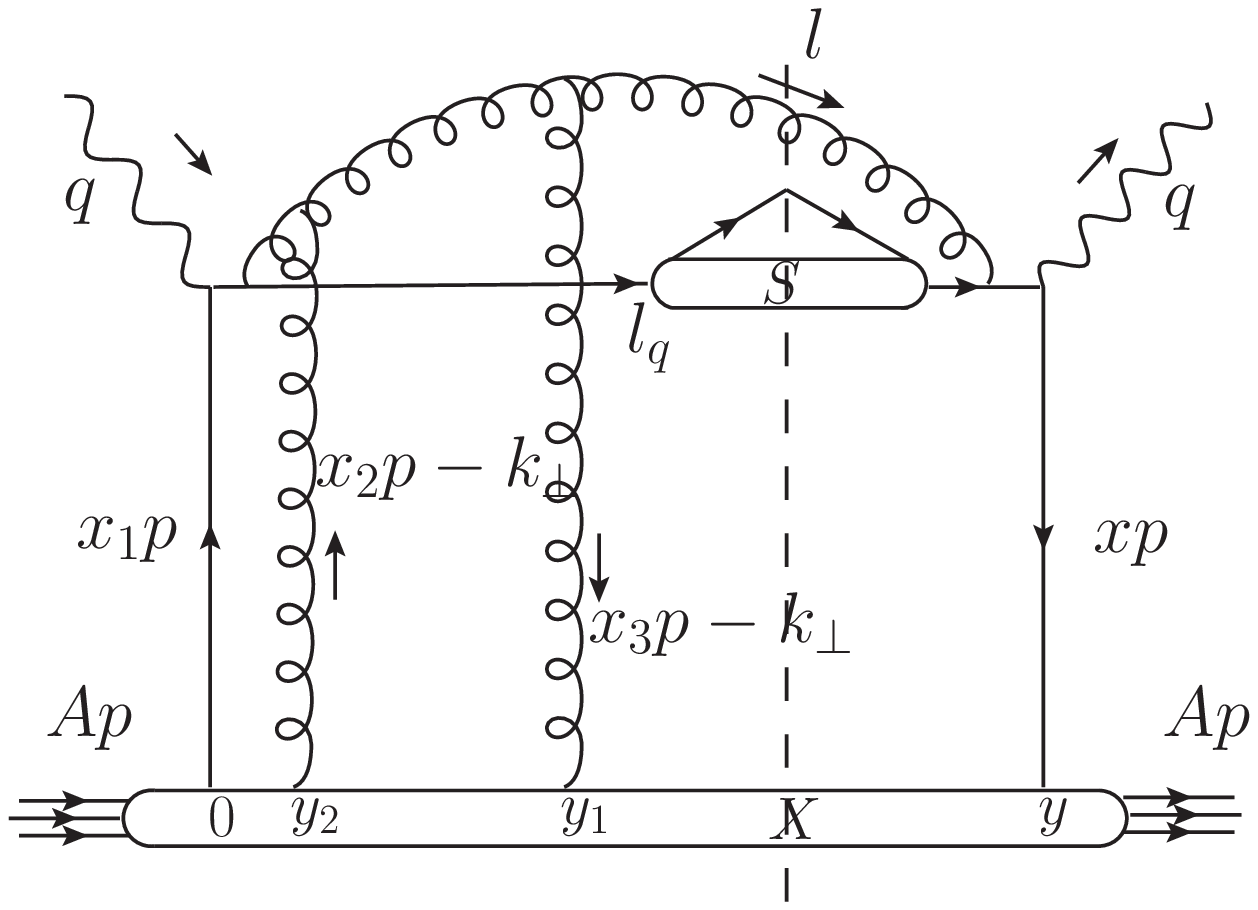}
  \caption{Right cut 7.}
  \label{fig:R7}
\end{figure}  
  
 \end{widetext}
    
 We reverse the sign of $\vec{k}_{\perp}$ for right-cut diagrams 4, 6, and 7.

  \subsection{Left cut diagram} 
  One can obtain contributions from left-cut diagrams from the hard parts for right-cut diagrams by the following variable changes,
  \begin{equation}
  \begin{split}
  H^D_{L} =  H^D_{R} (y_1^-  \rightarrow y^- - y_2^-, y_2^-  \rightarrow y^- - y_1^-).
  \end{split}
  \end{equation}

\section{Helicity Amplitude}
\label{append-helicity}

In this Appendix, we calculate the hadronic tensor within the helicity amplitude approach in which the helicity of the propagating quark is conserved in the scattering amplitude when the transverse momentum of the fast quark is neglected as compared to its longitudinal momentum. 
 
Assuming the dominant component of a fast quark's momentum is the minus component, $l_q \approx [0,l_q^-,\vec{0}_{\perp}] $, we have under the helicity amplitude approximation,
$$ \bar{u}^r (l_q) \gamma^{\mu} u^{s}(l_q^{'})  \approx  2\sqrt{l_q^- l_q^{'-}} \delta_{rs} \underline{n}^{\mu}   $$
where $r$ and $s$ are helicities of quarks, and $\underline{n} = [0,1,\vec{0}_{\perp}]$. 

\begin{figure}[h]
  \centering 
  \includegraphics[scale=0.6]{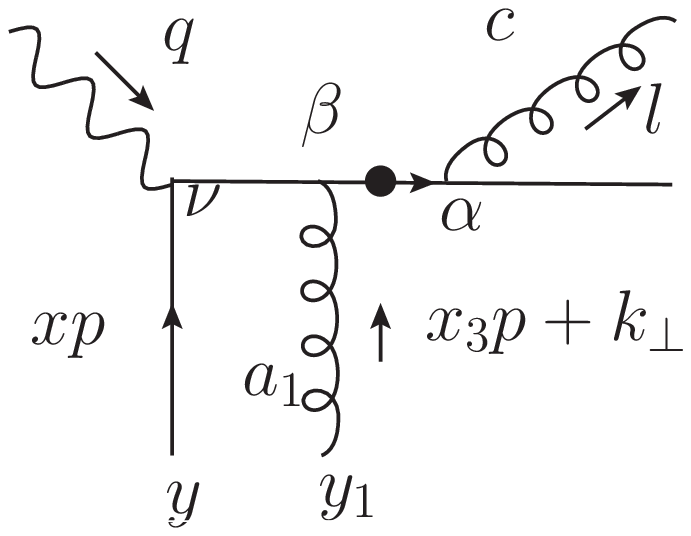}
  \caption{Double scattering 1 with $x= x_B$, $x_3 = x_L +x_D$. }
  \label{fig:double scattering amplitude}
\end{figure}

In the calculation of the scattering amplitude in this helicity amplitude approach, we assign initial quark and gluons from the nucleus as
\begin{equation}
\begin{split}
&\text{initial quark} \rightarrow u(p) \int \dfrac{dx_i}{2\pi} e^{ix_ip^+y_i^- - i \vec{k}_{\perp} \cdot \vec{y}_{i\perp}} ,  \\
&\text{initial gluon} \rightarrow p_{\sigma} \int \dfrac{dx_i}{2\pi} e^{ix_ip^+y_i^- - i \vec{k}_{\perp} \cdot \vec{y}_{i\perp}}.
\end{split}
\end{equation}
The normal Feynman rules apply in the rest of the calculation of scattering amplitude. We also take the soft radiative gluon approximation, $z \rightarrow 1$.

There are three kinds of diagrams for double scattering. We calculate the amplitude of the double scattering  in Fig.~\ref{fig:double scattering amplitude} in detail as an example. The black dot  in Fig.~\ref{fig:double scattering amplitude} denotes the off-shell parton before the radiation vertex. One can write down the scattering amplitude according to the Feynman rules defined for this helicity amplitude method,
 \begin{widetext}
\begin{equation}
\begin{split}
iM^{\nu}_{D1}(y,y_1)=&\int \dfrac{d x_3}{2\pi} e^{ix_3p^+y_1^--i\vec{k}_{\perp}\cdot \vec{y}_{1\perp}} \int \dfrac{dx}{2\pi} e^{ixp^+y^-}
\bar{u}^{s}[(x+x_3)p+q+k_{\perp}-l] \\
&\times (ig) \gamma^{\alpha} T_c \epsilon_{\alpha} \dfrac{i[(x+x_3)\slashed{p}+\slashed{q}+\slashed{k}_{\perp}]}{[(x+x_3)p+q+k_{\perp}]^2+i\epsilon} (ig) \gamma^{\beta} T_{a_1}   \dfrac{i(x\slashed{p}+\slashed{q})}{(xp+q)^2+i\epsilon} (-i\gamma^{\nu}) u^{s'}(p) .
\end{split}
\end{equation}
Using the approximation mentioned in the helicity amplitude approach and the final quark on-shell condition 
$\delta([(x+x_3)p+q+k_{\perp}-l]^2)=2zp^+q^-\delta(x+x_3-x_B-x_L-x_D)$, one can simplify the amplitude as
\begin{equation}
\begin{split}
M^{\nu}_{D1}(y,y_1) =& 2(\sqrt{z})^3 g \dfrac{\vec{\epsilon}_{\perp}\cdot \vec{l}_{\perp}}{ l_{\perp}^2} T_c T_{a_1} e^{i(x_B+x_L)p^+y^-} e^{ix_Dp^+y_1^-} e^{ix_Lp^+(y_1^- -y^-)}e^{-i\vec{k}_{\perp}  \cdot \vec{y}_{1\perp}}ig \theta(y_1^- - y^-)\\
&\times \frac{\bar{u}^{s}(x_Bp+q) \gamma^{\nu} u^{s'}(p)}{2\pi}.
\end{split}
\end{equation}
Under the soft gluon approximation $z\rightarrow 1$, one can rewrite the amplitude as
\begin{equation}
\begin{split}
M^{\nu}_{D1}(y,y_1) =&\int \frac{dx}{2\pi} \delta(x-x_B)\int dx_3  \delta (x_3 - x_L -x_D)  \bar{u}^{s}(xp+q) \gamma^{\nu} u^{s'}(p) \overline{M}_{D1}(y,y_1)   \\
\overline{M}_{D1}(y,y_1) = & 2 g \dfrac{\vec{\epsilon}_{\perp}\cdot \vec{l}_{\perp}}{ l_{\perp}^2} T_c T_{a_1} e^{ixp^+y^-} e^{ix_3p^+y_1^-} e^{-i\vec{k}_{\perp}  \cdot \vec{y}_{1\perp}}ig \theta(y_1^- - y^-) \\
= & 2 g \dfrac{\vec{\epsilon}_{\perp}\cdot \vec{l}_{\perp}}{ l_{\perp}^2} T_c T_{a_1} e^{ix_Bp^+y^-} e^{i(x_L+x_D)p^+y_1^-} e^{-i\vec{k}_{\perp}  \cdot \vec{y}_{1\perp}}ig \theta(y_1^- - y^-).
\end{split}
\end{equation}
Contribution to the hadronic tensor of SIDIS from the above amplitude is
\begin{equation}
\begin{split}
\frac{d W^{\mu \nu}_{D(1)q} }{d z_h}= & \int_{z_h}^1 \frac{dz}{z}  D_{q\rightarrow h} (z_h/z)  \int \frac{dy^-}{2\pi} dy_1^- dy_2^- \int d^2\vec{y}_{12\perp}  \int \frac{d^2 k_{\perp}}{(2\pi)^2}e^{i\vec{k}_{\perp}\cdot\vec{y}_{12\perp}} \\
& \times \langle A | \bar{\psi}(y^-)  \dfrac{\gamma^{+}}{2} A^{+}(y_1^-,y_{1\perp})  A^+(y_2^-, y_{2\perp}) \psi(0)| A \rangle H^{\mu \nu}_{D(1)q}, \\
\end{split}
\label{eq:hard_part}
\end{equation}
where the hard partonic part is
\begin{equation}
\begin{split}
H_{D(1)q}^{\mu \nu} =&  \frac{1}{2} \frac{1}{N_c(N_c^2-1)}\sum_\text{spin,color}   \int \dfrac{d^4 l}{(2\pi)^4} 2\pi \delta(l^2) M^{\mu}_{D1}(0,y_2) M^{\nu \dagger}_{D1}(y,y_1)   \\
 &\times \delta(x+x_3-x_1-x_2)  2\pi \delta([(x+x_3)p +q +k_{\perp}-l]^2)  \\
=&  \frac{1}{2} \frac{1}{N_c(N_c^2-1)}\sum_\text{spin,color} \int \dfrac{d^4 l}{(2\pi)^4} 2\pi \delta(l^2) \int \frac{dx}{2\pi} 2\pi \delta[(xp+q)^2]  \\
&\bar{u}^{s'}(p) \gamma^{\mu} u^s(x_1p+q) \bar{u}^s(xp+q)\gamma^{\nu} u^{s'}(p) \overline{M}_{D1}(0,y_2) \overline{M}^{\dagger}_{D1}(y,y_1)  \\
= &  \int dx H_{(0)}^{\mu\nu} \int \frac{dz}{1-z} \int \frac{d l_{\perp}^2 }{2(2\pi)^2} \frac{1}{2} \frac{1}{N_c(N_c^2-1)} \sum_\text{spin,color} \overline{M}_{D1}(0,y_2) \overline{M}^{\dagger}_{D1}(y,y_1)\\
=& \int dx H_{(0)}^{\mu\nu} \int dz \frac{2}{1-z}  \int d l_{\perp}^2  e^{-ix_Bp^+y^-}e^{-i(x_L+x_D)p^+(y_1^- -y_2^-)} \frac{1}{{l}_{\perp}^2} \frac{\alpha_s}{2\pi} C_F \frac{2\pi \alpha_s}{N_c}.
\end{split}
\label{eq:hard_part_helicity}
\end{equation}
 \end{widetext}
 
In the above calculation, the initial spin and color indices are averaged and final state spin/color indices are summed. 
In the soft gluon approximation $z \rightarrow 1$,
\begin{equation}
\begin{split}
& \frac{1+z^2}{1-z}   \approx \frac{2}{1-z},\\
& \frac{1}{[\vec{l}_{\perp} -(1-z)\vec{k}_{\perp}]^2}   \approx \frac{1}{{l}_{\perp}^2},
\end{split}
\end{equation}
the above contribution to the hadronic tensor in the helicity amplitude approach is the same as the complete result Eq.~(\ref{eq:central 11}) from the cut-diagram in Fig.~\ref{fig:C11w}. In this work, we use results from helicity amplitude approach to cross check the complete result calculated from cut diagrams. Below we list the helicity amplitude for single, double and triple scattering.

\begin{figure}[h]
  \centering 
  \includegraphics[scale=0.6]{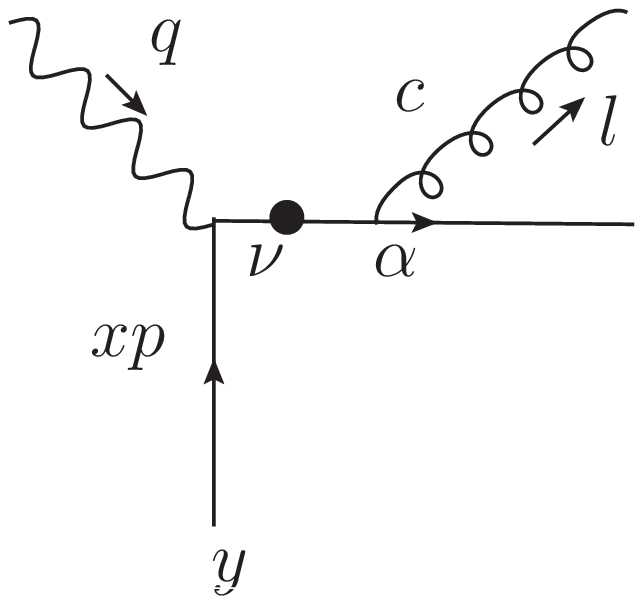}
  \caption{Single scattering with $x= x_B+x_L$.}
  \label{fig:single scattering amplitude}
\end{figure}

\subsection{Single scattering amplitude}

 The amplitude for  single scattering in Fig.~\ref{fig:single scattering amplitude}:
\begin{equation}
\begin{split}
 M^{\nu}_{S}(y)&= \int \dfrac{dx}{2\pi} \delta(x-x_B-x_L)\bar{u}^s(xp+q)\gamma^{\nu} u^{s'}(p)  \overline{M}_{S}(y), \\
 \overline{M}_{S}(y) &= 2 g  \dfrac{\vec{\epsilon}_{\perp}\cdot \vec{l}_{\perp}}{l_{\perp}^2}T_c e^{i(x_B+x_L)p^+y^-} .
\end{split}
\end{equation}


 \begin{widetext}
\subsection{Double scattering amplitude}

 (1) Double scattering 1 in Fig.~\ref{fig:double scattering amplitude}:
\begin{equation}
\begin{split}
M^{\nu}_{D1}(y,y_1) =&\int \frac{dx}{2\pi} \delta(x-x_B)\int dx_3  \delta (x_3 - x_L -x_D)  \bar{u}^{s}(xp+q) \gamma^{\nu} u^{s'}(p) \overline{M}_{D1}(y,y_1),   \\
\overline{M}_{D1}(y,y_1) 
= & 2 g \dfrac{\vec{\epsilon}_{\perp}\cdot \vec{l}_{\perp}}{ l_{\perp}^2} T_c T_{a_1} e^{ix_Bp^+y^-} e^{i(x_L+x_D)p^+y_1^-} e^{-i\vec{k}_{\perp}  \cdot \vec{y}_{1\perp}}ig \theta(y^- - y_1^-).
\end{split}
\end{equation}

\begin{figure*}[h] 
\centering
\subfloat[][Double scattering  2a with $x= x_B$, $x_3 = x_L +x_D$ ]{
\begin{minipage}[t]{0.5\linewidth} 
\centering 
\includegraphics[scale=0.55]{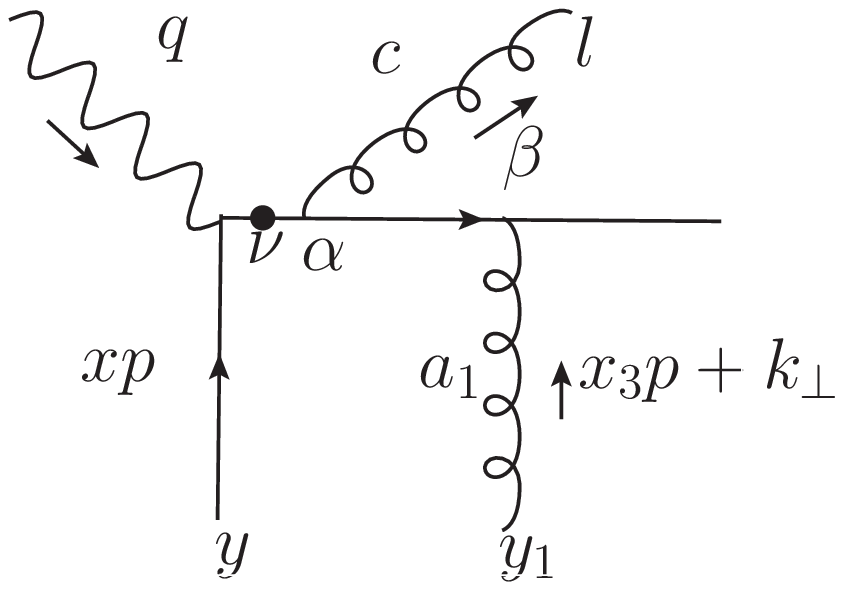} 
\end{minipage}
} 
\subfloat[][Double scattering 2b with $x= x_B+x_L$, $x_3 = x_D$]{
\begin{minipage}[t]{0.5\linewidth} 
\centering 
\includegraphics[scale=0.55]{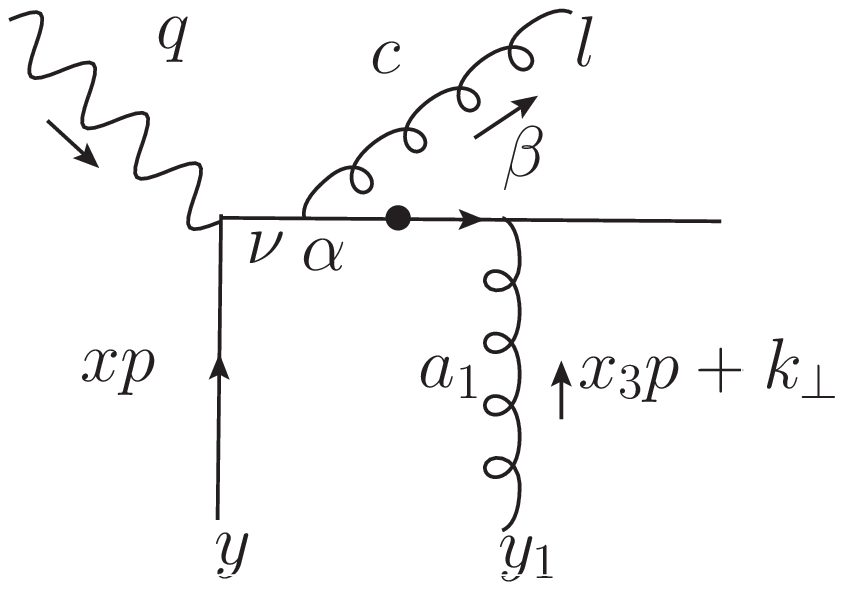} 
\end{minipage}
}
 \caption{Double scattering 2.}
  \label{fig:double scattering amplitude 2}
\end{figure*}

(2) Double scattering 2 in Fig.~\ref{fig:double scattering amplitude 2}:
\begin{equation}
\begin{split}
M^{\nu}_{D2a}(y,y_1) =&\int \frac{dx}{2\pi}  \delta(x-x_B) \int dx_3  \delta (x_3 - x_L -x_D)  \bar{u}^{s}(xp+q) \gamma^{\nu} u^{s'}(p) \overline{M}_{D2a}(y,y_1),   \\
\overline{M}_{D2a}(y,y_1) = &-2g \dfrac{\vec{\epsilon}_{\perp}\cdot\vec{l}_{\perp}}{l_{\perp}^2} T_{a_1} T_c e^{ix_Bp^+y^-}e^{i(x_L+x_D)p^+y_1^-}e^{-i\vec{k}_{\perp}  \cdot \vec{y}_{1\perp}}ig \theta(y^-  - y_1^-),\\
M^{\nu}_{D2b}(y,y_1) =&\int \frac{dx}{2\pi}  \delta(x-x_B-x_L) \int dx_3  \delta (x_3 -x_D)  \bar{u}^{s}(xp+q) \gamma^{\nu} u^{s'}(p) \overline{M}_{D2b}(y,y_1),   \\
\overline{M}_{D2b}(y,y_1) = &2g \dfrac{\vec{\epsilon}_{\perp}\cdot\vec{l}_{\perp}}{l_{\perp}^2} T_{a_1} T_c e^{i(x_B+x_L)p^+y^-}e^{ix_Dp^+y_1^-}e^{-i\vec{k}_{\perp}  \cdot \vec{y}_{1\perp}} ig \theta(y^-  - y_1^-) ,\\
\overline{M}_{D2}(y,y_1) =&\overline{M}_{D2a}(y,y_1) + \overline{M}_{D2b}(y,y_1) \\
=&2g \dfrac{\vec{\epsilon}_{\perp}\cdot\vec{l}_{\perp}}{l_{\perp}^2} T_{a_1} T_c [e^{i(x_B+x_L)p^+y^-}e^{ix_Dp^+y_1^-} - e^{ix_Bp^+y^-}e^{i(x_L+x_D)p^+y_1^-}]e^{-i\vec{k}_{\perp}  \cdot \vec{y}_{1\perp}} ig \theta(y^-  - y_1^-).\\
\end{split}
\end{equation}

\begin{figure*}[h] 
\centering

\subfloat[][Double scattering amplitude 3a with $x= x_B$, $x_3 = x_L +x_D$. ]{
\begin{minipage}[t]{0.5\linewidth} 
\centering 
\includegraphics[scale=0.55]{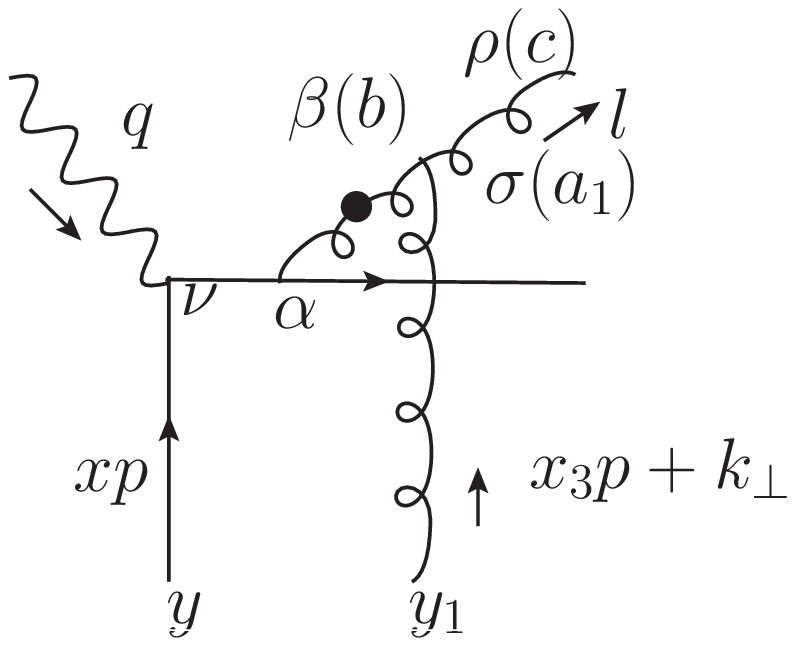} 
\end{minipage}
} 
\subfloat[][Double scattering amplitude 3b with $x= x_B+x_L+\frac{x_D}{1-z}$, $x_3 = -\frac{z}{1-z} x_D$.]{
\begin{minipage}[t]{0.5\linewidth} 
\centering 
\includegraphics[scale=0.55]{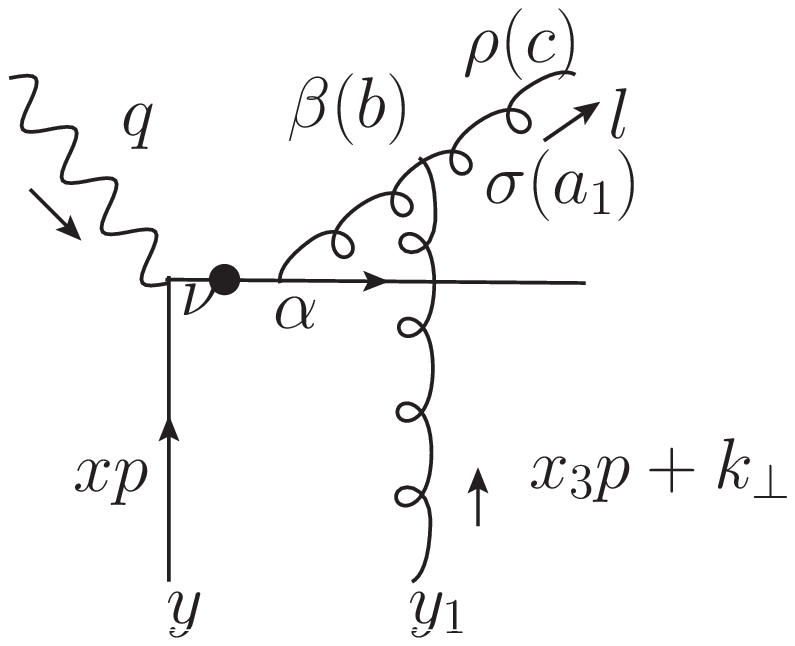} 
\end{minipage}
}
 \caption{Double scattering  3.}
  \label{fig:double scattering amplitude 3}
\end{figure*}

(3) Double scattering 3 in Fig.~\ref{fig:double scattering amplitude 3}:
\begin{equation}
\begin{split}
M^{\nu}_{D3a}(y,y_1) =&\int \frac{dx}{2\pi}  \delta(x-x_B) \int dx_3  \delta (x_3 - x_L -x_D)  \bar{u}^{s}(xp+q) \gamma^{\nu} u^{s'}(p) \overline{M}_{D3a}(y,y_1),   \\
\overline{M}_{D3a}(y,y_1) = &  2g \dfrac{\vec{\epsilon}_{\perp}\cdot(\vec{l}_{\perp}-\vec{k}_{\perp})}{(\vec{l}_{\perp}-\vec{k}_{\perp})^2}[T_{a_1},T_c] e^{ix_Bp^+y^-} e^{i(x_L+x_D)p^+y_1^-}e^{-i\vec{k}_{\perp}\cdot \vec{y}_{1\perp}}ig \theta(y^-  - y_1^-), \\
M^{\nu}_{D3b}(y,y_1) =&\int \frac{dx}{2\pi}  \delta(x-x_B-x_L-\frac{x_D}{1-z}) \int dx_3  \delta (x_3 +\frac{z}{1-z}x_D)  \bar{u}^{s}(xp+q) \gamma^{\nu} u^{s'}(p) \overline{M}_{D3b}(y,y_1),   \\
\overline{M}_{D3b}(y,y_1) = &-2g \dfrac{\vec{\epsilon}_{\perp}\cdot(\vec{l}_{\perp}-\vec{k}_{\perp})}{(\vec{l}_{\perp}-\vec{k}_{\perp})^2}[T_{a_1},T_c] e^{i(x_B+x_L+\frac{x_D}{1-z})p^+y^-} e^{-i\frac{z}{1-z} x_Dp^+y_1^-}e^{-i\vec{k}_{\perp}\cdot \vec{y}_{1\perp}}ig \theta( y^- -y_1^- ),\\
\overline{M}_{D3}(y,y_1) =&\overline{M}_{D3a}(y,y_1) + \overline{M}_{D3b}(y,y_1) \\
=&2g  \dfrac{\vec{\epsilon}_{\perp}\cdot(\vec{l}_{\perp}-\vec{k}_{\perp})}{(\vec{l}_{\perp}-\vec{k}_{\perp})^2}[T_{a_1},T_c]  [e^{ix_Bp^+y^-} e^{i(x_L+x_D)p^+y_1^-} - e^{i(x_B+x_L+\frac{x_D}{1-z})p^+y^-} e^{-i\frac{z}{1-z} x_Dp^+y_1^-}]\\
&\times e^{-i\vec{k}_{\perp}\cdot \vec{y}_{1\perp}} ig \theta(y^-  - y_1^-). \\
\end{split}
\end{equation}

The sum of double scattering amplitude is 
\begin{equation}
\begin{split}
\overline{M}_{D}(y,y_1) = \overline{M}_{D1}(y,y_1)+ \overline{M}_{D2}(y,y_1)+\overline{M}_{D3}(y,y_1).
\end{split}
\end{equation}

\subsection{Triple scattering amplitude}

(1) Triple scattering 1 in Fig.~\ref{fig:triple scattering amplitude 1}:

 \begin{figure}[h]
  \centering 
  \includegraphics[scale=0.65]{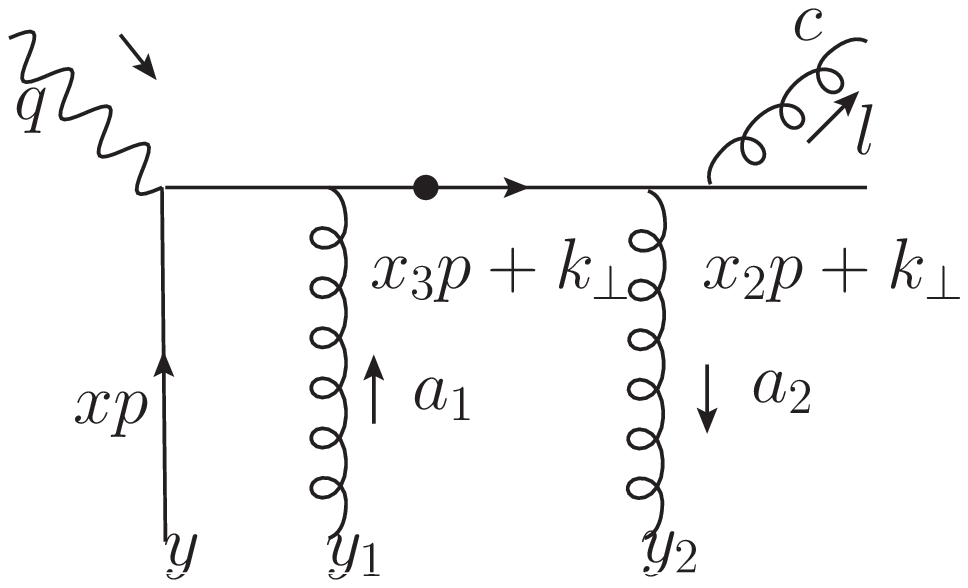}
  \caption{Triple scattering 1 with $x= x_B$, $x_3 = x_D^0$, $x_2 = x_D^0 -x_L$. }
  \label{fig:triple scattering amplitude 1}
\end{figure}

\begin{equation}
\begin{split}
M_{T1}^{\nu}(y,y_1,y_2) = & \int \dfrac{dx}{2\pi} \delta(x-x_B)  \int dx_3 \delta(x_3 - x_D^0) \int dx_2 \delta(x_2 - x_D^0 +x_L)\bar{u}^s(xp+q)\gamma^{\nu}u^{s'}(p) \overline{M}_{T1}(y,y_1,y_2), \\
\overline{M}_{T1}(y,y_1,y_2) = & 2g \dfrac{\vec{\epsilon}_{\perp}\cdot \vec{l}_{\perp}}{l_{\perp}^2}T_c T_{a_2} T_{a_1} e^{ix_Bp^+y^-}e^{ix_D^0p^+y_1^-}e^{-i(x_D^0-x_L)p^+y_2^-}e^{-i\vec{k}_{\perp}\cdot (\vec{y}_{1\perp}-\vec{y}_{2\perp})}\\
&\times (-g^2) \theta(y_1^- - y_2^-)\theta(y^- - y_1^-),\\ 
\end{split}
\end{equation}
where $x_D^0 =  \dfrac{k_{\perp}^2}{2p^+q^-} $

\begin{figure*}[h] 
\centering
\subfloat[][Triple scattering 2a with $x= x_B$, $x_3 =x_L+ x_D$, $x_2 = x_D$.]{
\begin{minipage}[t]{0.5\linewidth} 
\centering 
\includegraphics[scale=0.55]{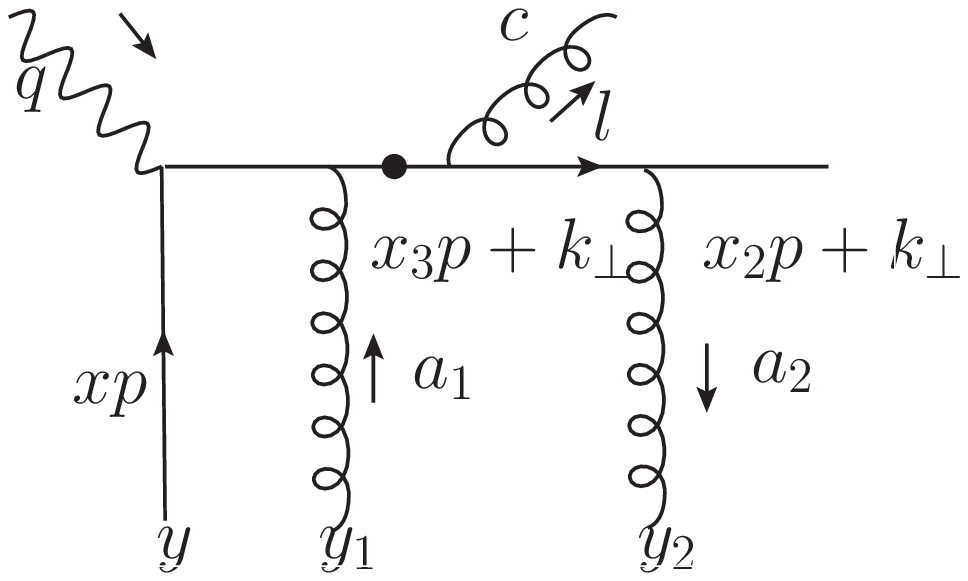} 
\end{minipage}
} 
\subfloat[][Triple scattering 2b with $x= x_B$, $x_3 = x_D^0$, $x_2 = x_D^0 -x_L$. ]{
\begin{minipage}[t]{0.5\linewidth} 
\centering 
\includegraphics[scale=0.55]{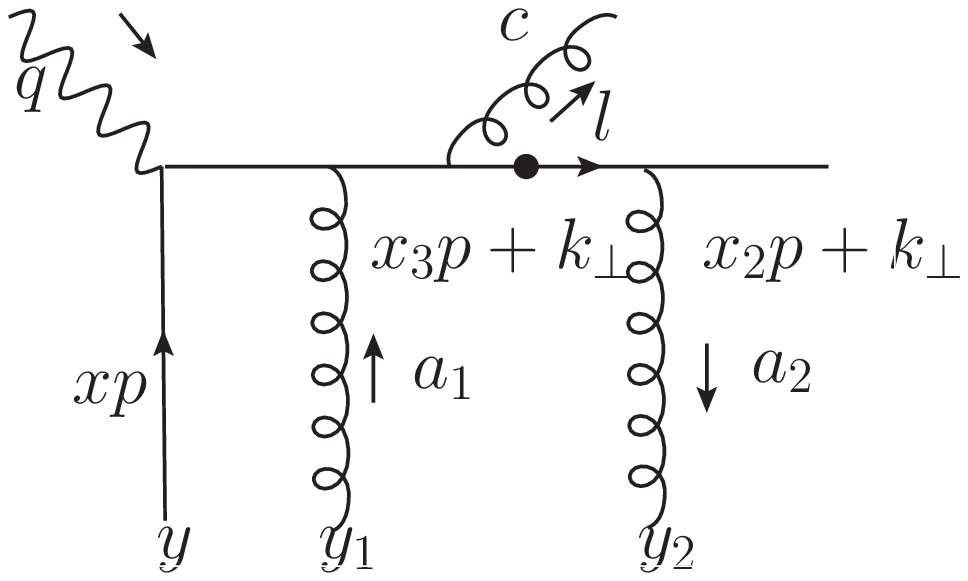} 
\end{minipage}
}
 \caption{Triple scattering 2.}
  \label{fig:triple scattering amplitude 2}
\end{figure*}

(2) Triple scattering 2 in Fig.~\ref{fig:triple scattering amplitude 2}:
\begin{equation}
\begin{split}
M_{T2a}^{\nu}(y,y_1,y_2) = & \int \dfrac{dx}{2\pi} \delta(x-x_B)  \int dx_3 \delta(x_3 - x_L-x_D) \int dx_2 \delta(x_2 - x_D) \bar{u}^s(xp+q)\gamma^{\nu}u^{s'}(p) \overline{M}_{T2a}(y,y_1,y_2), \\
\overline{M}_{T2a}(y,y_1,y_2) = & 2g \dfrac{\vec{\epsilon}_{\perp}\cdot \vec{l}_{\perp}}{l_{\perp}^2} T_{a_2}T_c T_{a_1} e^{ix_Bp^+y^-}e^{i(x_L+x_D)p^+y_1^-}e^{-ix_Dp^+y_2^-}e^{-i\vec{k}_{\perp}\cdot (\vec{y}_{1\perp}-\vec{y}_{2\perp})}\\
&\times(-g^2) \theta(y_1^- - y_2^-)\theta(y^- - y_1^-),\\ 
M_{T2b}^{\nu}(y,y_1,y_2) = & \int \dfrac{dx}{2\pi} \delta(x-x_B)  \int dx_3 \delta(x_3 - x_D^0) \int dx_2 \delta(x_2 - x_D^0 +x_L) \bar{u}^s(xp+q)\gamma^{\nu}u^{s'}(p) \overline{M}_{T2b}(y,y_1,y_2) ,\\
\overline{M}_{T2b}(y,y_1,y_2) = & - 2g \dfrac{\vec{\epsilon}_{\perp}\cdot \vec{l}_{\perp}}{l_{\perp}^2} T_{a_2}T_c T_{a_1} e^{ix_Bp^+y^-}e^{ix_D^0p^+y_1^-}e^{-i(x_D^0-x_L)p^+y_2^-}e^{-i\vec{k}_{\perp}\cdot (\vec{y}_{1\perp}-\vec{y}_{2\perp})}\\
&\times(-g^2) \theta(y_1^- - y_2^-)\theta(y^- - y_1^-)\\ 
\overline{M}_{T2}(y,y_1,y_2) = &\overline{M}_{T2a}(y,y_1,y_2) +   \overline{M}_{T2b}(y,y_1,y_2)  \\
& = 2g \dfrac{\vec{\epsilon}_{\perp}\cdot \vec{l}_{\perp}}{l_{\perp}^2} T_{a_2}T_c T_{a_1} [e^{ix_Bp^+y^-}e^{i(x_L+x_D)p^+y_1^-}e^{-ix_Dp^+y_2^-}-e^{ix_Bp^+y^-}e^{ix_D^0p^+y_1^-}e^{-i(x_D^0-x_L)p^+y_2^-}]\\
&\times e^{-i\vec{k}_{\perp}\cdot (\vec{y}_{1\perp}-\vec{y}_{2\perp})}(-g^2) \theta(y_1^- - y_2^-)\theta(y^- - y_1^-).\\
\end{split}
\end{equation}

\begin{figure*}[h] 

\centering

\subfloat[][Triple scattering 3a with $x= x_B+x_L$, $x_3 =x_D$, $x_2 = x_D$.]{
\begin{minipage}[t]{0.5\linewidth} 
\centering 
\includegraphics[scale=0.55]{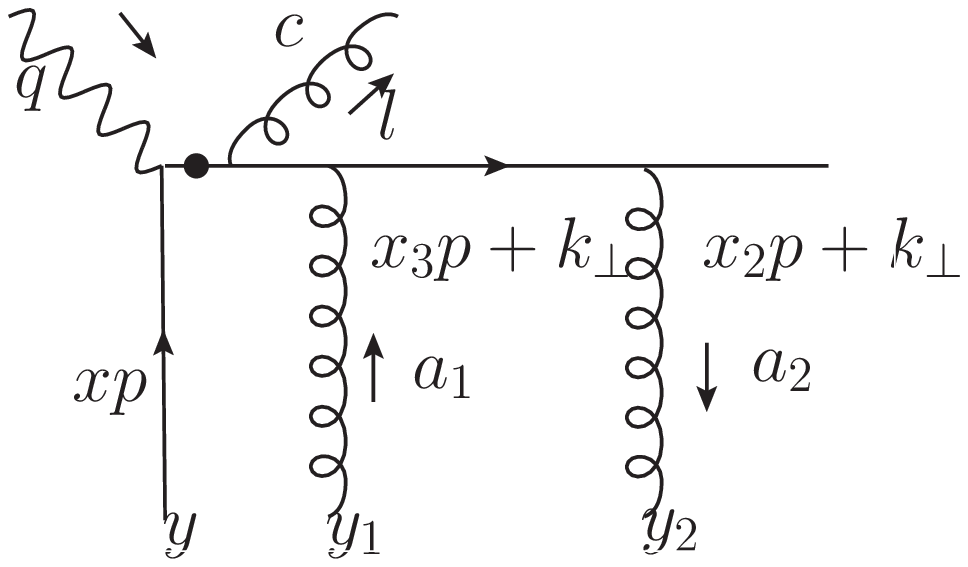} 
\end{minipage}
} 
\subfloat[][Triple scattering 3b with $x= x_B$, $x_3 = x_L+x_D$, $x_2 = x_D$.]{
\begin{minipage}[t]{0.5\linewidth} 
\centering 
\includegraphics[scale=0.55]{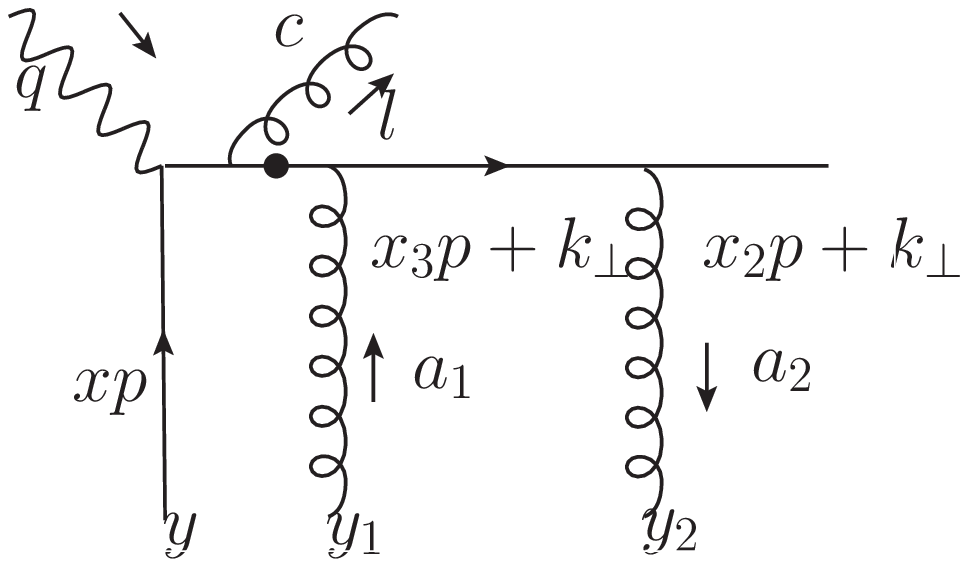} 
\end{minipage}
}
 \caption{Triple scattering 3}
  \label{fig:triple scattering amplitude 3.}
\end{figure*}

(3) Triple scattering  3 in Fig.~\ref{fig:triple scattering amplitude 3}:
\begin{equation}
\begin{split}
M_{T3a}^{\nu}(y,y_1,y_2) = & \int \dfrac{dx}{2\pi} \delta(x-x_B-x_L)  \int dx_3 \delta(x_3 - x_D) \int dx_2 \delta(x_2 - x_D) \bar{u}^s(xp+q)\gamma^{\nu}u^{s'}(p) \overline{M}_{T3a}(y,y_1,y_2) \\
\overline{M}_{T3a}(y,y_1,y_2) = & 2g \dfrac{\vec{\epsilon}_{\perp}\cdot \vec{l}_{\perp}}{l_{\perp}^2}T_{a_2} T_{a_1} T_c e^{i(x_B+x_L)p^+y^-}e^{ix_Dp^+y_1^- } e^{-ix_Dp^+y_2^-}
e^{-i\vec{k}_{\perp}\cdot (\vec{y}_{1\perp}-\vec{y}_{2\perp})}\\
&\times(-g^2) \theta(y_1^- - y_2^-)\theta(y^- - y_1^-)\\ 
M_{T3b}^{\nu}(y,y_1,y_2) = & \int \dfrac{dx}{2\pi} \delta(x-x_B)  \int dx_3 \delta(x_3 - x_L - x_D) \int dx_2 \delta(x_2 - x_D)\bar{u}^s(xp+q)\gamma^{\nu}u^{s'}(p) \overline{M}_{T3b}(y,y_1,y_2) \\
\overline{M}_{T3b}(y,y_1,y_2) = &- 2g \dfrac{\vec{\epsilon}_{\perp}\cdot \vec{l}_{\perp}}{l_{\perp}^2}T_{a_2} T_{a_1} T_c e^{ix_Bp^+y^-}e^{i(x_L+x_D)p^+y_1^-}e^{-ix_Dp^+y_2^-}
 e^{-i\vec{k}_{\perp}\cdot (\vec{y}_{1\perp}-\vec{y}_{2\perp})}\\
 &\times (-g^2)\theta(y_1^- - y_2^-)\theta(y^- - y_1^-)\\ 
\overline{M}_{T3}(y,y_1,y_2) = &\overline{M}_{T3a}(y,y_1,y_2) +   \overline{M}_{T3b}(y,y_1,y_2)  \\
&=  2g \dfrac{\vec{\epsilon}_{\perp}\cdot \vec{l}_{\perp}}{l_{\perp}^2} T_{a_2}T_{a_1}T_c  [e^{i(x_B+x_L)p^+y^-}e^{ix_Dp^+y_1^- } e^{-ix_Dp^+y_2^-} \\
&- e^{ix_Bp^+y^-}e^{i(x_L+x_D)p^+y_1^-}e^{-ix_Dp^+y_2^-}]e^{-i\vec{k}_{\perp}\cdot (\vec{y}_{1\perp}-\vec{y}_{2\perp})}\\
&\times (-g^2) \theta(y_1^- - y_2^-)\theta(y^- - y_1^-)\\
\end{split}
\end{equation}

\begin{figure*}[h] 
\centering

\subfloat[][Triple scattering  4a with $x= x_B$, $x_3 =x_D^0$, $x_2 = x_D^0-x_L$.]{
\begin{minipage}[t]{0.5\linewidth} 
\centering 
\includegraphics[scale=0.55]{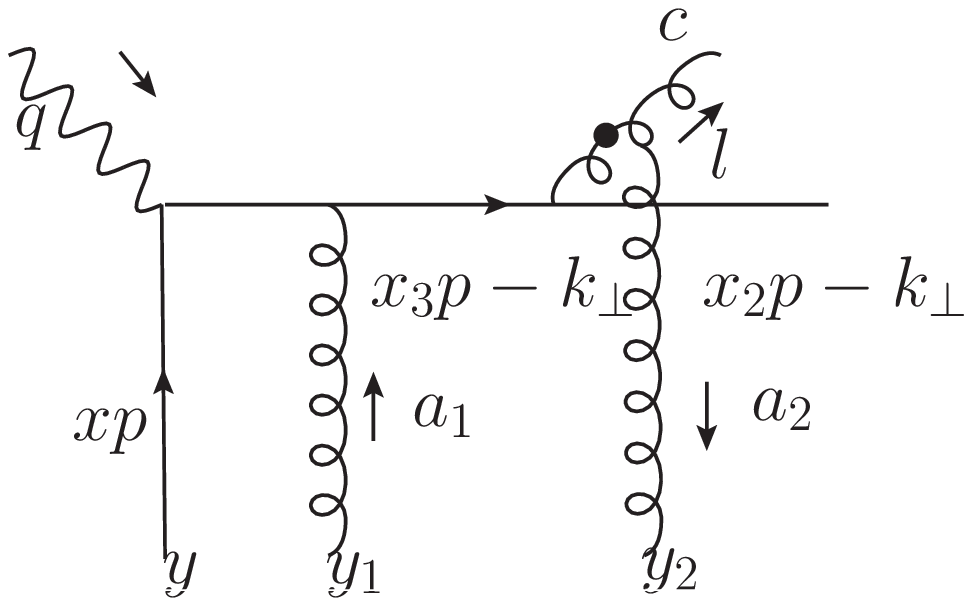} 
\end{minipage}
} 
\subfloat[][Triple scattering 4b with $x= x_B$, $x_3 = x_L+\frac{z}{1-z}x_D$, $x_2 =  \frac{z}{1-z}x_D$. ]{
\begin{minipage}[t]{0.5\linewidth} 
\centering 
\includegraphics[scale=0.55]{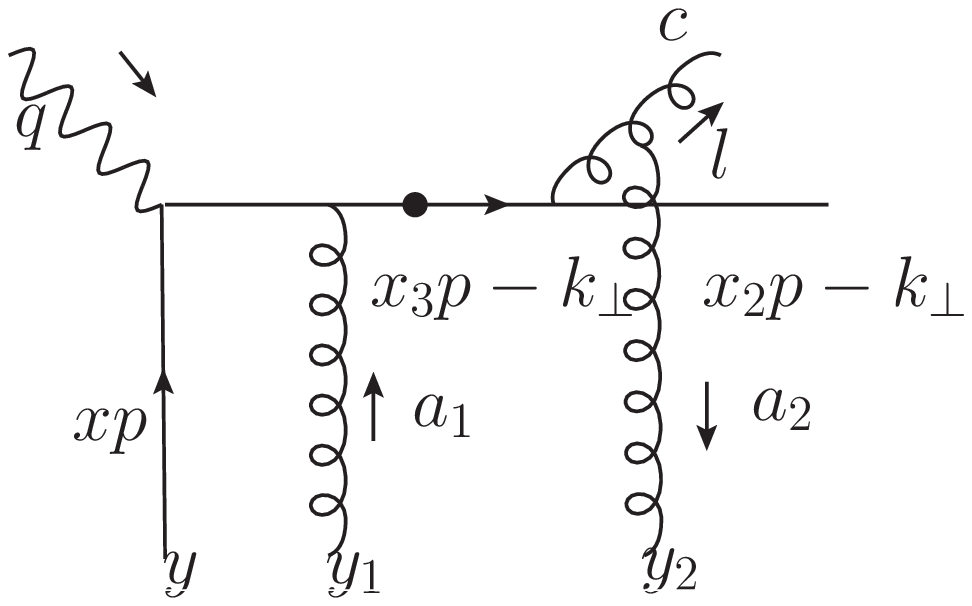} 
\end{minipage}
}
 \caption{Triple scattering 4.}
  \label{fig:triple scattering amplitude 4}
\end{figure*}

(4) Triple scattering 4 in Fig.~\ref{fig:triple scattering amplitude 4}:
\begin{equation}
\begin{split}
M_{T4a}^{\nu}(y,y_1,y_2) = & \int \dfrac{dx}{2\pi} \delta(x-x_B)  \int dx_3 \delta(x_3 - x_D^0) \int dx_2 \delta(x_2 - x_D^0+x_L) \bar{u}^s(xp+q)\gamma^{\nu}u^{s'}(p) \overline{M}_{T4a}(y,y_1,y_2), \\
\overline{M}_{T4a}(y,y_1,y_2) = & 2g \dfrac{\vec{\epsilon}_{\perp}\cdot (\vec{l}_{\perp}-\vec{k}_{\perp})}{(\vec{l}_{\perp}-\vec{k}_{\perp})^2}[T_{a_2},T_c] T_{a_1}  e^{ix_Bp^+y^-}e^{ix_D^0p^+y_1^- } e^{-i(x_D^0-x_L)p^+y_2^-}
e^{-i\vec{k}_{\perp}\cdot (\vec{y}_{1\perp}-\vec{y}_{2\perp})}\\
&\times(-g^2) \theta(y_1^- - y_2^-)\theta(y^- - y_1^-),\\ 
M_{T4b}^{\nu}(y,y_1,y_2) = & \int \dfrac{dx}{2\pi} \delta(x-x_B)  \int dx_3 \delta(x_3 - x_L-\frac{z}{1-z}x_D ) \int dx_2 \delta(x_2 - \frac{z}{1-z}x_D )\\
&\times \bar{u}^s(xp+q)\gamma^{\nu}u^{s'}(p) \overline{M}_{T4b}(y,y_1,y_2), \\
\overline{M}_{T4b}(y,y_1,y_2) = &- 2g \dfrac{\vec{\epsilon}_{\perp}\cdot (\vec{l}_{\perp}-\vec{k}_{\perp})}{(\vec{l}_{\perp}-\vec{k}_{\perp})^2}[T_{a_2},T_c] T_{a_1} e^{i(x_Bp^+y^-}e^{i(x_L+\frac{z}{1-z}x_D )p^+y_1^- }e^{-i\frac{z}{1-z}x_Dp^+y_2^-}e^{-i\vec{k}_{\perp}\cdot (\vec{y}_{1\perp}-\vec{y}_{2\perp})}\\
&\times (-g^2)\theta(y_1^- - y_2^-)\theta(y^- - y_1^-),\\ 
\overline{M}_{T4}(y,y_1,y_2) = &\overline{M}_{T4a}(y,y_1,y_2) +   \overline{M}_{T4b}(y,y_1,y_2)  \\
=&  2g \dfrac{\vec{\epsilon}_{\perp}\cdot (\vec{l}_{\perp}-\vec{k}_{\perp})}{(\vec{l}_{\perp}-\vec{k}_{\perp})^2}[T_{a_2},T_c] T_{a_1}[ e^{ix_Bp^+y^-}e^{ix_D^0p^+y_1^- } e^{-i(x_D^0-x_L)p^+y_2^-} \\
&- e^{ix_Bp^+y^-}e^{i(x_L+\frac{z}{1-z}x_D )p^+y_1^- }e^{-i\frac{z}{1-z}x_Dp^+y_2^-}]e^{-i\vec{k}_{\perp}\cdot (\vec{y}_{1\perp}-\vec{y}_{2\perp})}\\
&\times(-g^2) \theta(y_1^- - y_2^-)\theta(y^- - y_1^-).\\
\end{split}
\end{equation}
We have made variable change $\vec{k} \rightarrow -\vec{k}$ in the above.

\begin{figure*}[h] 
\centering

\subfloat[][Triple scattering 5a with $x= x_B$, $x_3 =x_L+x_D$, $x_2 = x_D$.]{
\begin{minipage}[t]{0.5\linewidth} 
\centering 
\includegraphics[scale=0.55]{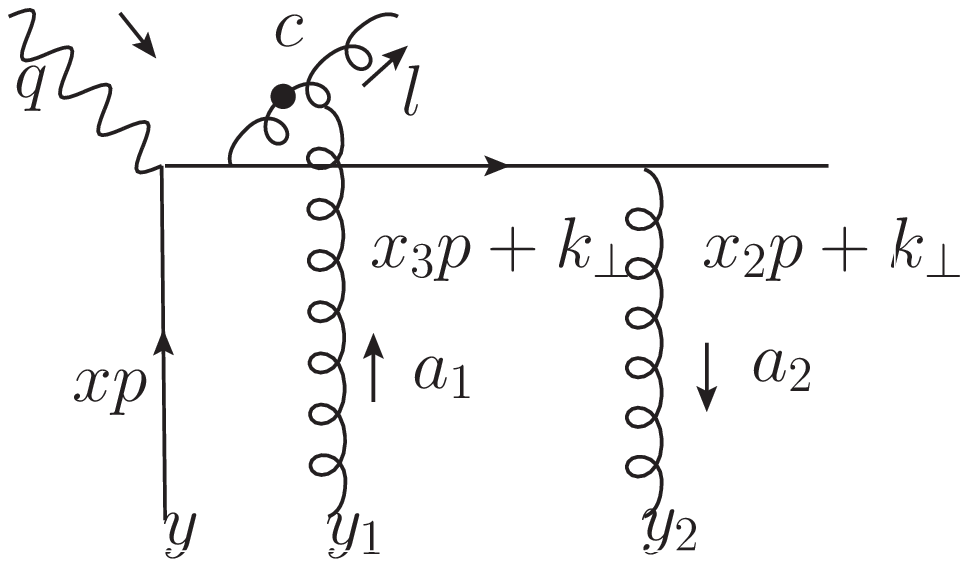} 
\end{minipage}
} 
\subfloat[][Triple scattering 5b with $x= x_B+x_L+\frac{x_D}{1-z}$, $x_3 = -\frac{z}{1-z}x_D$, $x_2 =  x_D$.]{
\begin{minipage}[t]{0.5\linewidth} 
\centering 
\includegraphics[scale=0.55]{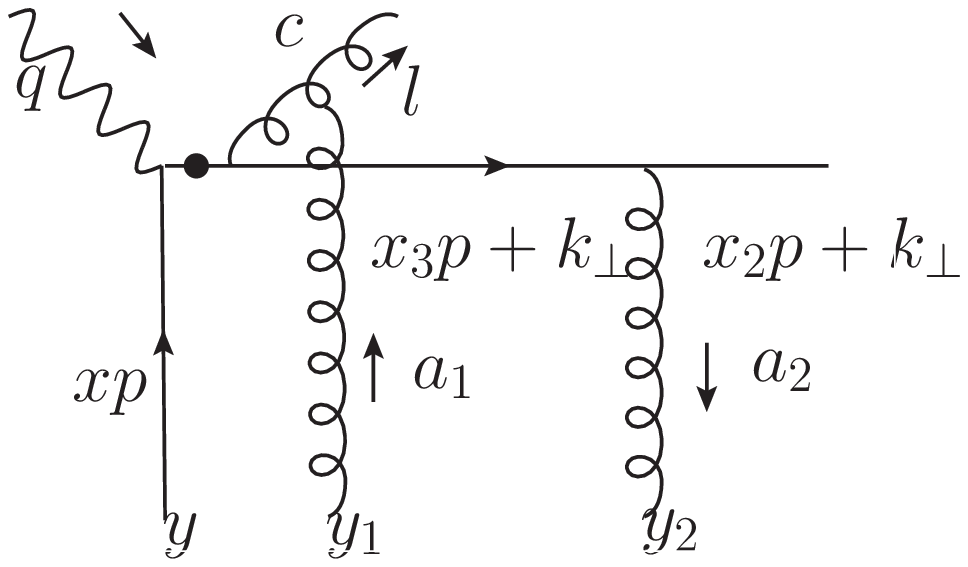} 
\end{minipage}
}
 \caption{Triple scattering 5.}
  \label{fig:triple scattering amplitude 5}
\end{figure*} 

(5) Triple scattering  5 in Fig.~\ref{fig:triple scattering amplitude 5}: 
\begin{equation}
\begin{split}
M_{T5a}^{\nu}(y,y_1,y_2) = & \int \dfrac{dx}{2\pi} \delta(x-x_B)  \int dx_3 \delta(x_3 - x_L-x_D) \int dx_2 \delta(x_2 - x_D) \bar{u}^s(xp+q)\gamma^{\nu}u^{s'}(p) \overline{M}_{T5a}(y,y_1,y_2), \\
\overline{M}_{T5a}(y,y_1,y_2) = & 2g \dfrac{\vec{\epsilon}_{\perp}\cdot (\vec{l}_{\perp}-\vec{k}_{\perp})}{(\vec{l}_{\perp}-\vec{k}_{\perp})^2}T_{a_2}[T_{a_1},T_c] e^{ix_Bp^+y^-}e^{i(x_L+x_D)p^+y_1^- } e^{-ix_Dp^+y_2^-} \\
&\times e^{-i\vec{k}_{\perp}\cdot (\vec{y}_{1\perp}-\vec{y}_{2\perp})}(-g^2) \theta(y_1^- - y_2^-)\theta(y^- - y_1^-),\\ 
M_{T5b}^{\nu}(y,y_1,y_2) = & \int \dfrac{dx}{2\pi} \delta(x-x_B-x_L-  \frac{x_D}{1-z} )  \int dx_3 \delta(x_3 +\frac{z}{1-z}x_D ) \int dx_2 \delta(x_2 - x_D )\\
&\times \bar{u}^s(xp+q)\gamma^{\nu}u^{s'}(p) \overline{M}_{T5b}(y,y_1,y_2), \\
\overline{M}_{T5b}(y,y_1,y_2) = &- 2g \dfrac{\vec{\epsilon}_{\perp}\cdot (\vec{l}_{\perp}-\vec{k}_{\perp})}{(\vec{l}_{\perp}-\vec{k}_{\perp})^2}T_{a_2}[T_{a_1},T_c] e^{i(x_B+x_L+  \frac{x_D}{1-z} )p^+y^-}e^{-i\frac{z}{1-z}x_D p^+y_1^- }e^{-ix_Dp^+y_2^-}\\
&\times e^{-i\vec{k}_{\perp}\cdot (\vec{y}_{1\perp}-\vec{y}_{2\perp})}(-g^2)\theta(y_1^- - y_2^-)\theta(y^- - y_1^-),\\ 
\overline{M}_{T5}(y,y_1,y_2) = &\overline{M}_{T5a}(y,y_1,y_2) +   \overline{M}_{T5b}(y,y_1,y_2)  \\
=&  2g \dfrac{\vec{\epsilon}_{\perp}\cdot (\vec{l}_{\perp}-\vec{k}_{\perp})}{(\vec{l}_{\perp}-\vec{k}_{\perp})^2}T_{a_2}[T_{a_1},T_c] [e^{ix_Bp^+y^-}e^{i(x_L+x_D)p^+y_1^- } e^{-ix_Dp^+y_2^-}\\
&- e^{i(x_B+x_L+  \frac{x_D}{1-z} )p^+y^-}e^{-i\frac{z}{1-z}x_D p^+y_1^- }e^{-ix_Dp^+y_2^-}  ]e^{-i\vec{k}_{\perp}\cdot (\vec{y}_{1\perp}-\vec{y}_{2\perp})}\\
&\times(-g^2) \theta(y_1^- - y_2^-)\theta(y^- - y_1^-).\\
\end{split}
\end{equation}

 \begin{figure*}[h] 
\centering

\subfloat[][Triple scattering amplitude 6a with $x= x_B$, $x_3 =x_L+\frac{z}{1-z}x_D$, $x_2 = \frac{z}{1-z}x_D$.]{
\begin{minipage}[t]{0.5\linewidth} 
\centering 
\includegraphics[scale=0.55]{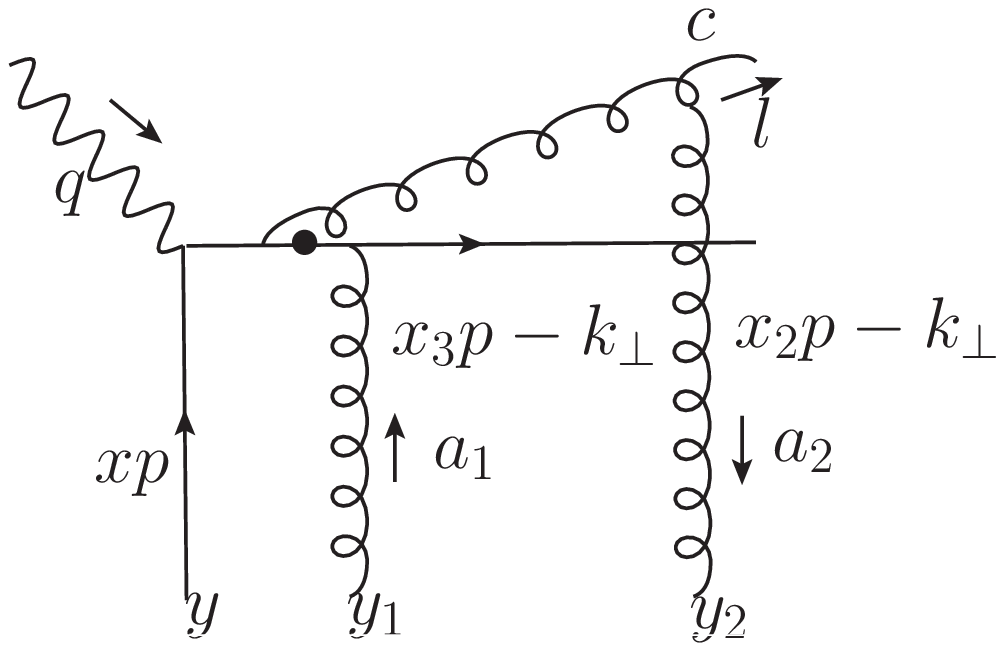} 
\end{minipage}
} 
\subfloat[][Triple scattering amplitude 6b with $x= x_B+x_L+\frac{x_D}{1-z}$, $x_3 = -x_D$, $x_2 = \frac{z}{1-z}x_D$.]{
\begin{minipage}[t]{0.5\linewidth} 
\centering 
\includegraphics[scale=0.55]{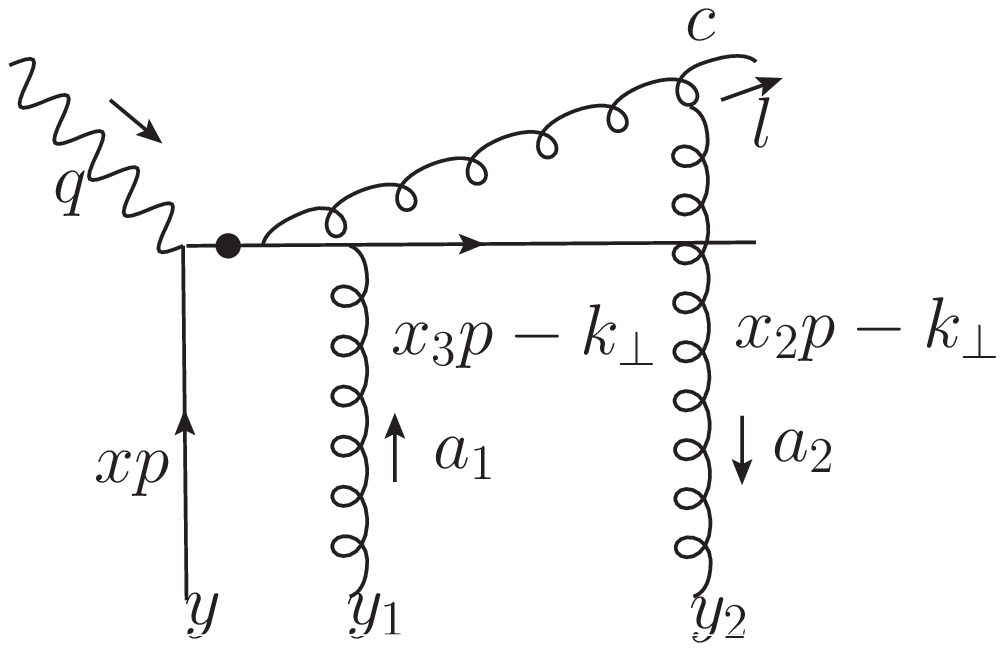} 
\end{minipage}
}
 \caption{Triple scattering amplitude 6.}
  \label{fig:triple scattering amplitude 6}
\end{figure*} 
(6) Triple scattering 6 in Fig.~\ref{fig:triple scattering amplitude 6}:
\begin{equation}
\begin{split}
M_{T6a}^{\nu}(y,y_1,y_2) = & \int \dfrac{dx}{2\pi} \delta(x-x_B)  \int dx_3 \delta(x_3 - x_L-\frac{z}{1-z}x_D) \int dx_2 \delta(x_2 - \frac{z}{1-z}x_D)\\
&\times \bar{u}^s(xp+q)\gamma^{\nu}u^{s'}(p) \overline{M}_{T6a}(y,y_1,y_2), \\
\overline{M}_{T6a}(y,y_1,y_2) = & 2g \dfrac{\vec{\epsilon}_{\perp}\cdot (\vec{l}_{\perp}-\vec{k}_{\perp})}{(\vec{l}_{\perp}-\vec{k}_{\perp})^2}T_{a_1}[T_{a_2},T_c] e^{ix_Bp^+y^-}e^{i(x_L+\frac{z}{1-z}x_D)p^+y_1^- } e^{-i\frac{z}{1-z}x_Dp^+y_2^-}\\
&\times e^{-i\vec{k}_{\perp}\cdot (\vec{y}_{1\perp}-\vec{y}_{2\perp})}(-g^2) \theta(y_1^- - y_2^-)\theta(y^- - y_1^-),\\ 
M_{T6b}^{\nu}(y,y_1,y_2) = & \int \dfrac{dx}{2\pi} \delta(x-x_B-x_L-  \frac{x_D}{1-z} )  \int dx_3 \delta(x_3 +x_D ) \int dx_2 \delta(x_2 - \frac{z}{1-z}
x_D )\\
&\times \bar{u}^s(xp+q)\gamma^{\nu}u^{s'}(p) \overline{M}_{T6b}(y,y_1,y_2), \\
\overline{M}_{T6b}(y,y_1,y_2) = &- 2g \dfrac{\vec{\epsilon}_{\perp}\cdot (\vec{l}_{\perp}-\vec{k}_{\perp})}{(\vec{l}_{\perp}-\vec{k}_{\perp})^2}T_{a_1}[T_{a_2},T_c] e^{i(x_B+x_L+  \frac{x_D}{1-z} )p^+y^-}e^{-ix_D p^+y_1^- }e^{-i\frac{z}{1-z}
x_Dp^+y_2^-}\\
 &\times e^{-i\vec{k}_{\perp}\cdot (\vec{y}_{1\perp}-\vec{y}_{2\perp})}(-g^2)\theta(y_1^- - y_2^-)\theta(y^- - y_1^-),\\ 
\overline{M}_{T6}(y,y_1,y_2) = &\overline{M}_{T6a}(y,y_1,y_2) +   \overline{M}_{T6b}(y,y_1,y_2)  \\
=& 2g \dfrac{\vec{\epsilon}_{\perp}\cdot (\vec{l}_{\perp}-\vec{k}_{\perp})}{(\vec{l}_{\perp}-\vec{k}_{\perp})^2}T_{a_1}[T_{a_2},T_c] [
e^{ix_Bp^+y^-}e^{i(x_L+\frac{z}{1-z}x_D)p^+y_1^- } e^{-i\frac{z}{1-z}x_Dp^+y_2^-}\\
&-e^{i(x_B+x_L+  \frac{x_D}{1-z} )p^+y^-}e^{-ix_D p^+y_1^- }e^{-i\frac{z}{1-z}
x_Dp^+y_2^-} ]e^{-i\vec{k}_{\perp}\cdot (\vec{y}_{1\perp}-\vec{y}_{2\perp})}\\
&\times(-g^2) \theta(y_1^- - y_2^-)\theta(y^- - y_1^-).\\
\end{split}
\end{equation}
We have made variable change $\vec{k}  \rightarrow  -\vec{k}$ in the above.

\begin{figure*}[h] 
\centering

\subfloat[][Triple scattering 7a with $x= x_B+x_L$, $x_3 =\frac{z}{1-z}x_D$, $x_2 = \frac{z}{1-z}x_D$.]{
\begin{minipage}[t]{0.5\linewidth} 
\centering 
\includegraphics[scale=0.55]{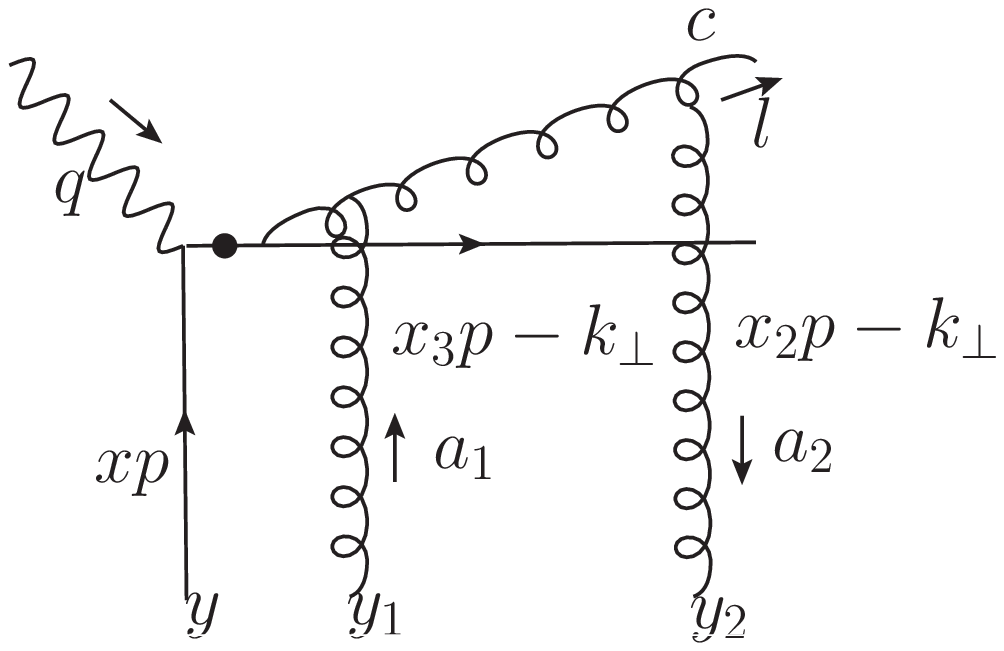} 
\end{minipage}
} 
\subfloat[][Triple scattering 7b with $x= x_B$, $x_3 =x_L+ \frac{z}{1-z}x_D$, $x_2 = \frac{z}{1-z}x_D$.]{
\begin{minipage}[t]{0.5\linewidth} 
\centering 
\includegraphics[scale=0.55]{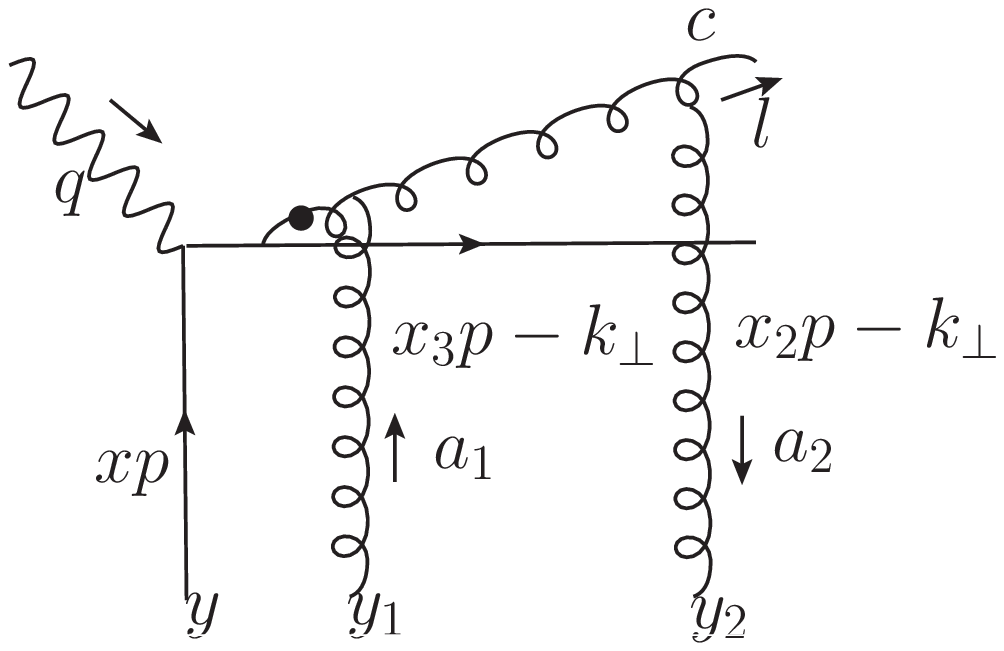} 
\end{minipage}
}
 \caption{Triple scattering 7.}
  \label{fig:triple scattering amplitude 7}
\end{figure*} 
(7) Triple scattering 7 in Fig.~\ref{fig:triple scattering amplitude 7}:
\begin{equation}
\begin{split}
M_{T7a}^{\nu}(y,y_1,y_2) = & \int \dfrac{dx}{2\pi} \delta(x-x_B-x_L)  \int dx_3 \delta(x_3 -\frac{z}{1-z}x_D) \int dx_2 \delta(x_2 - \frac{z}{1-z}x_D) \\
&\times \bar{u}^s(xp+q)\gamma^{\nu}u^{s'}(p) \overline{M}_{T7a}(y,y_1,y_2), \\
\overline{M}_{T7a}(y,y_1,y_2) = & 2g \dfrac{\vec{\epsilon}_{\perp}\cdot \vec{l}_{\perp}}{l_{\perp}^2}[T_{a_1},[T_{a_2},T_c]] e^{i(x_B+x_L)p^+y^-}e^{i\frac{z}{1-z}x_Dp^+y_1^- } e^{-i\frac{z}{1-z}x_Dp^+y_2^-}\\
&\times e^{-i\vec{k}_{\perp}\cdot (\vec{y}_{1\perp}-\vec{y}_{2\perp})}(-g^2) \theta(y_1^- - y_2^-)\theta(y^- - y_1^-),\\ 
M_{T7b}^{\nu}(y,y_1,y_2) = & \int \dfrac{dx}{2\pi} \delta(x-x_B )  \int dx_3 \delta(x_3 -x_L-\frac{z}{1-z}x_D ) \int dx_2 \delta(x_2 - \frac{z}{1-z}
x_D )\\
&\times \bar{u}^s(xp+q)\gamma^{\nu}u^{s'}(p) \overline{M}_{T7b}(y,y_1,y_2), \\
\overline{M}_{T7b}(y,y_1,y_2) = &- 2g \dfrac{\vec{\epsilon}_{\perp}\cdot \vec{l}_{\perp}}{l_{\perp}^2}[T_{a_1},[T_{a_2},T_c]] e^{ix_B p^+y^-}e^{i(x_L+\frac{z}{1-z}x_D) p^+y_1^- }e^{-i\frac{z}{1-z}
x_Dp^+y_2^-}\\
 &\times e^{-i\vec{k}_{\perp}\cdot (\vec{y}_{1\perp}-\vec{y}_{2\perp})}(-g^2)\theta(y_1^- - y_2^-)\theta(y^- - y_1^-),\\ 
\overline{M}_{T7}(y,y_1,y_2) = &\overline{M}_{T7a}(y,y_1,y_2) +   \overline{M}_{T7b}(y,y_1,y_2)  \\
=& 2g \dfrac{\vec{\epsilon}_{\perp}\cdot \vec{l}_{\perp}}{l_{\perp}^2}[T_{a_1},[T_{a_2},T_c]] [e^{i(x_B+x_L)p^+y^-}e^{i\frac{z}{1-z}x_Dp^+y_1^- } e^{-i\frac{z}{1-z}x_Dp^+y_2^-}\\
&-e^{ix_B p^+y^-}e^{i(x_L+\frac{z}{1-z}x_D) p^+y_1^- }e^{-i\frac{z}{1-z}
x_Dp^+y_2^-}]e^{-i\vec{k}_{\perp}\cdot (\vec{y}_{1\perp}-\vec{y}_{2\perp})}\\
&\times (-g^2) \theta(y_1^- - y_2^-)\theta(y^- - y_1^-).\\
\end{split}
\end{equation}
We have made variable change $\vec{k}  \rightarrow  -\vec{k}$ in the above.
 
The sum of triple scattering amplitudes is 
\begin{equation}
\begin{split}
\overline{M}_{T}(y,y_1,y_2) =& \overline{M}_{T1}(y,y_1,y_2)+ \overline{M}_{T2}(y,y_1,y_2)+\overline{M}_{T3}(y,y_1,y_2) + \overline{M}_{T4}(y,y_1,y_2)\\
&+\overline{M}_{T5}(y,y_1,y_2)+
\overline{M}_{T6}(y,y_1,y_2)+\overline{M}_{T7}(y,y_1,y_2).
\end{split}
\end{equation}

The hard partonic part of the hadronic tensor is
\begin{equation}
\begin{split}
H^{\mu\nu}_{D(1)q}= &  \int dx H_{(0)}^{\mu\nu} \int \frac{dz}{1-z} \int \frac{d l_{\perp}^2 }{2(2\pi)^2} \frac{1}{2} \frac{1}{N_c(N_c^2-1)} \sum_\text{spin,color} [\overline{M}_{D}(0,y_2) \overline{M}^{\dagger}_{D}(y,y_1)\\
& + \overline{M}_{T}(0,y_2,y_1) \overline{M}^{\dagger}_{S}(y) +  \overline{M}_{S}(0)\overline{M}^{\dagger}_{T}(y,y_1,y_2) ],
\end{split}
\end{equation}
where $\overline{M}_{D}(0,y_2) \overline{M}^{\dagger}_{D}(y,y_1)$ contains all central cut diagrams, while left cut diagrams are in $\overline{M}_{S}(0)\overline{M}^{\dagger}_{T}(y,y_1,y_2)$ and right cut diagrams in $\overline{M}_{T}(0,y_2,y_1) \overline{M}^{\dagger}_{S}(y)$.

 \end{widetext}

\bibliographystyle{apsrev4-1}
\bibliography{paper_GHT_PRD}
\end{document}